\documentclass[12pt]{iopart}
\pdfoutput=1
\eqnobysec
\usepackage{graphicx}
\begin{document}


\begin{center}
{\large{\it Rep. Prog. Phys.\/} {\bf 70} (2007) 2067-2148}
\end{center}

\vspace{-1.0cm}
\title[Symmetry breaking and quantum correlations]{Symmetry breaking 
and quantum correlations in finite systems: Studies of 
quantum dots and ultracold Bose gases and related nuclear and chemical methods}

\author{Constantine Yannouleas and Uzi Landman}

\address{School of Physics, Georgia Institute of Technology, Atlanta,
Georgia 30332-0430, USA} 
\ead{Constantine.Yannouleas@physics.gatech.edu}
\ead{Uzi.Landman@physics.gatech.edu}
\begin{abstract}
Investigations of emergent symmetry breaking phenomena occurring in small
finite-size systems are reviewed, with a focus on the strongly correlated 
regime of electrons in two-dimensional semicoductor quantum dots and trapped 
ultracold bosonic atoms in harmonic traps. Throughout the review we emphasize
universal aspects and similarities of symmetry breaking found in these systems, 
as well as in more traditional fields like nuclear physics and quantum 
chemistry, which are characterized by very different interparticle forces.
A unified description of strongly correlated phenomena
in finite systems of repelling particles (whether fermions or bosons) is 
presented through the development of a two-step method of symmetry breaking at
the unrestricted Hartree-Fock level and of subsequent symmetry restoration via 
post Hartree-Fock projection techniques. Quantitative and qualitative aspects
of the two-step method are treated and validated by exact diagonalization
calculations.

Strongly-correlated phenomena emerging from symmetry breaking include:\\
(I) Chemical bonding, dissociation, and entanglement (at zero and finite
magnetic fields) in quantum dot molecules and in pinned electron molecular 
dimers formed within a single anisotropic quantum dot,
with potential technological applications to solid-state quantum-computing
devices.\\
(II) Electron crystallization, with particle localization on the vertices of 
concentric polygonal rings, and formation of rotating electron molecules 
(REMs) in circular quantum dots. Such electron molecules exhibit ro-vibrational
excitation spectra, in analogy with natural molecules.\\
(III) At high magnetic fields, the REMs are described by parameter-free
analytic wave functions, which are an alternative to the Laughlin and 
composite-fermion approaches, offering a new point of view of the fractional
quantum Hall regime in quantum dots (with possible implications for the
thermodynamic limit).\\
(IV) Crystalline phases of strongly repelling bosons. In rotating traps and
in analogy with the REMs, such repelling bosons form rotating boson molecules 
(RBMs). For a small number of bosons, the RBMs are energetically favored 
compared to the Gross-Pitaevskii solutions describing vortex formation.

We discuss the present status concerning experimental signatures
of such strongly correlated states, in view of the promising outlook created
by the latest experimental improvements that are achieving
unprecedented control over the range and strength of interparticle 
interactions.  
\end{abstract}

\maketitle

\tableofcontents

\title[Symmetry breaking and quantum correlations]{}


\section{Introduction}

\subsection{Preamble}

Fermionic or bosonic particles confined in manmade devices, i.e., 
electrons in two-dimensional (2D) quantum dots (QDs, referred to also as
artificial atoms) or ultracold atoms in harmonic traps, can localize and 
form structures with molecular, or crystalline, characteristics. These 
molecular states of localized particles differ in an essential way from the 
electronic-shell-structure picture of delocalized electrons filling successive
orbitals in a central-mean-field potential (the Aufbau principle), 
familiar from the many-body theory of natural atoms and the Mendeleev
periodic table; they also present a different regime from that exhibited
by a Bose-Einstein condensate (BEC, associated often with the mean-field 
Gross-Pitaevskii equation). The molecular states originate from strong 
correlations between the constituent repelling particles and they are called 
electron (and often Wigner) or boson molecules. 

Such molecular states forming within a {\it single confining potential well\/}
constitute new phases of matter and allow for investigations of novel 
strongly-correlated phenomena arising in physical systems with a range of 
materials' characteristics unavailable experimentally (and theoretically 
unexplored) until recently. One example is the range of values of the socalled
Wigner parameter (denoted as $R_W$ for charged particles and $R_\delta$ for 
neutral ones, see \sref{wigparssb}) which expresses the relative strength of
the two-body repulsion and the one-particle kinetic energy, reflecting and
providing a measure of the strength of correlations in the system under study. 
For the two-dimensional systems which we discuss here, these values are often 
larger than the corresponding ones for natural atoms and molecules.

Other research opportunities offered by the quantum-dot systems are related to
their relatively large (spatial) size (arising from a small electron 
effective mass and large dielectric constant), which allows the full range of 
orbital magnetic effects to be covered for magnetic fields that are readily 
attained in the laboratory (less than 40 T). 
In contrast, for natural atoms and molecules, magnetic fields of sufficient 
strength (i.e., larger than $10^5$ T) to produce novel phenomena related to 
orbital magnetism (beyond the perturbative regime) are known to occur only in
astrophysical environments (e.g., on the surface of neutron stars) 
\cite{ruder}. For ultracold gases, a similar extraordinary physical regime can
be reached via the fast rotation of the harmonic trap.

In addition to the fundamental issues unveiled through investigations of
molecular states in quantum dots, these strongly-correlated states are of 
technological significance because of the potential use of manmade nanoscale 
systems for the implementation of qubits and quantum logic gates in quantum 
computers.

The existence of electron and boson molecules is supported by large-scale 
exact diagonalization (EXD) calculations, which provide the ultimate 
theoretical test. The discovery of these ``crystalline'' states has raised 
important fundamental aspects, including the nature of quantum phase 
transitions and the conceptual issues relating to spontaneous symmetry 
breaking (SSB) in small finite-size systems. 

The present report addresses primarily the physics, theoretical description,
and fundamental many-body aspects of molecular (crystalline) states
in small systems. For a comprehensive
description of the electronic-shell-structure regime (Aufbau-principle regime)
in quantum dots and of Bose-Einstein condensates in harmonic traps, see the 
earlier reviews by Kouwenhoven \etal \cite{kouw01} (QDs), 
Reimann and Manninen \cite{reim02} (QDs), Dalfovo \etal \cite{dalf99}
(BECs), and Leggett \cite{legg01} (BECs).    
Furthermore, in larger quantum dots, the symmetries of the external 
confinement that lead to shell structure are broken, and such dots exhibit
mesoscopic fluctuations and interplay between single-particle quantum chaos 
\cite{bohi90} and many-body correlations. For a 
comprehensive description of this mesoscopic regime in quantum dots, see 
the reviews by Beenakker \cite{been97} and Alhassid \cite{alha00}.

\subsection{Spontaneous symmetry breaking: confined geometries versus extended
systems}

Spontaneous symmetry breaking is a ubiquitous phenomenon in the 
macroscopic world. Indeed, there is an abundance of macroscopic systems and 
objects that are observed, or can be experimentally prepared, with effective 
many-body ground states whose symmetry is lower than the symmetry of the 
underlying many-body quantum-mechanical Hamiltonian; one says that in such 
cases the system lowers its energy through spontaneous symmetry breaking,
resulting in a state of lower symmetry and higher order. It is important to
stress that macroscopic SSB strongly suppresses quantum fluctuations and thus 
it can be described appropriately by a set of non-linear mean-field equations 
for the ``order parameter.'' The appearance of the order parameter is governed
by bifurcations associated with the non-linearity of the mean-field equations
and has led to the notion of ``emergent phenomena,'' a notion that helped
promote condensed-matter physics as a branch of physics on a par 
with high-energy particle physics (in reference to the fundamental nature of 
the pursuit in these fields; see the seminal paper by Anderson in Ref.\ 
\cite{and72}).

Our current understanding of the physics of SSB in the thermodynamic limit 
(when the number of particles $N \rightarrow \infty$) owes a great deal to the 
work of Anderson \cite{pwa}, who suggested that the broken-symmetry 
state can be safely taken as the effective ground state. In arriving at this
conclusion Anderson invoked the concept of (generalized) rigidity.
As a concrete example, one would expect a crystal to behave like a 
{\it macroscopic\/} body, whose Hamiltonian is that of a {\it heavy rigid 
rotor\/} with a low-energy excitation spectrum $L^2/2{\cal J}$ of
angular-momentum $(L)$ eigenstates, with the moment of
inertia ${\cal J}$ being of order $N$ (macroscopically large when 
$N \rightarrow \infty$). The low-energy excitation spectrum of this heavy 
rigid rotor above the ground-state ($L=0$) is essentially gapless (i.e., 
continuous). Thus although the formal ground state posseses continuous
rotational symmetry (i.e., $L=0$), ``there is a manifold of other states, 
degenerate in the $N \rightarrow \infty$ limit, which can be recombined to 
give a very stable wave packet with essentially the nature'' of the 
broken-symmetry state (see p 44 in Ref.\ \cite{pwa}). 

As a consequence of the ``macroscopic 
heaviness'' as $N \rightarrow \infty$, the relaxation of the system from the 
wave packet state (i.e., the broken-symmetry state) to the exact symmetrical 
ground state becomes exceedingly long. Consequently, in this limit, when
symmetry breaking occurs, there is practically no need to follow up with a
symmetry restoration step; that is the symmetry-broken state is admissible
as an effective ground state.

The present report addresses the much less explored question of symmetry
breaking in finite condensed-matter systems with a small number of particles.
For small systems, spontaneous symmetry breaking appears again at the level of
mean-field description [e.g., the Hartree-Fock (HF) level]. A major difference
from the $N \rightarrow \infty$ limit, however, 
arises from the fact that quantum fluctuations in 
small systems {\it cannot\/} be neglected. To account for the
large fluctuations, one has to perform a subsequent post-Hartree-Fock step
that restores the broken symmetries (and the linearity of the many-body
Schr\"{o}dinger equation). Subsequent to symmetry restoration,
the ground state obeys all the original 
symmetries of the many-body Hamiltonian; however, effects of the mean-field 
symmetry breaking do survive in the properties of the ground state of small 
systems and lead to emergent phenomena associated with formation of novel 
states of matter and with characteristic behavior in the excitation spectra. 
In the following, we will present an overview of the current understanding of 
SSB in small systems focusing on the essential theoretical aspects, as well as
on the contributions made by SSB-based approaches to the fast developing 
fields of two-dimensional semiconductor quantum dots and ultracold atomic 
gases in harmonic and toroidal traps.

\subsection{Historical background from nuclear physics and chemistry}

The mean field approach, in the form of the Hartree-Fock theory and of the
Gross-Pitaevskii (GP) equation, has been a useful tool in elucidating the 
physics of finite-size fermionic and bosonic systems, respectively. Its 
applications cover a wide range of systems, from natural atoms, natural 
molecules, and atomic nuclei, to metallic nanoclusters, and most recently  
two-dimensional quantum dots and ultracold 
gases confined in harmonic (parabolic) traps. Of
particular interest for the present review (due to spatial-symmetry-breaking
aspects) has been the mean-field description of deformed nuclei 
\cite{bm1,nil,naza94} and metal clusters \cite{cle,yl1995,yl99.b} (exhibiting 
ellipsoidal shapes). At a first level of description, deformation effects in 
these latter systems can be investigated via semi-empirical mean-field
models, like the particle-rotor model \cite{bm1} of Bohr and Mottelson
(nuclei), the anisotropic-harmonic-oscillator model of Nilsson
(nuclei \cite{nil} and metal clusters \cite{cle}), and the
shell-correction method of Strutinsky (nuclei \cite{str} and metal
clusters \cite{yl1995,yl99.b}). At the microscopic level, the mean field 
for fermions is often described \cite{rs,so}
via the self-consistent single-determinantal Hartree-Fock theory. At this
level, the description of deformation effects mentioned above
requires \cite{rs} consideration of unrestricted Hartree-Fock (UHF) wave
functions that break explicitly the rotational symmetries of the original
many-body Hamiltonian, but yield HF Slater determinants with lower energy 
compared to the symmetry-adapted restricted Hartree-Fock (RHF) 
solutions.\footnote[1]{
See in particular Ch 5.5 and Ch 11 in Ref.\ \cite{rs}.
However, our terminology (i.e., UHF vs. RHF) follows the practice in
quantum chemistry (see Ref.\ \cite{so}).}

In earlier publications \cite{yl99,yl00.1,yl01,yl02.1,yl02.2,yl03.1,yl03.2}, 
we have shown that, in the strongly correlated regime, UHF solutions that 
violate the rotational (circular) symmetry arise most naturally in the case of 
two-dimensional single quantum dots, for both the 
cases of zero and high magnetic field; for a UHF 
calculation in the lowest Landau level (LLL), see also Ref.\ \cite{mk96}. 
Unlike the case of atomic nuclei, however, where (due to the attractive 
interaction) symmetry breaking is associated primarily with quadrupole shape 
deformations (a type of Jahn-Teller distortion), spontaneous symmetry breaking
in 2D quantum dots induces electron localization 
(or ``crystallization'') associated 
with formation of {\it electron\/}, or {\it Wigner\/}, {\it molecules\/}). 
The latter name is used in honor of Eugene Wigner who predicted the formation 
of a classical {\it rigid\/} Wigner crystal for the 3D electron gas at very 
low densities \cite{wign}. We stress, however, that because of the finite 
size, Wigner molecules are most often expected 
to show a physical behavior quite different from the classical Wigner crystal.
Indeed, for finite $N$, Wigner molecules exhibit 
analogies closer to natural molecules, and the Wigner-crystal limit is expected
to be reached only for special limiting conditions.

For a small system the violation in the mean-field approximation of the 
symmetries of the original many-body Hamiltonian appears to be paradoxical at 
a first glance, and some times it has been described mistakenly as an 
``artifact'' (in particular in the context of 
density-functional theory \cite{hirwin}).
However, for the specific cases arising in Nuclear Physics and Quantum
Chemistry, two theoretical developments had already resolved this
paradox. They are: (1) the theory of restoration of broken symmetries via
projection techniques\footnote[2]{
For the restoration of broken rotational symmetries in atomic nuclei, see
Ref.\ \cite{py} and Ch 11 in Ref.\ \cite{rs}. For the restoration of broken 
spin symmetries in natural 3D molecules, see Ref.\ \cite{low55}.}
\cite{py,low55,low64}, and (2) the group theoretical analysis of
symmetry-broken HF orbitals and solutions in chemical reactions, initiated by
Fukutome and coworkers \cite{fuk} who used the symmetry groups associated with
the natural 3D molecules.
Despite the different fields, the general principles established in these
earlier theoretical developments in nuclear physics and quantum chemistry
have provided a wellspring of assistance in our investigations of symmetry 
breaking for electrons in quantum dots and bosons in harmonic traps. 
In particular, the
restoration of broken symmetries in QDs and ultracold atomic traps via 
projection techniques constitutes a main theme of the present report.

The theory of restoration of broken symmetries has been developed into a 
sophisticated computational approach in modern nuclear physics. Using the 
broken-symmetry solutions of the Hartree-Fock-Bogoliubov 
theory\footnote[3]{See Ch 7 in Ref.\ \cite{rs}.}
(that accounts for nuclear pairing and superfluidity), this approach has been 
proven particularly efficient in describing the competition between shape 
deformation and pairing in nuclei. For some recent papers in nuclear physics,
see, e.g., Refs.\ \cite{egid05,schm04,floc03,naza03,bend03,rein96}; 
for an application to 
superconducting metallic grains, see Ref.\ \cite{fern03}. Pairing effects 
arise only in the case of attractive interactions and they are not considered 
in this report, since we deal only with repulsive two-body interactions.

\subsection{Scope of the review} 

Having discussed earlier the general context and historical background
from other fields regarding symmetry breaking, we give here an outline of 
the related methodologies and of the newly discovered strongly correlated 
phenomena that are discussed in this report in the area of condensed-matter 
nanosystems.

In particular, a two-step method \cite{yl99,yl00.1,yl01,yl02.1,yl02.2,yl03.1} 
of {\it symmetry breaking\/} at the unrestricted
Hartree-Fock level and of subsequent post-Hartree-Fock {\it restoration 
of the broken symmetries\/} via projection techniques is reviewed for the case
of two-dimensional (2D) semiconductor quantum dots and ultracold bosons 
in rotating traps with a small number ($N$) of particles. The general 
principles of the two-step method can be traced to nuclear theory (Peierls and
Yoccoz, see the original Ref.\ \cite{py}, but also the recent Refs.\ 
\cite{egid05,schm04,floc03,naza03,bend03,rein96}) 
and quantum chemistry (L\"{o}wdin, see 
Ref.\ \cite{low55}); in the context of condensed-matter nanophysics and the 
physics of ultracold atomic gases, it constitutes a novel powerful
many-body approach that has led to unexpected discoveries in the area of 
strongly correlated phenomena. The successes of the method have generated a 
promising theoretical outlook, bolstered by the unprecedented experimental and
technological advances, pertaining particularly to control of system parameters
(most importantly of the strength and variety of two-body interactions), that 
can be achieved in manmade nanostructures. 

In conjunction with exact diagonalization 
calculations \cite{yl03.2,elle06,yl06.tnt,yl06.nac,ihn06} and recent 
experiments \cite{elle06,ihn06,zumb04}, it is shown that the two-step method 
can describe a wealth of novel strongly correlated phenomena in quantum dots 
and ultracold atomic traps. These include:

(I) Chemical bonding, dissociation, and entanglement in quantum dot molecules
\cite{yl99,yl01,yl02.3} and in electron molecular dimers formed within 
a single elliptic QD \cite{elle06,yl06.tnt,yl06.nac,ihn06}, 
with potential technological 
applications to solid-state quantum logic gates \cite{loss98,burk99,tayl05}.

(II) Electron crystallization, with localization on the vertices of concentric
polygonal rings, and formation of rotating electron molecules (REMs) in 
circular QDs. At zero magnetic field ($B$), the REMs can approach the limit of
a rigid rotor \cite{yl00.2,yl04.1}; at high $B$, the REMs  are highly floppy 
and ``supersolid''-like, that is, they exhibit \cite{yl04.1,yl04.2,li06} a 
{\it non-rigid\/} rotational inertia \cite{legg04}, with the rings rotating 
independently of each other \cite{yl04.2,li06}.

(III) At high magnetic fields and under the restriction of the many-body 
Hilbert space to the lowest Landau level, the two-step method yields 
fully analytic many-body wave functions \cite{yl02.2,yl03.2}, which are an 
alternative to the Jastrow/Laughlin (JL) \cite{laug83.1} and composite-fermion
(CF) \cite{jain89,jain90} approaches, offering a new point of view
of the fractional quantum Hall regime (FQHE) \cite{fqhe1,jainbook} in 
quantum dots (with possible implications for the thermodynamic limit). 

Large scale exact-diagonalization calculations \cite{yl03.2,yl04.2,li06}
support the results of the two-step method outlined in items II and III above.

(IV) The two-step method has been used \cite{roma04} to discover 
crystalline phases of strongly repelling ultracold bosons (impenetrable 
bosons/ Tonks-Girardeau regime \cite{tonks1936,gira60}) in 2D harmonic traps. 
In the case of rotating traps, such repelling bosons form rotating boson 
molecules (RBMs) \cite{roma06} that are energetically favorable compared to 
the Gross-Pitaevkii solutions, even for weak repulsion and, in particular, in 
the regime of GP vortex formation.

We will not discuss in this report specific applications of the two-step 
method to atomic nuclei. Rather, as the title conveys, the report aims at 
exploring the universal characteristics of quantum correlations arising from
symmetry breaking across various fields dealing with small finite systems, 
such as 2D quantum dots, trapped ultracold atoms, and nuclei -- and even 
natural 3D molecules. Such universal characteristics and similarities in 
related methodologies persist across the aforementioned fields in spite of the 
differences in the size of the physical systems and in the range, nature, and 
strength of the two-body interactions. For specific applications to atomic 
nuclei, the interested reader is invited to consult the nucler physics 
literature cited in this report.

\begin{figure*}[t]
\centering\includegraphics[width=12.0cm]{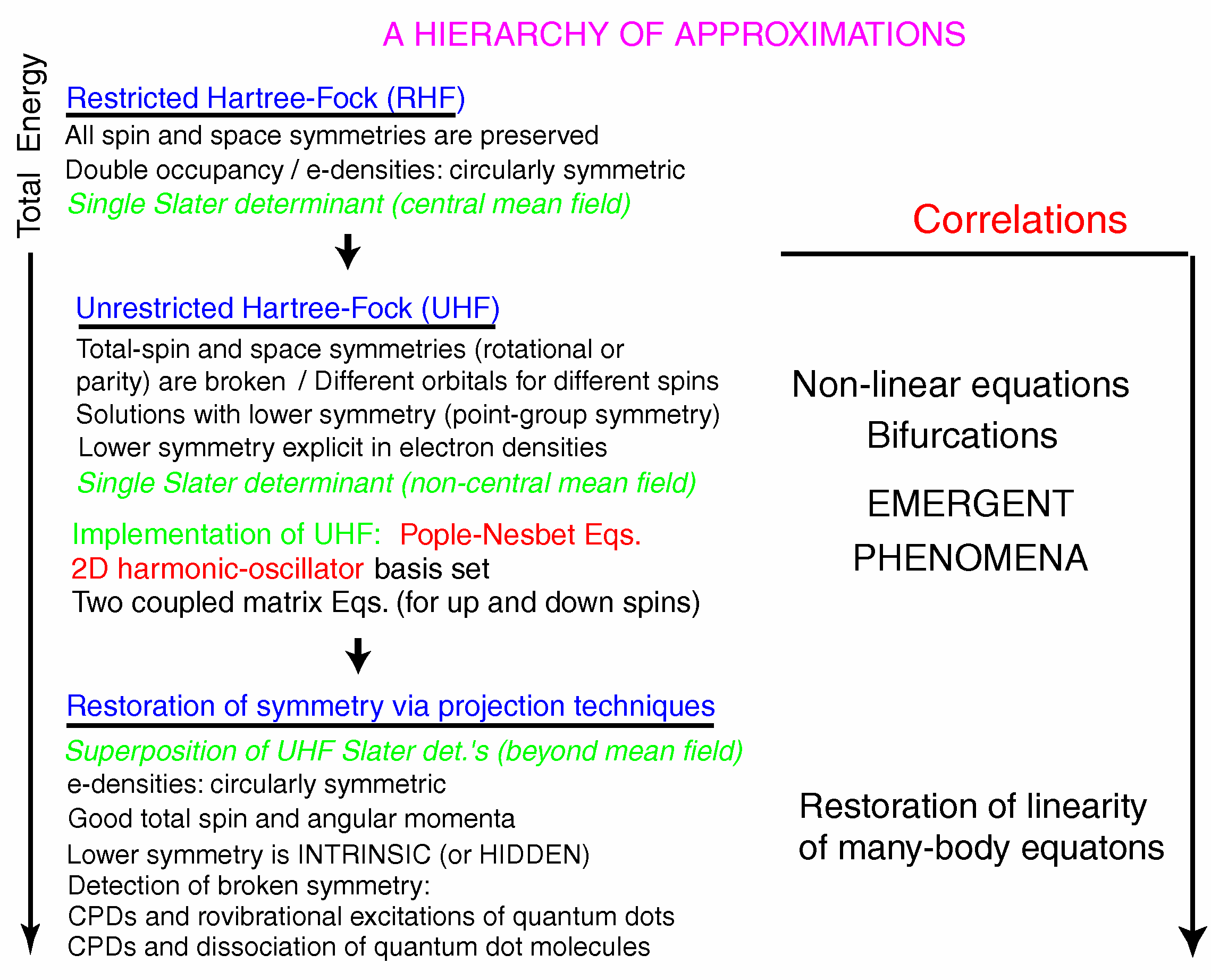}
\caption{(Color online) Synopsis of the method of hierarchical 
approximations (also referred to as the ``two-step method,'' emphasizing that 
symmetry breaking at the mean-field level must be accompanied by a subsequent 
post-Hartree-Fock step of symmetry restoration, with a subsequent further 
lowering of the energy). See text for a detailed description.   
}
\label{wdown}
\end{figure*}

\subsection{Using a hierarchy of approximations versus probing of exact 
solutions}
\label{synopsis}

\Fref{wdown} presents a synopsis
of the hierarchy of approximations associated with the two-step method, and in 
particular for the case of 2D quantum dots. 
(A similar synopsis can also be written for
the case of bosonic systems.) This method produces approximate wave functions 
with lower energy at each approximation level (as indicated by the downward 
vertical arrow on the left of the figure). 

At the lowest level of approximation (corresponding to higher energy with no
correlations included), one places the restricted Hartree-Fock, whose main 
restriction is the double occupancy (up and down spins) of each space orbital.
The many-body wave function is a single Slater determinant associated with a 
``central mean field.'' The RHF preserves all spin and space symmetries. For 
2D quantum dots, the single-particle density 
[also referred to as electron density ($e$-density)] is circularly symmetric.

The next approximation involves the unrestricted Hartree-Fock, which employs 
different space orbitals for the two different spin directions. The UHF 
preserves the spin projection, but allows the total-spin and space symmetries 
(i.e., rotational symmetries or parity) to be broken. The broken symmetry 
solutions, however, are not devoid of any symmetry; they exhibit 
characteristic lower symmetries (point-group symmetries) that are explicit in 
the electron densities. The UHF many-body wave function is a single Slater 
determinant associated with a ``non-central mean field.''

Subsequent approximations aim at restoring the broken symmetries via projection
techniques. The restoration-of-symmetry step goes beyond the mean field 
approximation and it 
provides a many-body wave function $|\Phi^{\rm PRJ}\rangle$ that is a linear 
superposition of Slater determinants (see detailed description in 
\sref{restsym} below). The projected (PRJ) many-body wave function 
$|\Phi^{\rm PRJ}\rangle$ 
preserves all the symmetries of the original many-body Hamiltonian; it
has good total spin and angular momentum quantum numbers, and as a
result the circular symmetry of the electron densities is restored. 

However, the lower (point-group) spatial symmetry found at the broken-symmetry
UHF level (corresponding to the first step in this method) does not disappear. 
Instead, it becomes {\it intrinsic\/} or {\it hidden\/}, and it can be 
revealed via an inspection of conditional probability distributions (CPDs), 
defined as (within a proportionality constant)
\begin{equation}
P({\bf r},{\bf r}_0) =
\langle \Phi^{\rm PRJ} |
\sum_{i \neq j}  \delta({\bf r}_i -{\bf r})
\delta({\bf r}_j-{\bf r}_0)
| \Phi^{\rm PRJ} \rangle,
\label{cpds}
\end{equation}
where $\Phi^{\rm PRJ} ({\bf r}_1, {\bf r}_2, \ldots, {\bf r}_N)$
denotes the projected many-body wave function under consideration.

If one needs to probe the intrinsic spin distribution of the localized 
electrons, one has to consider spin-resolved two-point correlation functions
(spin-resolved CPDs), defined as
\begin{equation}
P_{\sigma\sigma_0}({\bf r}, {\bf r}_0)=  \langle \Phi^{\rm{PRJ}} |
\sum_{i \neq j} \delta({\bf r} - {\bf r}_i) \delta({\bf r}_0 - {\bf r}_j)
\delta_{\sigma \sigma_i} \delta_{\sigma_0 \sigma_j}
|\Phi^{\rm{PRJ}}\rangle.
\label{sponcpd}
\end{equation}
The spin-resolved CPD gives the spatial probability distribution of
finding a second electron with spin projection $\sigma$ under the condition 
that a first electron is located (fixed) at ${\bf r}_0$ with spin projection
$\sigma_0$; $\sigma$ and $\sigma_0$ can be either up $(\uparrow$) or
down ($\downarrow$). The meaning of the space-only CPD in \eref{cpds} is 
analogous, but without consideration of spin.

Further signatures of the intrinsic lower symmetry occur in the excitation 
spectra of circular quantum dots that exhibit ro-vibrational character
related to the intrinsic molecular structure, or in the dissociation of quantum
dot molecules.

As the scheme in \fref{wdown} indicates, the mean-field HF 
equations are non-linear
and the symmetry breaking is associated with the appearance of bifurcations
in the total HF energies. The occurrence of such bifurcations cannot be
predicted {\it a priori\/} from a mere inspection of the many-body Hamiltonian
itself; it is a genuine many-body effect that belongs to the class of 
so-called emergent phenomena \cite{and72,laug99,land05} that may be revealed 
only through the solutions of the Hamiltonian themselves (if obtainable) or 
through experimental signatures. We note that the step of symmetry restoration 
recovers also the linear properties of the many-body Schr\"{o}dinger equation. 

The relation between quantum correlations and the two-step method (also
called the method 
of hierarchical approximations) is portrayed by the downward vertical arrow 
on the right of \fref{wdown}. Indeed, the correlation energy is defined 
\cite{low59} as the difference between the restricted Hartree-Fock and exact 
ground-state energies, i.e.,
\begin{equation}
E_{\rm corr}=E_{\rm RHF} - E_{\rm EXD}.
\label{ecorr}
\end{equation}
As seen from \fref{wdown}, starting with the broken-symmetry UHF solution, 
each further approximation captures successively a larger fraction of the 
correlation energy \eref{ecorr}; a specific example of this process is given 
in \fref{succapp} below (in \sref{restsym}). 
   
An alternative approach for studying the emergence of crystalline structures 
is the exact-diagonalizaion method that will be discussed in detail in 
\sref{exdmeth}. Like the projected wave functions, the EXD many-body wave 
functions preserve of course
all the symmetries of the original Hamiltonian. As a result, the intrinsic,
or hidden, point-group symmetry associated with particle localization and 
molecule formation is not explicit, but it is revealed through inspection of 
CPDs [one simply uses the exact-diagonalization wave function 
$\Phi^{\rm EXD}({\bf r}_1,{\bf r}_2, \ldots, {\bf r}_N)$ in \Eref{cpds}]
and \Eref{sponcpd}, or recognized via characteristic trends
in the calculated excitation spectra. When feasible, the EXD results provide
a definitive answer in terms of numerical accuracy, and as such they serve as 
a test to the results obtained through approximation methods (e.g., the above 
two-step method). However, the underlying physics of electron or boson
molecule formation is less transparent when analyzed with the 
exact-diagonalization method compared to the two-step approach. Indeed, many 
exact-diagonalization studies of 2D quantum dots and
trapped bosons in harmonic traps have focused simply on providing
high accuracy energetics and they omitted calculation of CPDs. However, the 
importance of using CPDs as a tool for probing the many-body wave functions
cannot be overstated. For example, while exact-diagonalization
calculations for {\it bosons\/} in the 
lowest Landau level have been reported rather
early \cite{wilk00,vief00,wilk01,regn03,regn05}, the analysis in these studies
did not include calculations of the CPDs, and consequently formation of
rotating boson molecules and particle ``crystallization'' was not recognized
(for further discussion of these issues, see Romanovsky \etal 
\cite{roma04,roma06} and Baksmaty \etal \cite{baks07}).

From the above, it is apparent that both methods, i.e., the two-step method
and the exact-diagonalization one, complement each other, and it is in this 
spirit that we use them in this report.

\subsection{Experimental signatures of quantum correlations}

Historically, the isolation of a small number ($N < 20$) of electrons down to
a single electron was experimentally realized in the so-called ``vertical'' 
quantum dots \cite{kouw01}. The name vertical QDs derives from the fact that 
the leads and voltage gates are located in a vertical arrangement, on top and 
below the two-dimensional dot. At zero magnetic field, experimental 
measurements \cite{kouw01,kouw96} of addition energies,
\begin{equation}
\Delta^2 E_N = \mu_{N+1}-\mu_{N},
\end{equation}
where the chemical potential $\mu_N=E_{N}-E_{N-1}$, indicated that correlation 
effects at zero and low $B$ are rather weak in such dots, a property 
that later was attributed to the strong screening of the Coulomb interaction 
in these devices. The measured addition energies
exhibited maxima at closed electronic shells ($N=2,6,12,\ldots$) and at 
mid-shells ($N=4,9,\ldots$) in agreement with a 2D-harmonic-oscillator 
central-mean-field model and the Hund's rules, and in analogy with the Aufbau
principle and the physics of natural 3D atoms. 
It was found that the measured {\it ground-state\/} energy 
spectra for low magnetic fields could be understood on the basis of a simple 
``constant-interaction'' model where the effect of the two-body 
Coulomb interaction is reduced phenomenologically to an overall classical 
capacitance, $C$, characterizing the charging energy $Z^2 e^2/(2C)$ of the
quantum dot.

As a result of screening, strong correlation effects and formation of Wigner 
molecules can be expected to occur in vertical 
dots particularly under the influence
of high magnetic fields. Evidence about the formation of Wigner molecules in
vertical quantum dots has been provided recently in Ref.\ \cite{maks06}, where 
measured ground-state spectra as a function of $B$ for $N=3e$ and $N=4e$ were 
reanalyzed with exact-diagonalization
calculations that included screening. At the time
of submission of this report, a second ground-state crossing at high $B$ due 
to strong correlations was also demonstrated experimentally in a 
two-electron vertical quantum dot with an external confinement 
that was smaller than the previously used ones \cite{nish07}. 

Early theoretical work \cite{yl99} at zero magnetic field using simply the 
symmetry broken UHF solutions suggested that an {\it unscreened\/} Coulomb 
repulsion may result in a violation of Hund's rules. However, following the
two-step method of Refs.\ \cite{yl99,yl00.1,yl01,yl02.1,yl02.2,yl03.1}, it has
been shown \cite{kram07} most recently that the companion step of symmetry 
restoration recovers the Hund's rules in the case of $N=4e$.      

In addition, the $B=0$ results of Ref.\ \cite{yl99} suggested that both the 
maxima of the addition energies at closed shells and at mid-shells become 
gradually weaker (and they eventually disappear) as the strength of the 
Coulomb interaction (and consequently the strength of correlations) 
increases, leading to formation of ``strong'' Wigner molecules.
The qualitative trend of formation of strong 
Wigner molecules obtained from a relatively 
simple UHF calculation at $B=0$ has been confirmed later by more accurate EXD
\cite{yl00.2,ront06} and quantum Monte Carlo \cite{umri06} calculations, as 
well as through symmetry restoration calculations \cite{yl02.1,kram07}, 
although its experimental demonstration remains still a challenge.

A more favorable experimental configuration for the development and 
observation of strong interelectron correlations is the so-called 
``lateral'' dot, where the leads and gates are located on the sides of the 
dot and thus screening effects are reduced. Tunability of these dots down to a
single electron has been achieved only in the last few years \cite{cior00}.
Most recently, continually improving experimental techniques have allowed 
precise measurements of excitation spectra of $2e$ lateral (and anisotropic) 
quantum dots at zero and low magnetic fields 
\cite{elle06,zumb04,kyri02}. As discussed 
in detail in \sref{stcor2e}, the behavior of these excitation spectra
\cite{elle06,zumb04} as a function of $B$ provides unambiguous signatures for 
the presence of strong correlations and the formation of Wigner molecules.

Experimentally observed behavior of two electrons in lateral 
{\it double\/} QDs \cite{taru05} provides further evidence for strong 
correlation phenomena. Indeed, instead of successively populating delocalized 
states over both QDs according to a molecular-orbital scheme, the two 
electrons localize on the individual dots according to a Heitler-London 
picture \cite{hl}. Theoretically, such strongly correlated phenomena in double
quantum dots were described in Refs.\ \cite{yl99,yl01,yl02.3}; 
see \sref{uhfelpu} below.

Correlations are expected to influence not only the spectral properties of
quantum dots, but also to effect transport characteristics. Indeed  
correlation effects may underlie the behavior of the transmission 
amplitudes (magnitude and phase) of an electron tunneling through a quantum
dot. Such transmission measurements have been performed using 
Aharonov-Bohm interferometry \cite{heim05}, and an interpretation involving
strongly correlated states in the form of Wigner molecules has been
proposed recently \cite{gurv07}. The quantity that links transport experiments
with many-body theory of electrons in QDs is the overlap between many-body
states with $N-1$ and $N$ electrons, i.e., 
$\langle \Phi(N-1)|c_j|\Phi(N) \rangle$, where $c_j$ annihilates the $j$th
electron.

The strength of correlations in quantum dots at 
zero $B$ can be quantified by the 
Wigner parameter $R_W$, which is the ratio between the strength of the Coulomb 
repulsion and the one-electron kinetic energy  
(see \sref{wigparssb}). Naturally, for the case of neutral repelling bosons, 
the corresponding parameter is the ratio between
the strength of the contact interaction and the one-particle kinetic 
energy in the harmonic trap, and it is denoted as $R_\delta$. Larger
values of these parameters ($R_W$ or $R_\delta$) result in stronger 
correlation effects. 

Progress in the ability to experimentally control the above parameters has been
particularly impressive in the case of ultracold trapped bosons. Indeed,
realizations of continuous tunability of $R_\delta$ over two orders of
magnitude (from 1 to 5 \cite{wei} and from 5 to 200 \cite{par}) has 
been most recently reported in quasi-linear harmonic traps. Such high values 
of $R_\delta$ allowed experimental realization of novel strongly correlated 
states drastically different from a Bose-Einstein condensate.
This range of high values of $R_\delta$ is known as the Tonks-Girardeau regime
and the corresponding states are one-dimensional analogues of molecular
structures made out of localized bosons. In two dimenional traps, it has
been predicted that such large values of $R_\delta$ lead to the emergence of
crystalline phases \cite{roma04,roma06}.

The high experimental control of optical lattices has also been exploited for
the creation \cite{grei} of novel phases of ultracold bosons analogous to Mott 
insulators; such phases are related to the formation of electron puddles 
discussed in \sref{uhfelpu} and to the fragmentation of Bose-Einstein 
condensates \cite{muel06}.
 
\subsection{Plan of the report}

The plan of the report can be visualized through the table of contents. 
Special attention has been given to the Introduction, which offers a general
presentation of the subject of symmetry breaking and quantum correlations in 
confined geometries -- including a discussion of the differences with the case
of extended systems, a historical background from other fields, and a 
diagrammatic synopsis of the two-step method of symmetry breaking/symmetry 
restoration.  

The theoretical framework and other technical methodological background are 
presented in Section 2 (symmetry breaking/symmetry restoration in quantum 
dots), Section 3 (symmetry breaking/symmetry restoration for trapped ultracold
bosons), and Section 4 (exact-diagonalization approaches). Section 4
includes also a commentary on quantum Monte Carlo methods.    

For the case of semiconductor quantum dots, the main results and description
of the strongly correlated regime are presented in Sections 5, 6, and 7, with
Section 5 focusing on the case of two electrons and its historical 
significance. Section 8 is devoted to a description of the strongly-correlated 
regime of trapped repelling bosons.

Finally, a summary is given in Section 9, and the Appendix offers an outline of
the Darwin-Fock single-particle spectra for a two-dimensional isotropic 
oscillator under a perpendicular magnetic field or under rotation.  

We note that the sections on trapped bosons (Section 3 and Section 8) can be 
read independently from the sections on quantum dots. 


\section{Symmetry breaking and subsequent symmetry
restoration for electrons in confined geometries: Theoretical framework}
\label{symbreak}

The many-body Hamiltonian describing $N$ electrons confined in a 
two-dimensional QD and interacting via a Coulomb repulsion is written as
\begin{equation}
{\cal H} =\sum_{i=1}^N H(i) +
\sum_{i=1}^N \sum_{j>i}^N \frac{e^2}{\kappa r_{ij}}.
\label{mbhn}
\end{equation}
In \Eref{mbhn}, $\kappa$ is the dielectric constant of the semiconducting
material and $r_{ij}=|{\bf r}_i - {\bf r}_j|$. The single-particle Hamiltonian
in a perpendicular external magnetic field $B$ is given by
\begin{equation}
H=\frac{({\bf p}-e{\bf A}/c)^2}{2m^*} + V(x,y) +
\frac{g^* \mu_B}{\hbar} {\bf B \cdot s},
\label{hsp}
\end{equation}
where the external confinement is denoted by $V(x,y)$, the
vector potential ${\bf A}$ is given in the symmetric gauge by
\begin{equation}
{\bf A}({\bf r})=\frac{1}{2}{\bf B} \times {\bf r} =\frac{1}{2}(-By,Bx,0),
\label{vectp}
\end{equation}
and the last term in \eref{hsp} is the Zeeman interaction with $g^*$
being the effective Land\'{e} factor, $\mu_B$ the Bohr magneton, ${\bf s}$
the spin of an individual electron and $m^*$ is the effective electron mass. 
The external potential confinement $V(x,y)$ can assume various parametrizations
in order to model a single circular or elliptic quantum dot, or a quantum dot
molecule. Of course, in the case of an elliptic QD, one has
\begin{equation}
V(x,y) = \frac{1}{2} m^* (\omega_x^2 x^2 + \omega_y^2 y^2),
\label{vxyell}
\end{equation}
which reduces to the circular QD potential when $\omega_x=\omega_y=\omega_0$.
The appropriate parametrization of $V(x,y)$ in the case of a double QD is
more complicated. In our work, we use a parametrization based on a 2D
two-center oscillator with a smooth necking. This latter parametrization
is described in detail in Refs.\ \cite{yl02.1,yl02.3}, where readers 
are directed for further details. In contrast with other
parametrizations based on two displaced inverted Gaussians \cite{melni06}, the 
advantage of the two-center oscillator is that the height of the interdot
barrier, the distance between the dots, the ellipticity of each dot, and
the gate potentials of the two dots (i.e., the relative potential wells in the
neighboring dots) can be varied independently of each other.  

A prefactor multiplying the Coulomb term in \Eref{mbhn} (being either an 
overall constant $\gamma$ as in \sref{2eelldot} below, or having an appropriate
position-dependent functional form \cite{yl06.tnt,yl06.nac}) is used to 
account for the reduction of the Coulomb interaction due to the finite 
thickness of the electron layer and to additional screening (beyond
that produced by the dielectric constant of the material) arising from the
formation of image charges in the gate electrodes \cite{hall96}.

\subsection{Mean-field description and unrestricted Hartree-Fock} 

Vast literature is available concerning mean-field studies of electrons in
quantum dots. Such publications are divided mainly into 
applications of density functional theory 
\cite{reim02,kosk97,mart98,yann99,lebu02,nie04,borg05} 
and the use of Hartree-Fock methods 
\cite{yl99,yl03.1,mk96,yann99,pfan93,fuji96,bedn99,ront99,gra01,
sba03,szaf04,lipp06}.
The latter include treatments according to the restricted Hartree-Fock
\cite{pfan93}, unrestricted Hartree-Fock with spin, but not space, symmetry 
breaking \cite{fuji96,bedn99,ront99}, unrestricted Hartree-Fock with spin
and/or space symmetry breaking \cite{yl99,yl03.1,mk96,yann99,gra01,sba03,
szaf04}, and the so-called Brueckner Hartree-Fock \cite{lipp06,lippbook}.

From the several Hartree-Fock variants mentioned above, only the UHF with
consideration of both spin and space symmetry unrestrictions has been
able to describe formation of Wigner molecules, and in the following we will 
exclusively use this unrestricted version of Hartree-Fock theory.
The inadequacy of the density-functional theory in describing Wigner
molecules will be discussed in \sref{symdft}.

\subsubsection{The self-consistent Pople-Nesbet equations.}
\label{selfpopnes}

The unrestricted Hartree-Fock equations used by us are an adaptation of the
Pople-Nesbet \cite{pn} equations described in detail in Ch 3.8 of Ref.\ 
\cite{so}. For completeness, we present here a brief description of these
equations, along with pertinent details of their computational implementation 
by us to the 2D case of semiconductor QDs.

We start by requesting that the unrestricted Hartree-Fock many-body 
wave function for $N$ electrons is represented by a single Slater determinant 
\begin{equation}
\Psi_{\rm UHF} ({\bf x}_1, \ldots, {\bf x}_N) = \frac{1}{\sqrt{N!}}
{\rm det}[\chi_1({\bf x}_1), \chi_2({\bf x}_2), \ldots, \chi_N({\bf x}_N],
\label{psiuhf}
\end{equation}
where $[\chi_i({\bf x})]$ are a set of $N$ spin orbitals, with the index 
${\bf x}$ denoting both the space and spin coordinates. 
Furthermore, we take $\chi_i({\bf x})=\psi_i({\bf r})\alpha$ for a spin-up 
electron and $\chi_i({\bf x})=\psi_i({\bf r})\beta$ for a spin-down
electron. As a result, the UHF determinants in this report are eigenstates
of the projection of the total spin with eigenvalue $S_z=(N^\alpha-N^\beta)/2$,
where $N^{\alpha(\beta)}$ denotes the number of spin up (down) electrons.
However, these Slater determinants are not eigenstates of the square of the 
total spin, ${\bf S}^2$, except in the fully spin polarized case.

According to the variational principle, the best spin orbitals must
minimize the total energy $\langle \Psi_{\rm UHF} | {\cal H} | \Psi_{\rm UHF}
\rangle$. By varying the spin orbitals $[\chi_i({\bf x})]$ under the
constraint that they remain {\it orthonormal\/}, one can derive the UHF 
Pople-Nesbet equations described below.

A key point is that electrons with $\alpha$ (up) spin will be described by one 
set of spatial orbitals $\{ \psi^\alpha_j| j=1,2,\dots,K\}$, while electrons with
$\beta$ (down) spin are described by a different set of spatial orbitals
$\{ \psi^\beta_j| j=1,2,\ldots,K\}$; of course in the restricted Hartree-Fock
$\psi^\alpha_j = \psi^\beta_j=\psi_j$. Next, one introduces a set of basis 
functions
$\{ \varphi_\mu| \mu=1,2,\ldots,K\}$ (constructed to be
{\it orthonormal\/} in our 2D case), and expands the UHF orbitals as
\begin{equation}
\psi^\alpha_i = \sum_{\mu=1}^K C_{\mu i}^\alpha \varphi_\mu,~~~i=1,2,\ldots,K,
\label{expa}
\end{equation}
\begin{equation}
\psi^\beta_i = \sum_{\mu=1}^K C_{\mu i}^\beta \varphi_\mu,~~~i=1,2,\ldots,K.
\label{expb}
\end{equation}

The UHF equations are a system of two coupled matrix eigenvalue problems 
resolved according to up and down spins,
\begin{equation}
{\bf F}^{\alpha\beta} {\bf C}^\alpha = {\bf C}^\alpha {\bf E}^\alpha
\label{uhfa}
\end{equation}
\begin{equation}
{\bf F}^{\beta\alpha} {\bf C}^\beta = {\bf C}^\beta {\bf E}^\beta,
\label{uhfb}
\end{equation}
where ${\bf F}^{\alpha\beta(\beta\alpha)}$ are the Fock-operator matrices and
${\bf C}^{\alpha(\beta)}$ are the vectors formed with the coefficients in the
expansions \eref{expa} and \eref{expb}. The matrices
${\bf E}^{\alpha(\beta)}$ are {\it diagonal\/}, and as a result equations
(\ref{uhfa}) and (\ref{uhfb}) are {\it canonical\/} (standard). Notice that
noncanonical forms of HF equations are also possible (see Ch 3.2.2 of Ref.\
\cite{so}). Since the self-consistent iterative solution of the HF
equations can be computationally implemented only in their canonical form,
canonical orbitals and solutions will always be implied, unless
otherwise noted explicitly. We note that the coupling between the two UHF
equations (\ref{uhfa}) and (\ref{uhfb}) is given explicitly in the
expressions for the elements of the Fock matrices below [\eref{famn}
and \eref{fbmn}].

Introducing the density matrices ${\bf P}^{\alpha(\beta)}$ for $\alpha(\beta)$
electrons,
\begin{equation}
P^\alpha_{\mu\nu} = \sum_{a}^{N^\alpha}
C^\alpha_{\mu a} (C^\alpha_{\nu a})^*
\label{ppa}
\end{equation}
\begin{equation}
P^\beta_{\mu\nu} = \sum_{a}^{N^\beta}
C^\beta_{\mu a} (C^\beta_{\nu a})^*,
\label{ppb}
\end{equation}
where $N^\alpha + N^\beta = N$,
the elements of the Fock-operator matrices are given by
\begin{equation}
F^{\alpha\beta}_{\mu \nu} = H_{\mu \nu} +
\sum_\lambda \sum_\sigma P^\alpha_{\lambda \sigma}
[(\mu \sigma | \nu \lambda ) - (\mu \sigma | \lambda \nu)] 
+ \sum_\lambda \sum_\sigma P^\beta_{\lambda \sigma}
(\mu \sigma | \nu \lambda )
\label{famn}
\end{equation}
\begin{equation}
F^{\beta\alpha}_{\mu \nu} = H_{\mu \nu} +
\sum_\lambda \sum_\sigma P^\beta_{\lambda \sigma}
[(\mu \sigma | \nu \lambda ) - (\mu \sigma | \lambda \nu)]
+ \sum_\lambda \sum_\sigma P^\alpha_{\lambda \sigma}
(\mu \sigma | \nu \lambda ),
\label{fbmn}
\end{equation}
where $H_{\mu \nu}$ are the elements of the single electron Hamiltonian
(with an external magnetic field $B$ and an appropriate potential confinement),
and the Coulomb repulsion is expressed via the two-electron integrals
\begin{equation}
(\mu \sigma | \nu \lambda) = 
 \frac{e^2}{\kappa}
\int \rmd{\bf r}_1 \rmd{\bf r}_2 \varphi^*_\mu({\bf r}_1) 
\varphi^*_\sigma({\bf r}_2)
\frac{1}{|{\bf r}_1 - {\bf r}_2|}
\varphi_\nu({\bf r}_1) \varphi_\lambda({\bf r}_2),
\label{r12}
\end{equation}
with $\kappa$ being the dielectric constant of the semiconductor material.
Of course, the Greek indices $\mu$, $\nu$, $\lambda$, and $\sigma$ run from
1 to $K$.

The system of the two coupled UHF matrix equations \eref{uhfa} and \eref{uhfb}
is solved selfconsistently through iteration cycles.
For obtaining the numerical solutions, we have
used a set of $K$ basis states $\varphi_i$'s
that are chosen to be the product wave functions formed from the
eigenstates of one-center (single QD) and/or two-center \cite{yl01,yl02.3} 
(double QD) one-dimensional oscillators along the $x$ and $y$ axes. 
Note that for a circular QD a value $K=78$ corresponds to all the
states of the associated 2D harmonic oscillator up to and including the 12th
major shell.

The UHF equations preserve at each iteration step the symmetries
of the many-body Hamiltonian, if these symmetries happen to
be present in the input (initial) electron density of the iteration (see 
section 5.5 of Ref.\ \cite{rs}). The input densities into the iteration
cycle are controlled by the values of the $P^\alpha_{\lambda\sigma}$ and 
$P^\beta_{\lambda\sigma}$ matrix elements. Two cases arise
in practice: (i) Symmetry adapted RHF solutions 
are extracted from \eref{uhfa} and \eref{uhfb} by using as input
$P^\alpha_{\lambda\sigma}=P^\beta_{\lambda\sigma}$=0 for the case of closed
shells (with or without an infinitesimally small $B$ value). For open shells,
one needs to use an infinitesimally small value of $B$. With these choices,
the output of the first iteration (for either closed or open shells) is the
single-particle spectrum and corresponding electron densities at $B=0$ 
associated with the Hamiltonian in \eref{hsp} (the small value of $B$ 
mentioned above guarantees that the single-particle total and orbital 
densities are circular). (ii) For obtaining broken-symmetry UHF solutions, the 
input densities must be different in an essential way from the ones mentioned 
above. We have found that the choice $P^\alpha_{\lambda\sigma}=1$ and 
$P^\beta_{\lambda\sigma}=0$ usually produces broken-symmetry solutions (in the
regime where symmetry breaking occurs).

Having obtained the selfconsistent solution, the total UHF energy is
calculated as
\begin{equation}
E_{\rm UHF}= \frac{1}{2} \sum_\mu \sum_\nu
[ (P^\alpha_{\nu\mu}+P^\beta_{\nu\mu}) H_{\mu\nu}
   + P^\alpha_{\nu\mu} F^{\alpha\beta}_{\mu\nu}
   + P^\beta_{\nu\mu} F^{\beta\alpha}_{\mu\nu}].
\label{euhf}
\end{equation}

We note that the Pople-Nesbet UHF equations
are primarily employed in Quantum Chemistry for studying the ground states of
open-shell molecules and atoms. Unlike our studies of QDs, however, such
chemical UHF studies consider mainly the breaking of the total spin
symmetry, and not that of the space symmetries. As a result, for purposes of
emphasis and clarity, we have often used (see, e.g., our previous papers)
prefixes to indicate the specific unrestrictions (that is removal of
symmetry restrictions) involved in our UHF solutions, i.e., the prefix 
s- for the total-spin and the prefix S- for the space unrestriction.

The emergence of broken-symmetry solutions is associated with instabilities 
of the restricted HF solutions, i.e., the restricted HF energy is an 
{\it extremum\/} whose nature as a minimum or maximum depends on the positive 
or negative value of the second derivative of the HF energy. The importance of
this instability problem was first highlighted in a paper by Overhauser
\cite{over60}. Soon afterwards, the general conditions for 
the appearance of such instabilities (analyzed within linear response and 
the random-phase approximation) were discussed by Thouless in the 
context of nuclear physics \cite{thou61}. Subsequently, 
the Hartree-Fock stability/instability conditions were re-examined 
\cite{adam62,pald67}, using a language from (and applications to) the field of
quantum chemistry. For comprehensive reviews of mean-field symmetry breaking 
and the Hartree-Fock methods and instabilities in the context of quantum 
chemistry, see the collection of papers in Ref.\ \cite{carb90}.

\subsubsection{The Wigner parameter and classes of spontaneous symmetry
breaking solutions.}
\label{wigparssb}

Using the self-consistent (spin-and-space) unrestricted Hartree-Fock equations
presented in the previous section, we found \cite{yl99}, for zero and low 
magnetic fields, three classes of spontaneous symmetry breakings in circular 
single QDs and in lateral quantum dot molecules (i.e., formation of ground 
states of lower symmetry than that of the confining potentials). These include
the following: 

(I) Wigner molecules in both QDs and quantum dot molecules, i.e., (spatial) 
localization of individual electrons within a single QD or within each QD 
comprising the quantum dot molecule.

(II) formation of electron puddles in quantum dot molecules, that 
is, localization of the electrons on each of the individual dots comprising 
the quantum dot molecule, but without localization within each dot, and 

(III) pure spin-density waves (SDWs) which are not accompanied by spatial 
localization of the electrons \cite{kosk97}. 

\begin{figure}[t]
\centering\includegraphics[width=6.5cm]{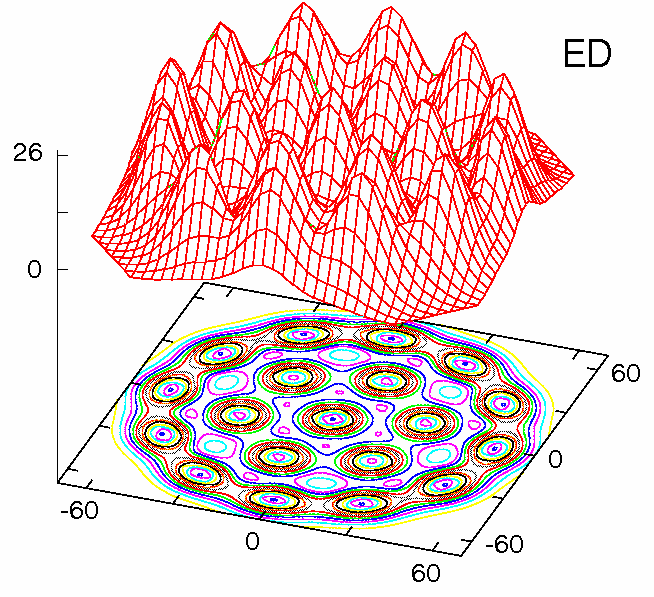}
\caption{(Color online) 
UHF electron density in a parabolic QD for $N=19$ and $S_z=19/2$, exhibiting
breaking of the circular symmetry at $R_W=5$ and $B=0$.
The choice of the remaining parameters is: $\hbar \omega_0=5$ meV
and $m^*=0.067 m_e$.
Distances are in nanometers and the electron density in $10^{-4}$ nm$^{-2}$.
}
\label{uhfwm}
\end{figure}

It can be shown that a central-mean-field description (associated
with the RHF) at zero and low magnetic fields may apply in the case of a
circular QD only for low values of the Wigner parameter 
\begin{equation}
R_{\rm W} \equiv Q/\hbar \omega_0,
\label{wigpar}
\end{equation}
where $Q$ is the Coulomb interaction strength and $\hbar \omega_0$ is the
energy quantum of the harmonic potential confinement (being proportional to
the one-particle kinetic energy); $Q=e^2/(\kappa l_0)$,
with $\kappa$ being the dielectric constant, $l_0=(\hbar/(m^*\omega_0))^{1/2}$
the spatial extension of the lowest state's wave function in the 
harmonic (parabolic) confinement, and $m^*$ the effective electron mass.

Furthermore, we find that Wigner molecules
(SSB class I) occur in both QDs and quantum dot molecules 
for $R_{\rm W} > 1$. Depending on the value of $R_W$, one may distinguish 
between ``weak'' (for smaller $R_W$ values) and ``strong'' (for larger $R_W$ 
values) Wigner molecules, with the latter termed sometimes as ``Wigner 
crystallites'' or ``electron crystallites.'' The appearance of such crystalline
structures may be regarded as a quantum phase transition of the electron 
liquid upon increase of the parameter $R_W$. Of course, due to the finite
size of QDs, this phase transition is not abrupt, but it develops gradually 
as the parameter $R_W$ varies.

For quantum dot molecules with $R_{\rm W} < 1$, Wigner molecules do not 
develop and instead electron puddles may form (SSB class II). 
For single QDs with $R_{\rm W} < 1$, we
find in the majority of cases that the ground-states  exhibit a
central-mean-field behavior without symmetry breaking; however, at several
instances (see an example below), a pure SDW (SSB class III) may
develop. 

\subsubsection{Unrestricted Hartree-Fock solutions representing Wigner 
molecules.}

As a typical example of a Wigner-molecule solution 
that can be extracted from the UHF 
equations, we mention the case of $N=19$ electrons for $\hbar \omega_0 = 5$ 
meV, $R_W=5$ $(\kappa=3.8191)$, and $B=0$. \Fref{uhfwm} displays the total 
electron density of the broken-symmetry UHF solution for these
parameters, which exhibits breaking of the rotational symmetry. In accordance
with electron densities for smaller dot sizes published by us earlier 
\cite{yl99,yl00.1} the electron density in \fref{uhfwm} is highly suggestive 
of the formation of a Wigner molecule, with a (1,6,12) ring structure in the 
present case; the notation $(n_1,n_2,\ldots,n_r)$ signifies the number of 
electrons in each ring: $n_1$ in the first, $n_2$ in the second, and so on.
This polygonal ring structure agrees with the classical one 
(that is the most stable arrangement of 19 point charges in a 2D circular
harmonic confinement \cite{lozo87,beda94,kong02}\footnote[4]{
These references presented extensive studies pertaining to the geometrical
arrangements of classical point charges in a two-dimensional harmonic
confinement.}), and it is sufficiently complex to 
instill confidence that the Wigner-molecule interpretation is valid. 
The following question, however, arises naturally at this point: is such 
molecular interpretation limited to the intuition provided by the landscapes 
of the total electron densities, or are there
deeper analogies with the electronic structure of natural 3D molecules? The
answer to the second part of this question is in the affirmative. Indeed,
it was found \cite{yl03.1} that SSB results in the replacement of a 
{\it higher\/} symmetry by a {\it lower\/} one. As a result, the molecular UHF
solutions exhibit point-group spatial symmetries that are amenable to a 
group-theoretical analysis in analogy with the case of 3D natural molecules. 

\begin{figure}[t]
\centering\includegraphics[width=12.0cm]{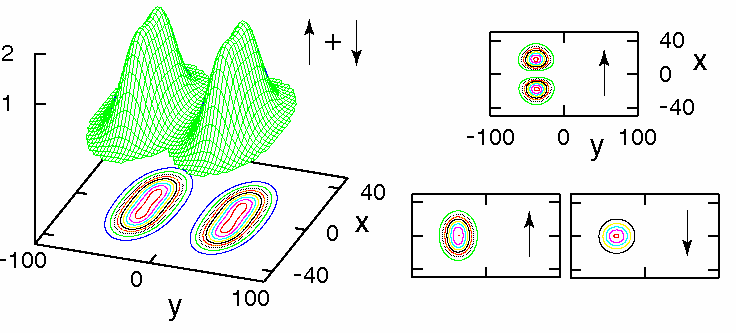}
\caption{(Color online) 
UHF ground-state of a 6$e$ quantum dot molecule (double dot), 
with parameters resulting in formation of two
{\it non-crystallized\/} electron puddles (akin to dissociation of the 
quantum dot molecule in two QDs with 3 electrons each).
Left: total electronic density. Right: contour plots of the densities 
(orbital squares) of the three individual orbitals localized on the left dot,
with spin polarization of the orbitals as indicated.
The choice of parameters is: $\hbar \omega_0=5$ meV (harmonic confinement of
each dot), $d=70$ nm (distance berween dots), $V_b=10$ meV (interdot barrier),
$m^*=0.067m_e$ (electron effective mass), and $\kappa=20$ 
(dielectric constant). Lengths ($x$ and $y$ axes) in nm, density distribution
(vertical axis) in 10$^{-3}$ nm$^{-2}$. 
}
\label{figqdm}
\end{figure}

\subsubsection{Unrestricted Hartree-Fock solutions representing electron 
puddles.}
\label{uhfelpu}

An example of formation of electron puddles in quantum dot molecules, that 
is, localization of the electrons on each of the individual dots comprising 
the quantum dot molecule, but without localization within each dot, is 
presented in \fref{figqdm}. We consider the case of $N=6$ electrons in a 
double dot under field-free conditions ($B=0$); 
with parameters $\hbar \omega_0=5$ meV (harmonic confinement of each dot), 
$d=70$ nm (distance berween dots), $V_b=10$ meV (interdot barrier) and
$m^*=0.067m_e$ (electron effective mass). Reducing the $R_W$ value 
(with reference to each constituent QD) to 0.95 (i.e., for a dielectric 
constant $\kappa=20$) guarantees that the ground-state of the $6e$ quantum dot
molecule consists of electron puddles [SSB of type II, \fref{figqdm}]. 
In this case, each of the electron puddles (on the left and right dots)
is spin-polarized with total spin projection $S_z=1/2$ on the left QD and
$S_z=-1/2$ on the right QD. As a result, the singlet and triplet states
of the whole quantum dot molecule are essentially degenerate. 
Note that the orbitals
on the left and right dots [see, e.g., those on the left dot in \fref{figqdm} 
(right)] are those expected from a central-mean-field treatment of each
individual QD, but with slight (elliptical) distortions due to the interdot 
interaction and the Jahn-Teller distortion associated with an open shell
of three electrons (in a circular harmonic confinement).
Note the sharp contrast between these central-mean-field
orbitals and corresponding electron density (\fref{figqdm}) with the electron 
density and the three orbitals associated with formation of a Wigner molecule 
inside a single QD [see, e.g., \fref{n3sz12} in \sref{groupstruc} below].   

The formation of electron puddles described above can be also seen as a form 
of dissociation of the quantum dot molecule. 
We found that only for much lower values of $R_W$
($< 0.20$, i.e., $\kappa > 90.0$) the electron orbitals do extend over both 
the left and right QDs, as is usually the case with 3D natural molecules
(molecular-orbital theory). Further examples and details of these two
regimes (dissociation versus molecular-orbital description) can be found
in Refs.\ \cite{yl01,yl02.3}.

\begin{figure}[t]
\centering\includegraphics[width=12.0cm]{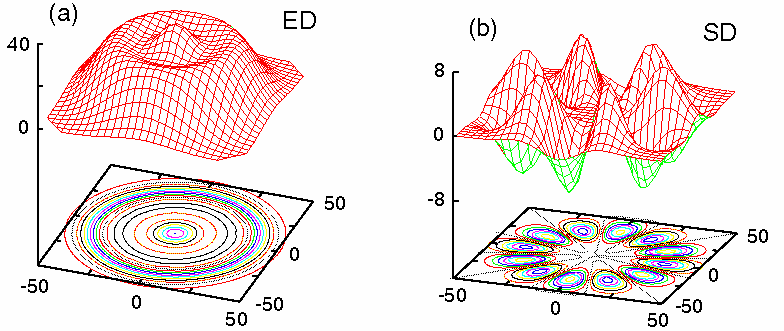}
\caption{(Color online) 
UHF solution in a parabolic QD exhibiting a pure spin density wave for $N=14$,
$S_z=0$, $R_{\rm W}=0.8$, and $B=0$. (a) The total electron density exhibiting
circular symmetry; (b) The spin density exhibiting azimuthal modulation
(note the 12 humps whose number is smaller than the number of electrons;
on the contrary in the case of a Wigner molecule, the 
number of humps in the electron density is always equal to $N$).
The choice of the remaining parameters is: $\hbar \omega_0=5$ meV
and $m^*=0.067 m_e$.
Distances are in nanometers and the electron (ED) and spin (SD) densities in
$10^{-4}$ nm$^{-2}$.
}
\label{uhfsdw}
\end{figure}

\subsubsection{Unrestricted Hartree-Fock solutions representing pure spin 
density waves within a single quantum dot.}

Another class of broken-symmetry solutions that can appear in single QDs
are the spin density waves. The SDWs are unrelated to electron localization 
and thus are quite distinct from the Wigner molecules \cite{yl99}; 
in single QDs, they were obtained
\cite{kosk97} earlier within the framework of spin density functional theory.
To emphasize the different nature of spin density waves and 
Wigner molecules, we present in 
\fref{uhfsdw} an example of a SDW obtained with the UHF approach [the 
corresponding parameters are: $N=14$, $S_z=0$, $R_{\rm W}=0.8$ 
($\kappa=23.8693$), and $B=0$]. Unlike the case of Wigner molecules, 
the SDW exhibits a 
circular electron density [see \fref{uhfsdw}(a)], and thus it does not break 
the rotational symmetry. Naturally, in keeping with its name, the SDW breaks 
the total spin symmetry and exhibits azimuthal modulations in the spin
density [see \fref{uhfsdw}(b); however, the number of humps is
smaller than the number of electrons]. 

We mention here that the possibility of ground-state configurations with 
uniform electron density, but nonuniform spin density, was first discussed for
3D bulk metals using the HF method in Ref.\ \cite{over}.

The SDWs in single QDs appear for $R_{\rm W} \leq 1$ and are of lesser 
importance; thus in the following we will 
exclusively study the case of Wigner molecules.
However, for $R_{\rm W} \leq 1$, formation of a special class of SDWs (often 
called electron puddles, see \sref{uhfelpu}) plays an important role 
in the coupling and dissociation of quantum dot molecules (see Ref.\ 
\cite{yl01} and Ref.\ \cite{yl02.3}).

\begin{figure}[t]
\centering\includegraphics[width=14.5cm]{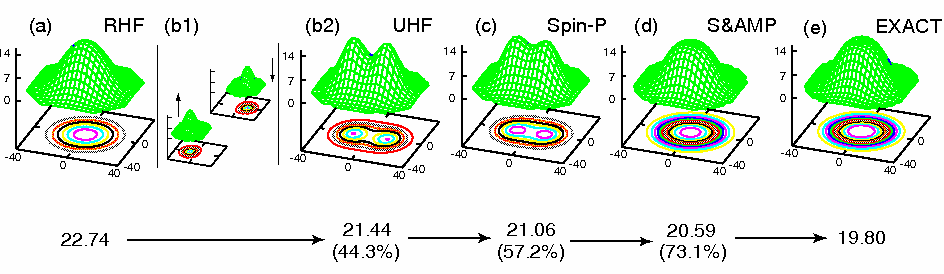}\\
~~~~~\\
\caption{(Color online) 
Various approximation levels for the lowest singlet state of a
field-free two-electron QD with $R_W=2.40$. The corresponding energies 
(in meV) are shown at the bottom of the figure.
(a): Electron density of the RHF solution, exhibiting circular 
symmetry (due to the imposed symmetry restriction). The correlation energy 
$E_{\rm corr} = 2.94$ meV, is defined as the difference between the energy of 
this state and the exact solution [shown in frame (e)]. (b1) and (b2): The two 
occupied orbitals (modulus square) of the symmetry-broken ``singlet'' UHF 
solution (b1), with the corresponding total electron density exhibiting 
non-circular shape (b2). The energy of the UHF solution shows a gain of 
44.3\% of the correlation energy. (c): Electron density of the spin-projected 
singlet (Spin-P), showing broken spatial symmetry, but with an additional gain
of correlation energy. (d): the spin-and-angular-momentum projected state 
(S\&AMP) exhibiting restored circular symmetry with a 73.1\% gain of the 
correlation energy. The choice of parameters is: dielectric constant 
$\kappa = 8$, parabolic confinement $\hbar \omega_0 = 5$ meV, and effective 
mass $m^* = 0.067m_e$. Distances are in nanometers and the densities in 
$10^{-4}$ nm$^{-2}$. 
}
\label{succapp}
\end{figure}

\subsection{Projection techniques and post-Hartree-Fock restoration of 
broken symmetries}
\label{restsym}

As discussed in \sref{synopsis}, for finite systems the symmetry 
broken UHF solutions are only an intermediate approximation. A subsequent 
step of post-Hartree-Fock symmetry restoration is needed. Here we present the 
essentials of symmetry restoration while considering for simplicity the case
of two electrons in a circular parabolic QD.

Results obtained for various approximation levels for a two-electron QD with
$B=0$ and $R_W=2.40$ (that is, in the Wigner-molecule regime) are displayed in
\fref{succapp}. In these calculations \cite{yl02.1}, the spin projection was 
performed following reference \cite{low55}, i.e., one constructs the wave 
function
\begin{equation}
\Psi_{\rm Spin-P}(s)={\cal P}_{\rm spin}(s) \Psi_{\rm UHF},
\label{psis}
\end{equation}
where $\Psi_{\rm UHF}$ is the original symmetry-broken UHF determinant
(which is already by construction an eigenstate of the projection $S_z$ of the
total spin). 
In \eref{psis}, the spin projection operator (projecting into a state which
is an eigenstate of the square of the total spin) is given by
\begin{equation}
{\cal P}_{\rm spin}(s) \equiv \prod_{s^\prime \neq s}
\frac{\hat{\bf S}^2 - s^\prime(s^\prime + 1)}
{s(s+1) - s^\prime(s^\prime + 1)},
\label{prjp}
\end{equation}
where the index $s^\prime$ runs over the quantum numbers associated with
the eigenvalues $s^\prime(s^\prime+1)$ of $\hat{\bf S}^2$
(in units of $\hbar^2$), with $\hat{\bf S}$ being the total spin operator.
For two electrons, the projection operator reduces to 
${\cal P}_{\rm spin}^{s,t}=1 \mp \varpi_{12}$, where the
operator $\varpi_{12}$ interchanges the spins of the two electrons; the upper 
(minus) sign corresponds to the singlet ($s$ supersript), and the lower (plus) 
sign corresponds to the triplet ($t$ superscript) state.

The angular momentum projector (projecting into a state with total 
angular momentum $L$) is given by
\begin{equation}
2\pi {\cal P}_L \equiv \int_0^{2\pi} \rmd \gamma 
\exp[-\rmi \gamma (\hat{L}-L)],
\label{amp}
\end{equation}
where $\hat{L}=\hat{l}_1+\hat{l}_2$ is the total angular momentum
operator. As seen from \eref{amp}, application of the projection
operator ${\cal P}_L$ to the spin-restored state $\Psi_{\rm Spin-P}(s)$
corresponds to a continuous configuration interaction expansion of the wave
function that uses, however, {\it non-orthogonal\/} orbitals (compare 
\sref{exdmeth}).

The application of the projection operator ${\cal P}_L$ to the state 
$\Psi_{\rm Spin-P}(s)$ generates a whole rotational band of states with good 
angular momenta (yrast band). The energy of the projected state with total
angular momentum $L$ is given by
\begin{equation}
E_{\rm PRJ}(L) = \left. \int_0^{2\pi} h(\gamma) \rme^{\rmi\gamma L} 
\rmd \gamma \right/
 \int_0^{2\pi} n(\gamma) \rme^{\rmi \gamma L} \rmd\gamma,
\label{eprj}
\end{equation}
with $h(\gamma)=
\langle \Psi_{\rm Spin-P}(s;0)|{\cal H}|\Psi_{\rm Spin-P}(s;\gamma)\rangle$ 
and $n(\gamma)=
\langle \Psi_{\rm Spin-P}(s;0)|\Psi_{\rm Spin-P}(s;\gamma)\rangle$, where 
$\Psi_{\rm Spin-P}(s;\gamma)$ is the spin-restored (i.e., spin-projected) wave 
function rotated by an azimuthal angle $\gamma$ and ${\cal H}$ is the many-body 
Hamiltonian. We note that the UHF energies are simply given by 
$E_{\rm UHF}=h(0)/n(0)$.

In the following we focus on the ground state of the 
two-electron system, i.e., $L = 0$. 
The electron densities corresponding to the initial RHF approximation [shown in
\fref{succapp}(a)] and the final spin-and-angular-momentum projection (S\&AMP)
[shown in \fref{succapp}(d)], are circularly symmetric, while those 
corresponding to the two intermediate approximations, i.e., the UHF and 
spin-projected solutions [\fref{succapp}(b2) and \fref{succapp}(c), 
respectively] break the circular symmetry. This
behavior illustrates graphically the meaning of the term ``restoration of
symmetry,'' and the interpretation that the UHF broken-symmetry solution refers
to the intrinsic (rotating) frame of reference of the electron molecule. In
light of this discussion the final projected state is called a {\it rotating
electron\/} or ({\it Wigner\/}) {\it molecule\/}.

Expressions \eref{amp} and \eref{eprj} apply directly to REMs having a single
polygonal ring of $N$ localized electrons, with 
$\hat{L}=\sum_{i=1}^N {\hat l}_i$. For a generalization to electron molecules 
with multiple concentric polygonal rings, see \sref{remtheo} below.

For restoring the total spin, an alternative method to the projection 
formula \eref{prjp} can be found in the literature \cite{fuk}. We do not make 
use of this alternative formulation in this report, but we briefly describe it 
here for the sake of completeness. Based on the 
formal similarity between the 3D angular momentum and the total spin, one can 
apply the formula by Peierls and Yoccoz \cite{py} and obtain the projection 
operator
\begin{equation}
{\cal P}^s_{{S_z}{q}} = \frac{2s+1}{8 \pi^2}
\int \rmd \Gamma D^{s*}_{{S_z}{q}}(\Gamma) {\cal R}(\Gamma),
\label{prjppy}
\end{equation}
where $D^{s*}_{{S_z}{q}}(\Gamma)$ are the 3D Wigner $D$ functions 
\cite{wignbook}, $\Gamma$ is a shorthand notation for the set of the
three Euler angles $(\phi,\theta,\psi)$, and 
\begin{equation}
{\cal R}(\Gamma) = \rme^{-\rmi \phi \hat{S}_z} 
\rme^{-\rmi \theta \hat{S}_y} \rme^{-\rmi \psi \hat{S}_z}
\label{rotop}
\end{equation}
is the rotation operator in spin space. In \eref{prjppy}, the indices of
the Wigner $D$ functions are $s$, $S_z$, and $q$.

The operator ${\cal P}^s_{{S_z}{q}}$ extracts from the symmetry broken 
wave function a state with a total spin $\hat{\bf S}$ and projection $S_z$ 
along the laboratory $z$ axis. However, $q$ is not a good quantum number
of the many-body Hamiltonian, and the most general symmetry restored state is 
written as a superposition over the components of $q$, i.e.,
\begin{equation}
\Psi_{\rm Spin-P}(s,S_z;i)=\sum_q g_q^i {\cal P}^s_{{S_z}{q}} 
\Psi_{\rm UHF},
\label{wfspin}
\end{equation}
where the coefficients $g_q^i$ are determined through a diagonalization of
the many-body Hamiltonian in the space spanned by the nonorthogonal
${\cal P}^s_{{S_z}{q}} \Psi_{\rm UHF}$ (see also Refs.\ \cite{hash82,igaw95}).
In \eref{wfspin}, the index $i$ reflects the possible degeneracies of
spin functions with a given good total-spin quantum number $s$ \cite{paunbook},
which is not captured by \eref{prjp}.

The Peierls-Yoccoz formulation for recovering spin-corrected wave functions
applies also in the case when the UHF determinants violate in addition the 
conservation of spin projection \cite{fuk}, unlike the projector 
${\cal P}_{\rm spin}(s)$ [see \eref{prjp}] which acts on UHF determinants
having a good $S_z=(N^\alpha-N^\beta)/2$ according to the Pople-Nesbet
theory presented in \sref{selfpopnes}.

In the literature \cite{rs}, there are two distinguishable implementations of 
symmetry restoration: variation before projection (VBP) and variation after
projection (VAP). In the former, which is the one that we mostly use this 
report, mean-field solutions with broken symmetry are first constructed and 
then the symmetry is restored via projection techniques as described above. In
the latter, the projected wave function is used as the trial wave function 
directly in the variational principle (in other words the trial function is
assured to have the proper symmetry).

The VAP is in general more accurate, but more difficult to implement 
numerically, and it has been used less often in the nuclear-physics 
literature. In quantum chemistry, the generalized valence bond method 
\cite{godd73}, or the spin-coupled valence bond method \cite{gerr68}, 
describing covalent bonding between pairs of electrons, employ a variation 
after projection.

For quantum dots, the variation after projection looks promising for reducing 
the error of the VBP techniques in the transition region from mean-field to
Wigner-molecule behavior, where this error is the largest. In fact, it has been
found that the discrepancy between variation-before-projection techniques and 
exact solutions is systematically reduced \cite{yl02.1,kram07,degi07} for 
stronger symmetry breaking (increasing $R_W$ and/or increasing magnetic field).

Moreover, in the case of an applied magnetic field (quantum dots) or
a rotating trap (Bose gases), our VBP implementation corresponds to 
projecting cranked symmetry-unrestricted Slater determinants \cite{doba07}.
This is because of the ``cranking'' terms $-\hbar \omega_c L/2$ or 
$-\hbar \Omega L$ that contribute to the many-body Hamiltonian 
${\cal H}$, respectively, with $\omega_c=eB/(m^*c)$ being the cyclotron frequency
and $\Omega$ the rotational frequency of the trap; these terms arise in the 
single-particle component of ${\cal H}$ 
[see \Eref{hsp} in \sref{symbreak} and \Eref{hkbosrot} in \sref{bosmol}].
The cranking form of the many-body Hamiltonian is particularly advantageous
to the variation {\it before\/} projection, since the cranking method 
provides a first-order approximation to the variation-{\it after\/}-projection
restoration of the total angular-momentum $\hat{L}$ \cite{kaml} 
(see also Ch 11.4.4 in Ref.\ \cite{rs}). 

\subsubsection{The REM microscopic method in medium and high magnetic field.}
\label{remtheo}

In our method of hierarchical approximations, we begin with a 
{\it static\/} electron molecule, described by an unrestricted 
Hartree-Fock determinant that violates the circular symmetry 
\cite{yl99,yl02.1,yl03.1}.
Subsequently, the {\it rotation\/} of the electron molecule is described by a
post-Hartree-Fock step of restoration of the broken circular symmetry via
projection techniques \cite{yl01,yl02.1,yl02.2,yl03.1,yl03.2,yl04.1,li06}.
Since we focus here on the case of strong $B$, we can approximate
the UHF orbitals (first step of our procedure) by (parameter free) displaced
Gaussian functions; that is, for an electron localized at ${\bf R}_j$
($Z_j$), we use the orbital \cite{li06}
\begin{equation}
u(z,Z_j) = \frac{1}{\sqrt{\pi} \lambda}
\exp \left( -\frac{|z-Z_j|^2}{2\lambda^2} - \rmi \varphi(z,Z_j;B) \right),
\label{uhfo}
\end{equation}
with $\lambda = \tilde{l} \equiv \sqrt{\hbar /m^* \tilde{\omega}}$;
$\tilde{\omega}=\sqrt{\omega_0^2+\omega_c^2/4}$, 
where $\omega_c=eB/(m^*c)$ is the
cyclotron frequency and $\omega_0$ specifies the external parabolic 
confinement. We have used complex numbers to represent the position
variables, so that $z=x+ \rmi y$, $Z_j = X_j + \rmi Y_j$.
The phase guarantees gauge invariance in the presence of
a perpendicular magnetic field and is given in the symmetric gauge by
$\varphi(z,Z_j;B) = (x Y_j - y X_j)/2 l_B^2$, with $l_B = \sqrt{\hbar c/ e B}$.

For an extended 2D system, the $Z_j$'s form a triangular lattice
\cite{jainbook,yosh}. For finite $N$, however, the $Z_j$'s coincide 
\cite{yl02.2,yl03.2,yl04.1,yl04.2,li06}
with the equilibrium positions [forming $r$ concentric
regular polygons denoted as ($n_1, n_2,\ldots,n_r$)] of $N=\sum_{q=1}^r n_q$
classical point charges inside an external parabolic confinement \cite{kong02}.
In this notation, $n_1$ corresponds to the innermost ring with $n_1 > 0$. 
For the case of a single polygonal ring, the notation $(0,N)$ is often used;
then it is to be understood that $n_1=N$.

The wave function of the {\it static\/} electron molecule is a
{\it single\/} Slater determinant $|\Psi^{\rm{UHF}} [z] \rangle$ made out of
the single-electron wave functions $u(z_i,Z_i)$, $i = 1,\ldots,N$.
Correlated many-body states with good total angular momenta $L$
can be extracted \cite{yl02.2,yl03.2,yl04.1,li06} (second step) from the UHF 
determinant using projection operators. The projected rotating electron 
molecule state is given by
\begin{eqnarray}
\hspace{-2.0cm} |\Phi^{\rm{REM}}_L \rangle = 
\int_0^{2\pi} \ldots \int_0^{2\pi}
\rmd \gamma_1 \ldots \rmd \gamma_r |\Psi^{\rm{UHF}}
(\gamma_1, \ldots, \gamma_r) \rangle
\exp \left( \rmi \sum_{q=1}^r \gamma_q L_q \right). \nonumber\\
\label{wfprj1}
\end{eqnarray}
Here $L=\sum_{q=1}^r L_q$ and
$|\Psi^{\rm{UHF}}[\gamma] \rangle$ is the original Slater determinant with 
{\it all the single-electron wave functions of the $q$th ring\/} rotated
(collectively, i.e., coherently) by the {\it same\/} azimuthal angle 
$\gamma_q$. Note that \eref{wfprj1} can be written as a product of projection
operators acting on the original Slater determinant [i.e., on
$|\Psi^{\rm{UHF}}(\gamma_1=0, \ldots, \gamma_r=0) \rangle$].
Setting $\lambda = l_B \sqrt{2}$ restricts the single-electron wave function in
\eref{uhfo} to be entirely in the lowest Landau level (see Appendix in
Ref.\ \cite{li06}). 
The continuous-configuration-interaction form of the projected wave functions
[i.e., the linear superposition of determimants in \eref{wfprj1}]
implies a highly entangled state. We require here that $B$ is sufficiently
strong so that all the electrons are spin-polarized and that the ground-state 
angular momentum $L \geq L_0 \equiv \sum_{i=0}^{N-1} i = N(N-1)/2$ 
(or equivalently that the fractional filling factor $\nu \equiv L_0/L \leq 1$).
The state corresponding to $L_0$ is a single Slater determinant in the lowest
Landau level and is called the ``maximum density droplet'' \cite{mac93}.
For high $B$, the calculations in this paper do not include
the Zeeman contribution, which, however, can easily be added (for a fully
polarized dot, the Zeeman contribution to the total energy is 
$Ng^* \mu_B B/2$, with $g^*$ being the effective Land\'{e} factor and 
$\mu_B$ the Bohr magneton).

Due to the point-group symmetries of each polygonal ring of electrons
in the UHF wave function, the total angular momenta $L$ of the rotating
crystalline electron molecule are restricted to the so-called {\it magic} 
angular momenta, i.e.,
\begin{equation}
L_m = L_0 + \sum_{q=1}^r k_q n_q,
\label{lmeq}
\end{equation}
where the $k_q$'s are non-negative integers (when $n_1=1$, $k_1=0$).

Magic angular momenta associated with multiple rings have been discussed in
Refs.\ \cite{yl02.2,yl03.2,yl04.1,yl04.2,li06}. For the simpler cases of 
$(0,N)$ or $(1,N-1)$ rings, see, e.g., Ref.\ \cite{rua95} and Ref.\ 
\cite{maks96}. 

The partial angular momenta associated with the $q$th ring,
$L_q$ [see \eref{wfprj1}], are given by
\begin{equation}
L_q = L_{0,q} + k_q n_q,
\label{lmpar}
\end{equation}
where $L_{0,q}=\sum_{i=i_q+1}^{i_q+n_q} (i-1)$ with
$i_q = \sum_{s=1}^{q-1} n_s$ $(i_1=0)$, and
$L_0 = \sum_{q=1}^r L_{0,q}$.

The energy of the REM state \eref{wfprj1} is given 
\cite{yl02.2,yl04.1,yl04.2,li06} by
\begin{equation}
E^{\rm{REM}}_L = \left. { \int_0^{2\pi} h([\gamma]) 
\rme^{\rmi [\gamma] \cdot [L]} \rmd [\gamma] } \right/%
{ \int_0^{2\pi} n([\gamma]) \rme^{\rmi [\gamma] \cdot [L]} \rmd [\gamma]},
\label{eproj}
\end{equation}
with the Hamiltonian and overlap matrix elements $h([\gamma]) =
\langle \Psi^{\rm{UHF}}([0]) | {\cal H} | \Psi^{\rm{UHF}}([\gamma]) \rangle$ 
and $n([\gamma]) =
\langle \Psi^{\rm{UHF}}([0]) | \Psi^{\rm{UHF}}([\gamma]) \rangle$, 
respectively, and
$[\gamma] \cdot [L] = \sum_{q=1}^r \gamma_q L_q$.
The UHF energies are simply given by $E_{\rm{UHF}} = h([0])/n([0])$.

The crystalline polygonal-ring arrangement $(n_1,n_2,\ldots,n_r)$ of classical
point charges is portrayed directly in
the electron density of the broken-symmetry UHF, since the latter consists of
humps centered at the localization sites $Z_j$'s ({\it one hump} for each
electron). In contrast, the REM has good angular momentum and thus its electron
density is circularly uniform. To probe the crystalline character of the REM,
we use the conditional probability distribution (CPD) defined in \eref{cpds}.
$P({\bf r},{\bf r}_0)$ is proportional to the conditional probability of
finding an electron at ${\bf r}$, given that another electron is assumed at 
${\bf r}_0$. This procedure subtracts the collective rotation of the electron 
molecule in the laboratory frame of referenece, and, as a result, the
CPDs reveal the structure of the many body state in the intrinsic (rotating)
reference frame.

\subsubsection{Group structure and sequences of magic angular momenta.}
\label{groupstruc}

It has been demonstrated \cite{yl03.1} that the broken-symmetry UHF 
determinants and orbitals describe 2D electronic molecular stuctures (Wigner 
molecules) in close analogy with the case of natural 3D molecules. However, 
the study of Wigner molecules at the UHF level 
restricts their description to the 
{\it intrinsic\/} (nonrotating) frame of reference. Motivated by the case of 
natural atoms, one can take a subsequent step and address the properties of
{\it collectively\/} rotating Wigner molecules in the 
laboratory frame of reference. As is
well known, for natural atoms, this step is achieved by writing the total wave
function of the molecule as the product of the electronic and ionic partial 
wave functions. In the case of the purely electronic 
Wigner molecules, however, such a 
product wave function requires the assumption of complete decoupling between 
intrinsic and collective degrees of freedom, an assumption that 
might be justifiable in limiting cases only. The simple product wave function 
was used in earlier treatments of Wigner molecules; see,
e.g., Ref.\  \cite{maks96}. The projected wave functions
employed here are integrals over such product wave functions, and thus they
account for quantal fluctuations in the rotational degrees of freedom.
The reduction of the projected wave functions to the limiting case of a single
product wave function is discussed in Ch 11.4.6.1 of Ref.\ \cite{rs}.

As was discussed  earlier, in the framework of the broken-symmetry UHF 
solutions, a further step is needed -- and this companion step can be 
performed by using the post-Hartree-Fock method of {\it restoration of broken 
symmetries\/} via projection techniques (see \sref{restsym}). In this section,
we use this approach to illustrate through
a couple of concrete examples how certain universal properties of the exact 
solutions, i.e., the appearance of magic angular momenta
\cite{rua95,maks96,girv83,mc90,sek96,maks00,haw98} in the exact rotational 
spectra, relate to the symmetry broken UHF solutions.
Indeed, {\it we demonstrate that the magic angular momenta are a 
direct consequence of the symmetry breaking at the UHF level and that they are
determined fully by the molecular symmetries of the UHF
determinant\/}.

\begin{figure}[t]
\centering\includegraphics[width=10.5cm]{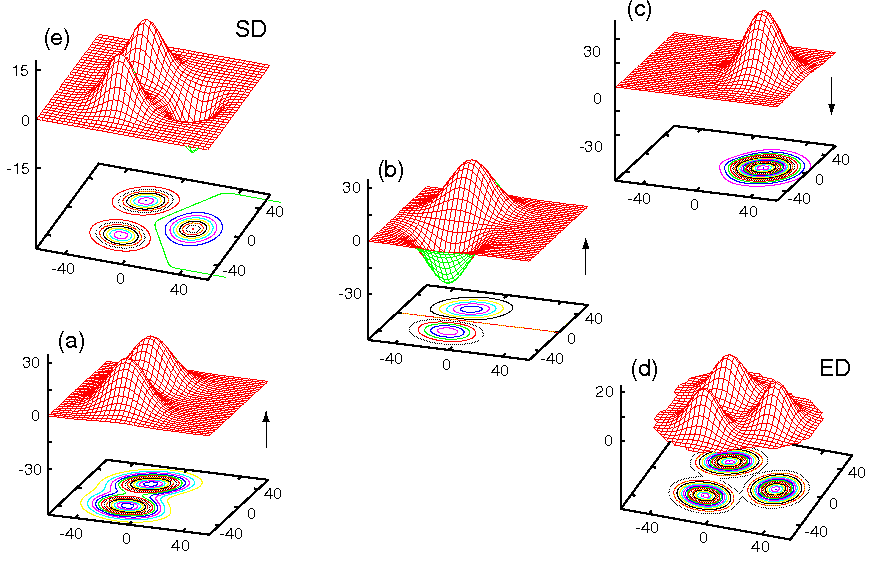}\\
\caption{(Color online) 
The UHF solution exhibiting breaking of the circular symmetry for $N=3$ and 
$S_z=1/2$ at $R_W=10$ and $B=0$. (a-b): real orbitals for the two 
spin-up electrons. (c): real orbital for the single spin-down electron. 
(d): total electron density (ED). (e): spin density (SD, difference of the 
spin-up minus the spin-down partial electron densities).
The choice of the remaining parameters is: $\hbar \omega_0=5$ meV 
and $m^*=0.067 m_e$. Distances are in nanometers. The real orbitals are in
10$^{-3}$ nm$^{-1}$ and the densities (electron density and spin density) 
in 10$^{-4}$ nm$^{-2}$. The arrows indicate the spin direction.
}
\label{n3sz12}
\end{figure}
As an illustrative example, we have chosen the relatively simple, but non
trivial case, of $N=3$ electrons. For $B=0$, both the $S_z=1/2$ and $S_z=3/2$
polarizations can be considered. We start with the $S_z=1/2$ polarization,
whose broken-symmetry UHF solution \cite{yl03.1} is portayed in \fref{n3sz12}
and which exhibits a breaking of the total spin symmetry in addition to the 
rotational symmetry. Let us denote the corresponding UHF determinant 
[made out of the three spin orbitals in \fref{n3sz12}(a), \fref{n3sz12}(b), and
\fref{n3sz12}(c)] as $|\downarrow \uparrow \uparrow \rangle$.
We first proceed with the restoration of the total spin by noticing that
$|\downarrow \uparrow \uparrow \rangle$ has a lower point-group symmetry (see 
Ref.\ \cite{yl03.1}) than the $C_{3v}$ symmetry of an equilateral triangle.
The $C_{3v}$ symmetry, however, can be readily restored by applying the
projection operator \eref{amp} to $|\downarrow \uparrow \uparrow \rangle$ and 
by using the character table of the cyclic $C_3$ group (see Table I in
Ref.\ \cite{yl03.1}). Then for the intrinsic part of the many-body wave 
function, one finds two different three-determinantal combinations, namely
\begin{equation}
\Phi_{\rm{intr}}^{E^\prime} (\gamma_0)=
|\downarrow \uparrow \uparrow \rangle
+ \rme^{2\pi \rmi/3} |\uparrow \downarrow \uparrow \rangle
+ \rme^{-2\pi \rmi/3} |\uparrow \uparrow \downarrow \rangle,
\label{3dete1}
\end{equation}
and
\begin{equation}
\Phi_{\rm{intr}}^{E^{\prime\prime}} (\gamma_0)=
|\downarrow \uparrow \uparrow \rangle
+ \rme^{-2\pi \rmi/3} |\uparrow \downarrow \uparrow \rangle
+ \rme^{2\pi \rmi/3} |\uparrow \uparrow \downarrow \rangle,
\label{3dete2}
\end{equation}
where $\gamma_0=0$ denotes the azimuthal angle of the vertex of the equilateral
triangle associated with the original spin-down orbital in 
$|\downarrow \uparrow \uparrow \rangle$. We note that, unlike the intrinsic 
UHF Slater determinant, the intrinsic wave functions 
$\Phi_{\rm{intr}}^{E^\prime}$ and $\Phi_{\rm{intr}}^{E^{\prime\prime}}$ here 
are eigenstates of the square of the total spin operator ${\hat{\bf S}}^2$  
($\hat{\bf S} = \sum_{i=1}^3 \hat{\bf s}_i$) with quantum number $s=1/2$. 
This can be verified directly by applying ${\hat {\bf S}}^2$ to them.
\footnote[5]{For the appropriate expression of ${\bf S}^2$, see equation (6) in
Ref.\ \cite{yl02.3}.}

To restore the circular symmetry in the case of a (0,N) ring arrangement, one
applies the projection operator \eref{amp}.
Note that the operator ${\cal P}_L$ is a direct generalization of the 
projection operators for finite point-groups discussed in Ref.\ \cite{yl03.1} 
to the case of the continuous cyclic group $C_\infty$ [the phases 
$\exp(i \gamma L)$ are the characters of $C_\infty$].

The symmetry-restored projected wave function, $\Psi_{\rm{PRJ}}$, (having both
good total spin and angular momentum quantum numbers) is of the form,
\begin{equation}
2 \pi \Psi_{\rm{PRJ}} = \int^{2\pi}_0 \rmd \gamma
\Phi_{\rm{intr}}^E (\gamma) \rme^{\rmi \gamma L},
\label{rbsi}
\end{equation}
where now the intrinsic wave function [given by \eref{3dete1} or
\eref{3dete2}] has an arbitrary azimuthal orientation $\gamma$. We note
that, unlike the phenomenological Eckardt-frame model
\cite{maks96,maks00} where only a single product term is involved, the 
PRJ wave function in \eref{rbsi} is an average over all azimuthal 
directions over an infinite set of product terms. These terms are formed by 
multiplying the intrinsic part $\Phi_{\rm{intr}}^{E}(\gamma)$ with the 
external rotational wave function  $\exp(i \gamma L)$ (the latter is properly
characterized as ``external'', since it is an eigenfunction of the total
angular momentum $\hat{L}$ and depends exclusively on the azimuthal
coordinate $\gamma$).\footnote[6]{
Although the wave functions of the Eckardt-frame model are inaccurate compared
to the PRJ ones [see \eref{rbsi}], they are able to yield the proper
magic angular momenta for $(0,N)$ rings. This result is intuitively
built in this model from the very beginning via the phenomenological
assumption that the intrinsic wave function, which is never specified,
exhibits $C_{Nv}$ point-group symmetries.}

The operator ${\hat{R}(2\pi/3) \equiv \exp (-\rmi 2\pi{\hat L}/3})$ can be
applied onto $\Psi_{\rm{PRJ}}$ in two different ways, namely either on
the intrinsic part $\Phi_{\rm{intr}}^{E}$ or the external part 
$\exp(\rmi \gamma L)$. 
Using \eref{3dete1} and the property $\hat{R}(2\pi/3)
\Phi_{\rm{intr}}^{E^\prime} =\exp (-2\pi \rmi/3)\Phi_{\rm{intr}}^{E^\prime}$,
one finds,
\begin{equation}
\hat{R}(2\pi/3) \Psi_{\rm{PRJ}} = \exp (-2\pi \rmi/3) \Psi_{\rm{PRJ}},
\label{r1rbs}
\end{equation}
from the first alternative, and
\begin{equation}
\hat{R}(2\pi/3) \Psi_{\rm{PRJ}} = \exp (-2\pi L \rmi/3) \Psi_{\rm{PRJ}},
\label{r2rbs}
\end{equation}
from the second alternative. Now if $\Psi_{\rm{PRJ}} \neq 0$, the only
way that equations (\ref{r1rbs}) and (\ref{r2rbs}) 
can be simultaneously true is
if the condition $\exp [2\pi (L-1) \rmi/3]=1$ is fulfilled. This leads to a 
first sequence of magic angular momenta associated with total spin $s=1/2$, 
i.e.,
\begin{equation}
L = 3 k +1,\; k=0,\pm 1, \pm 2, \pm 3,\ldots
\label{i1}
\end{equation}

Using \eref{3dete2} for the intrinsic wave function, and following
similar steps, one can derive a second sequence of magic angular momenta
associated with good total spin $s=1/2$, i.e.,
\begin{equation}
L = 3 k -1,\; k=0,\pm 1, \pm 2, \pm 3,\ldots
\label{i2}
\end{equation}

\begin{figure}[t]
\centering\includegraphics[width=8.5cm]{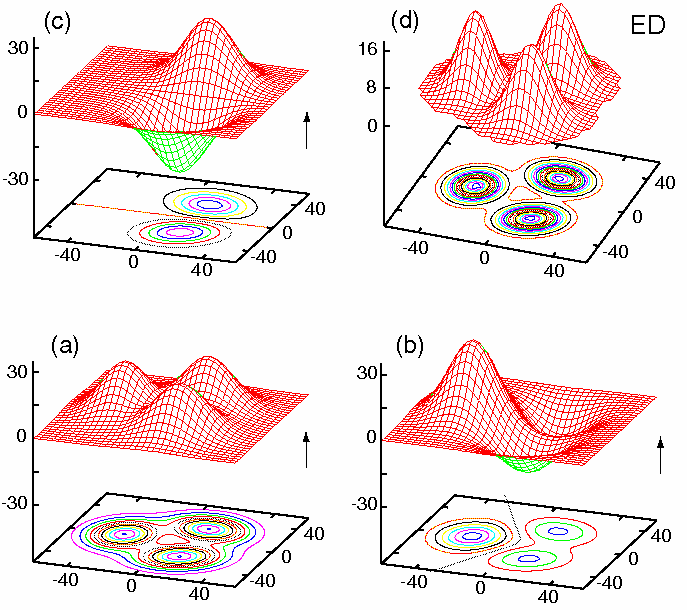}\\
\caption{(Color online) 
The UHF case exhibiting breaking of the circular symmetry for $N=3$ and 
$S_z=3/2$ at $R_W=10$ and $B=0$. (a-c): real orbitals. (d): the corresponding 
electron density (ED). The choice of the remaining parameters is: 
$\hbar \omega_0=5$ meV and $m^*=0.067 m_e$.
Distances are in nanometers. The real orbitals are in
10$^{-3}$ nm$^{-1}$ and the total electron density in 10$^{-4}$ nm$^{-2}$.
The arrows indicate the spin direction.
}
\label{n3sz32}
\end{figure}

In the fully spin-polarized case, the UHF determinant is portrayed in 
\fref{n3sz32}. This UHF determinant, which we denote as 
$|\uparrow \uparrow \uparrow\;\rangle$, is already an eigenstate of 
$\hat{\bf S}^2$ with quantum number
$s=3/2$. Thus only the rotational symmetry needs to be restored, that is,
the intrinsic wave function is simply $\Phi^A_{\rm{intr}}(\gamma_0) =
|\uparrow \uparrow \uparrow \;\rangle$. Since
$\hat{R}(2\pi/3) \Phi^A_{\rm{intr}} = \Phi^A_{\rm{intr}}$, the condition
for the allowed angular momenta is $\exp [-2\pi L \rmi/3]=1$, which yields
the following magic angular momenta,
\begin{equation}
L = 3 k,\; k=0,\pm 1, \pm 2, \pm 3,\ldots
\label{i3}
\end{equation}

We note that in high magnetic fields only the fully polarized case is
relevant and that only angular momenta with $k > 0$ enter in \eref{i3}
(see Ref.\ \cite{yl02.2}). In this case, in the thermodynamic limit, the
partial sequence with $k=2q+1$, $q=0,1,2,3,\ldots$ is directly related to the odd
filling factors $\nu=1/(2q+1)$ of the fractional quantum Hall effect
[via the relation $\nu = N(N-1)/(2L)$]. This suggests that the observed
hierarchy of fractional filling factors in the quantum Hall effect may be
viewed as a signature originating from the point group symmetries of the
intrinsic wave function $\Phi_{\rm{intr}}$, and thus it is a manifestation
of symmetry breaking at the UHF mean-field level.

We further note that the {\it discrete\/} rotational (and more generally
rovibrational) collective spectra associated with symmetry-breaking in a QD
may be viewed as finite analogs to the Goldstone modes accompanying symmetry
breaking transitions in extended media (see Ref.\ \cite{pwa}). 
Recently there has been some interest in studying Goldstone-mode analogs in
the framework of symmetry breaking in trapped BECs with attractive interactions
\cite{muel06}.

\subsection{The symmetry breaking dilemma and density functional theory}
\label{symdft}

Density functional theory (and its extension for cases with a magnetic
field known as current density functional theory) was initially considered
\cite{reim02} (and was extensively applied \cite{reim02,mart98,lebu02}) as a 
promising method for studying 2D semiconductor QDs. However, it soon became
apparent \cite{yl01,yl02.1,yl03.1,yl02.3} that density functional approaches 
exhibited severe drawbacks when applied to the regime of strong correlations 
in QDs, where the underlying physics is associated with symmetry breaking 
leading to electron localization and formation of Wigner molecules. The 
inadequacies of density functional approaches in the field of QDs have by now 
gained broad recognition \cite{elle06,lebu05,lebu06}.  

In particular, unlike the Hartree-Fock case for which a consistent theory
for the restoration of broken symmetries has been developed (see, e.g., the
earlier Refs.\ \cite{rs,py,low55,low64,fuk}; for developments in the area
of quantum dots, see the more recent Refs.\ 
\cite{yl01,yl02.1,yl02.2,yl03.1,yl02.3}), the breaking of space symmetry 
within the spin-dependent density functional theory poses \cite{perd95} a 
serious dilemma. This dilemma has not been resolved \cite{savi96} to date; 
several remedies are being proposed, but none of them appears to be completely
devoid of inconsistencies. In particular, a theory for symmetry restoration of
broken-symmetry solutions \cite{lebu05,lebu06} within the framework of density 
functional theory has not been developed as yet. This puts the density 
functional methods in a clear disadvantage with regard to the modern fields of
quantum information and quantum computing; for example, the description of
{\it quantum entanglement\/} (see \sref{secentan} below) requires the ability 
to calculate many-body wave functions exhibiting good quantum numbers, and thus
it lies beyond the reach of density functional theory. 

Moreover, due to the unphysical self-interaction error, the density-functional
theory becomes erroneously more resistant to space symmetry breaking 
\cite{baue96} compared to the UHF (which is free from such an error), and thus
it fails to describe a whole class of broken symmetries involving electron 
localization, e.g., the formation at $B=0$ of Wigner molecules in quantum dots
\cite{yl99,yl02.3} and in thin quantum wires \cite{tosi07}, 
the hole trapping at Al impurities in silica \cite{laeg01}, or the interaction
driven localization-delocalization transition in $d$- and $f$- electron 
systems, like Plutonium \cite{savr01}.

Recently, the shortcomings of the density functional theory to properly 
describe magnetic phenomena (such as exchange coupling constants associated 
with symmetry breaking of the total spin) has attracted significant attention 
in the quantum chemistry literature (see, e.g., Refs.\ 
\cite{adam06,rong06,illa06}).

\subsection{More on symmetry restoration methods}

In the framework of post-Hartree-Fock hierarchical approximations, projection 
techniques are one of methods used to treat correlations beyond the 
unrestricted Hartree-Fock. Two other methods are briefly discussed in this 
section, i.e., the method of symmetry restoration via random
phase approximation (RPA) and the generator coordinate method (GCM).  

\subsubsection{Symmetry restoration via random phase approximation.}

This method introduces energy correlations by considering the effect of
the zero-point motion of normal vibrations associated with the small
amplitude motion of the time-dependent-Hartree-Fock mean field (which is
equivalent to the RPA). In the case of space symmetry breaking, one of the
RPA vibrational frequencies vanishes, and the corresponding motion is 
associated with the rotation of the system as a whole (rotational Goldstone 
mode), with a moment of inertia given by the so-called Thouless-Valatin
expression \cite{thou62}.

The method has been used to calculate correlation energies of atomic nuclei 
\cite{hagi01,nazm02} and most recently to restore the broken symmetry in 
circular quantum dots \cite{serr03} (mainly for the case of two electrons at
zero magnetic field). As discussed in Ref.\ \cite{serr03}, restoration of the 
total spin cannot be treated within RPA.

\subsubsection{The generator coordinate method.}
\label{gencm}

The projection techniques by themselves do not take into account quantum 
correlation effects arising from the vibrations and other large-amplitude
intrinsic collective distortions of the Wigner molecule. For the inclusion of 
the effects of such collective motions, a natural extension beyond projection 
techniques is the generator coordinate method (see Ch 10 in Ref.\ 
\cite{rs}). Unlike the RPA, the GCM can treat large-amplitude collective 
motion in combination with the retoration of the total spin. Indeed, it has 
been shown that the RPA harmonic vibrations are a limiting small-amplitude 
case of the large-amplitude collective motion described via the generator 
coordinate method \cite{rs}.

The GCM represents an additional step in the hierarchy of approximations 
described in \sref{synopsis} and its use will result in a further reduction of
the difference from the exact solutions.
The GCM is complicated and computationally more expensive compared to 
projection techniques. Recent computational advances, however, have allowed 
rather extensive applications of the method in nuclear physics (see, e.g., 
Ref.\ \cite{bend03}). As yet, applications of the GCM to quantum dots or 
trapped atomic gases have not been reported.

The GCM employs a very general form for the trial many-body wave functions
expressed as a continuous superposition of determinants $|\Psi[a]\rangle$ (or 
permanents for bosons), i.e., 
\begin{equation}
|\Phi\rangle = \int \rmd [a] f[a] |\Psi[a]\rangle,
\label{psigcm}
\end{equation}
where $[a]=(a_1,a_2,\ldots,a_k)$ is a set of collective parameters depending 
on the physics of the system under consideration. 
An example of such parameters are the azimuthal angles 
$(\gamma_1, \gamma_2,\ldots, \gamma_r)$ in the REM trial wave function 
\eref{wfprj1}. Of course the crucial difference between the REM wave function
\eref{wfprj1} and the general GCM function \eref{psigcm} is the fact that the
weight coefficients $f[a]$ in the former are known in advance (they coincide
with the characters of the underlying symmetry group), while in the latter
they are calculated numerically via the Hill-Wheeler-Griffin equations
\cite{hill53,grif57}
\begin{equation}
\int \rmd [a^\prime] h(a,a^\prime)
f[a^\prime] = 
E \int \rmd [a^\prime] n(a,a^\prime) f[a^\prime], 
\label{hwgeqs}
\end{equation}
where $E$ are the eigenenergies, and 
\begin{equation}
h(a,a^\prime)=\langle \Psi[a] | {\cal H} |\Psi[a^\prime]\rangle,
\label{haap}
\end{equation}
\begin{equation}
n(a,a^\prime)=\langle \Psi[a] |\Psi[a^\prime]\rangle
\label{naap}
\end{equation}
are the Hamiltonian and overlap kernels. The Hill-Wheeler-Griffin 
equation \eref{hwgeqs} is usually solved numerically by 
discretization; then one can describe it as a diagonalization of the many-body 
hamiltonian in a nonorthogonal basis formed with the determinants
$|\Psi[a]\rangle$. 

An example of a potential case for the application of the GCM is an
anisotropic quamtum dot ($\zeta< 1$, with $\zeta=\omega_x / \omega_y$)
In this case, one cannot use projection techniques to restore the total
angular momentum, since the external 
confinement does not possess circular symmetry. Application of the GCM, 
however, will produce numerical values for the expansion coefficients 
$f[\gamma]$, and these values will reduce to 
$\exp [ \rmi \sum_{q=1}^r \gamma_q L_q]$ for
the circular case $\zeta=1$ [while the GCM wave function will reduce to the
REM wave function \eref{wfprj1}]. It is apparent that the GCM many-body wave 
function changes continuously with varying anisotropy $\zeta$, although the 
symmetry properties of the confinement potential change in an abrupt way at 
the point $\zeta \rightarrow 1$.

\section{Symmetry breaking and subsequent symmetry restoration for neutral 
and charged bosons in confined geometries: Theoretical framework}

\subsection{Symmetry breaking for bosons, Gross-Pitaevskii wave functions, 
and permanents}

Mean-field symmetry breaking for bosonic systems is transparent in the context
of two-component condensates, where each species is necessarily associated with
a different space orbital \cite{oehb98,esr99}. For one species of bosons, 
symmetry breaking can be considered through a generalization of the UHF method
of {\it different orbitals for different spins\/} known from the case of 
electrons in quantum chemistry and in quantum dots (\sref{selfpopnes}). Indeed,
as shown in Refs.\ \cite{roma04,roma06,alon05}, one can allow each bosonic 
particle to occupy a different orbital $\phi_i({\bf r}_i)$. 
Then the {\it permanent\/}
$|\Phi_N \rangle = {\rm Perm}[\phi_1({\bf r}_1), \ldots, \phi_N({\bf r}_N)]$
serves as the many-body wave function of an {\it unrestricted\/}
Bose-Hartree-Fock (UBHF) approximation. This wave function
reduces to the Gross-Pitaevskii form with the {\it restriction\/}
that all bosons occupy the same orbital $\phi_0({\bf r})$,
i.e., $|\Phi^{\rm{GP}}_N \rangle =\prod_{i=1}^N \phi_0({\bf r}_i)$, and
$\phi_0({\bf r})$ is determined self-consistently at the restricted
Bose-Hartree-Fock (RBHF) level via the equation \cite{esr97}
\begin{equation}
 [ H_0({\bf r}_1) + (N-1) \int d{\bf r}_2 \phi^*_0({\bf r}_2)
v({\bf r}_1,{\bf r}_2) \phi_0({\bf r}_2)] \phi_0({\bf r}_1) 
= \varepsilon_0 \phi_0({\bf r}_1).
\label{gpe}
\end{equation}
Here $v({\bf r}_1,{\bf r}_2)$ is the two-body repulsive interaction,
which is taken to be a contact potential,
$v_{\delta}= g\delta({\bf r}_1 -{\bf r}_2)$, for neutral bosons, or the
Coulomb repulsion $v_C=e^2Z^2/(\kappa |{\bf r}_1 -{\bf r}_2|)$ for charged
bosons. The single-particle Hamiltonian is given by 
\begin{equation}
H_0({\bf r}) = -\hbar^2 \nabla^2 /(2m) + m \omega_0^2 {\bf r}^2/2, 
\end{equation}
where $\omega_0$ characterizes the circular harmonic confinement, and where we
have considered a non-rotating trap.

Compared to the fermionic case (see \sref{selfpopnes}), the self-consistent
determination of the UBHF orbitals is rather complicated and numerically highly 
demanding (see section 2.5.3 in Ref.\ \cite{romathesis}). In fact,
the self-consistent UHBF equations cannot be put into
a standard (canonical) eigenvalue-problem form due to two reasons:(i) unitary 
transformations cannot be used to simplify the equations, since the permanent of
the product of two matrices does not factorize into a product of two simpler
terms (unlike the electronic case where the determinant of the product of two
matrices is equal to the product of the corresponding determinants) and (ii) 
as a result of boson statistics, the bosonic orbitals cannot be assumed to be 
(and remain) orthogonal, which leads to additional coupling terms between
the non-orthogonal orbitals. 

In the literature \cite{cede03}, an attempt has been made to derive unrestricted
self-consistent equations for bosons by disregarding point (ii) mentioned 
above and invoking the assumption of orthonormal orbitals. Such equations of 
course are not of general validity, although they appear to be useful for 
describing fragmentation of Bose condensates in double wells.

The difficulties of the self-consistent treatment can be bypassed and 
the UBHF problem can be simplified through consideration of explicit analytic 
expressions for the space orbitals $\phi_i({\bf r}_i)$. In particular, for 
repulsive interactions, {\it the bosons must avoid each other  
in order to minimize their mutual repulsion\/}, and thus, in analogy with the
case of electrons in QDs, one can take all the orbitals to be of the form of 
displaced Gaussians, namely, 
\begin{equation}
\phi_i({\bf r}_i) = \pi^{-1/2} \sigma^{-1}
\exp[-({\bf r}_i - {\bf a}_i)^2/(2 \sigma^2)]. 
\end{equation}
The positions ${\bf a}_i$ describe the vertices of concentric regular 
polygons, with both the width
$\sigma$ and the radius $a=|{\bf a}_i|$ of the regular polygons
determined variationally through minimization of the total energy
\begin{equation}
E_{\rm{UBHF}} = \langle \Phi_N | {\cal H} | \Phi_N \rangle
/\langle \Phi_N | \Phi_N \rangle, 
\end{equation}
where
\begin{equation}
{\cal H} = 
\sum_{i=1}^N H_0({\bf r}_i) + \sum_{i < j}^{N} v( {\bf r}_i,{\bf r}_j)
\end{equation}
is the many-body Hamiltonian.

With the above choice of localized orbitals the unrestricted permanent
$|\Phi_N \rangle$ breaks the continuous rotational symmetry. However,
the resulting energy gain becomes substantial for stronger repulsion.
Controlling this energy gain (the strength of correlations) is the ratio
$R_\delta$ (or $R_W$) between the strength of the 
repulsive potential and the zero-point kinetic energy. Specifically, for a 2D
trap, one has 
\begin{equation}
R_{\delta} = gm/(2\pi\hbar^2)
\end{equation}
for a contact potential (for $R_W$, see \sref{wigparssb}). Note that in this 
section, we refer to the case of a non-rotating trap; the generalization to 
rotating traps is presented later in \sref{bosmol}. 

\subsection{Repelling bosons in harmonic traps: Restoration of broken 
symmetry.}
\label{repbos}

Although the optimized UBHF permanent $|\Phi_N \rangle$ performs very well 
regarding the total energies of the trapped bosons, in particular in 
comparison to the resctricted wave functions (e.g., the GP anzatz), it is 
still incomplete. Indeed, due to its localized orbitals, $|\Phi_N \rangle$
does not preserve the circular (rotational) symmetry of the 2D many-body 
Hamiltonian $H$. Instead, it exhibits a lower point-group symmetry, i.e., a 
$C_2$ symmetry for $N=2$ and a $C_5$ one for the $(1,5)$ structure of $N=6$ 
(see \sref{bosmol} below). As a result, $|\Phi_N \rangle$ does not have a good
total angular momentum. In analogy with the case of electrons in quantum dots,
this paradox is resolved through a
post-Hartree-Fock step of {\it restoration\/} of broken symmetries via 
projection techniques \cite{yl02.1,yl02.2,yl03.1,roma04}, yielding a new wave
function $|\Psi_{N,L}^{\rm{PRJ}} \rangle$ with a definite angular
momentum $L$, that is
\begin{equation}
2 \pi |\Psi_{N,L}^{\rm{PRJ}} \rangle = \int^{2\pi}_0 \rmd \gamma
|\Phi_{N}(\gamma) \rangle \rme^{\rmi\gamma L},
\label{wfprj2}
\end{equation}
where $|\Phi_{N}(\gamma) \rangle$ is the original UBHF permanent having each
localized orbital rotated by an azimuthal angle $\gamma$, with $L$ being
the total angular momentum. The projection yields wave functions for a whole
rotational band. Note that the projected wave function
$|\Psi_{N,L}^{\rm{PRJ}} \rangle$ in \eref{wfprj2} may be regarded
as a superposition of the rotated permanents $|\Phi_{N}(\gamma) \rangle$,
thus corresponding to a ``continuous-configuration-interaction'' solution.

The energies of the projected states are given by
\begin{equation}
E_L^{\rm{PRJ}} = \langle \Psi_{N,L}^{\rm{PRJ}}  | {\cal H} |
\Psi_{N,L}^{\rm{PRJ}}  \rangle /
\langle \Psi_{N,L}^{\rm{PRJ}}  | \Psi_{N,L}^{\rm{PRJ}}
\rangle.
\label{eprjl}
\end{equation}

\section{Other many-body methods}

\subsection{Exact diagonalization methods: Theoretical framework}
\label{exdmeth}

We will discuss the essential elements of the exact-diagonalization 
method here by considering 
the special, but most important case of a many-body Hilbert space defined via 
the restriction that the single-particle states belong exclusively to the 
lowest Landau level. For 2D electrons, this LLL restriction is appropriate in 
the case of very high $B$. For rotating bosons in a harmonic trap, the LLL
restriction is appropriate for $\Omega \sim \omega_0$ and a very weak 
repulsive contact potential. The particulars of the EXD method for quantum dots
in the case of field-free (and/or low $B$) conditions will be discussed below 
in \sref{exdell}.

For sufficiently high magnetic field values (i.e., in the fractional 
quantum Hall effect, regime), the electrons are fully spin-polarized 
and the Zeeman term (not shown here) does not need to be considered. 
In the $B \rightarrow \infty$ limit, the external confinement $V(x,y)$ can 
be neglected, and the many-body $H$ can be restricted to operate in the lowest
Landau level, reducing to the form \cite{yl02.2,yl03.2,yl04.2,li06}
\begin{equation}
{\cal H}_{\rm LLL} = N \frac{\hbar \omega_c}{2} +
\sum_{i=1}^N \sum_{j>i}^N \frac{e^2}{\kappa r_{ij}},
\label{hlll}
\end{equation}
where $\omega_c = eB/(m^*c)$ is the cyclotron frequency. Namely, one needs
to diagonalize the interaction Hamiltonian only. 

For the case of rotating bosons in the LLL, one needs to replace in \eref{hlll}
the Coulomb interaction by $g \delta({\bf r}_i - {\bf r}_j)$ and the
cyclotron frequency by $2\Omega$ (see the Appendix for the details of the
equivalence between magnetic field $B$ and rotational frequency $\Omega$;
in short $\omega_c \rightarrow 2 \Omega$).

For a given total angular momentum $L=\sum_{k=1}^N l_k$, the 
exact-diagonalization $N$-body wave function is a linear superposition of 
Slater determinants for fermions (or permanents for bosons) $\Psi(J)$ made out
of lowest-Landau-level single-particle wave functions (see the Appendix),
\begin{equation}
\phi_l(z)= \frac{1}{\Lambda \sqrt{\pi l!}} \left( \frac{z}{\Lambda} \right)^l
e^{-zz^*/(2\Lambda^2)},
\label{splll}
\end{equation}
where $\Lambda=\sqrt{2\hbar c/(eB)}=l_B \sqrt{2}$ for the case of electrons in 
QDs ($l_B$ being the magnetic length), and $\Lambda=\sqrt{\hbar/(m\omega_0)}$
for the case of bosons in rotating traps. In \eref{splll}, we used complex 
coordinates $z=x+ \rmi y$, instead of the usual vector positions 
${\bf r}=(x,y)$; below we will use either notation interchangeably as needed.

Thus, the many-body EXD wave function is written as
\begin{equation}
\Phi^{\rm EXD}(z_1,z_2,\ldots,z_N) = \sum_{J=1}^K C_J \Psi(J)
\label{exdwf}
\end{equation}
with the index $J$ denoting any set of $N$ single-particle angular
momenta $\{l_1,l_2,\ldots,l_N\}$ such that
\begin{equation}
l_1 < l_2 <  \ldots < l_N
\label{sldcf}
\end{equation}
for fermions and
\begin{equation}
l_1 \leq l_2 \leq \ldots \leq l_N
\label{sldcb}
\end{equation}
for bosons, the absence or presence of the equal signs being determined 
by the different statistics between fermions and bosons, respectively. 

Using the expansion \eref{exdwf}, one transforms the many-body Schr\"{o}dinger
equation
\begin{equation}
{\cal H} \Phi^{\rm EXD}(z_1,z_2,\ldots,z_N) = E \Phi^{\rm EXD}(z_1,z_2,\ldots,z_N),
\label{mbsch}
\end{equation}
into a matrix eigenvalue problem. When the single-particle states are 
orthonormal [like the LLL ones in \eref{splll}], the matrix elements 
$\langle \Psi(I)|{\cal H}|\Psi(J) \rangle$ 
between two Slater determinants are calculated
using the so-called Slater rules (see, e.g., Ch 2.3.3 of Ref.\ \cite{so}). 
For the case of bosons, the correpsonding rules for the matrix elements 
$\langle \Psi(I)|{\cal H}|\Psi(J) \rangle$ between two permanents are given 
in the appendix of Ref.\ \cite{baks07}.

We remark here that the calculation of energies associated with projected
wave functions [see, e.g., \Eref{eprjl}] requires calculation of similar matrix
elements between two Slater determinants (or permanents) with 
{\it non-orthogonal\/} orbitals; the corresponding formulas for the case of 
fermions can be found in Ch 6.3. of Ref.\ \cite{hurl}, and for the case of 
bosons in Ref.\ \cite{romathesis}. 

Of course a necessary ingredient for the application of the above rules is
the knowledge of the matrix elements of the two-body interaction 
$v({\bf r}_1, {\bf r}_2)$, i.e.,
\begin{equation}
v_{\alpha\beta\gamma\Delta} \equiv
\int \rmd{\bf r}_1 \rmd{\bf r}_2 \phi^*_\alpha(1) \phi^*_\beta(2) v(1,2) 
\phi_\gamma(1) \phi_\Delta(2).
\label{tbme}
\end{equation}

In the general case, these two-body matrix elements need to be calculated
numerically. For the simpler case specified by the LLL orbitals \eref{splll},
the two-body matrix elements are given by analytic expressions. In particular,
for the Coulomb repulsion, see Refs.\ \cite{yl03.2,tsip02}; for a contact 
potential, see Ref.\ \cite{pape99}.  

The  dimension ${\cal D}$ of the Hilbert space increases very fast with the
number of particles $N$ and the value of the total angular momentum $L$, and
is is controlled by the maximum allowed single-particle
angular momentum $l_{\rm{max}}$, such that $l_k \leq l_{\rm{max}}$,
$1 \leq k \leq N$. 
By varying $l_{\rm{max}}$, we can check that this choice produces well
converged numerical results.

For the solution of the large scale, but sparse, EXD matrix eigenvalue problem
associated with the special Hamiltonian $H_{\rm{LLL}}$ [or the general one
in \eref{mbhn}], we use the ARPACK computer code \cite{arpack}. 

The availabilty of analytic expressions for the two-body interaction has
greatly facilitated exact-diagonalization calculations in the 
lowest Landau level (appropriate for quantum dots at high 
$B$), and in this case (starting with Refs.\ \cite{yl03.2,tsip01}) 
diagonalization of large matrices of dimensions of order 500,000 x 500,000 has 
become a commom occurrence. For circular quantum dots, similar analytic 
expressions for the matrix elements of the Coulomb 
interaction between general Darwin-Fock orbitals \cite{darw,fock}
(i.e., the single-particle
orbitals of a circular 2D harmonic oscillator under a perpendicular magnetic
field $B$) are also available \cite{girv83,anis98}, 
but they are not numerically as stable as Tsiper's expressions \cite{tsip02} 
in the lowest Landau level. 

Exact-diagonalization calculations for field-free (and/or low $B$) conditions 
have been presented in several papers. Among them, we note 
the exact-diagonalization calculations of Refs.\ \cite{ront06,pfan93,
eto97,haw93,mik02.3,mik02.4,tave03,hawr04}. EXD calculations employing
Coulombic two-body matrix elements that are calculated {\it numerically\/} 
have also been reported for elliptic quantum dots (see \sref{exdell} below).  
Furthermore, some authors have used the method of hyperspherical harmonics 
\cite{rua99} for circular quantum dots, while others have carried out 
exact-diagonalization calculations for quantum dots with a 
polygonal external confinement \cite{cref99}.
 
Concerning EXD calculations in the lowest Landau level, we 
mention Refs.\ \cite{yl03.2,nie04,maks96,girv83,mc90,sek96,maks00,laug83.2,
jain96,man00,jeon07} for the case of quantum dots (high $B$) and
Refs.\ \cite{wilk00,vief00,wilk01,baks07,barb06,mann06} for the case of 
bosons in rapidly rotating traps. A version of EXD in the LLL uses a 
correlated basis constructed out of composite-fermion wave functions
\cite{jeon07}, while another exact-diagonalization version used non-orthogonal
floating Gaussians in the place of the usual single-particle states 
\eref{splll} in the LLL. 

For two electrons in a single quantum dot, exact calculations have been 
carried out through separation into center-of-mass and relative coordinates
\cite{yl00.2,wagn92,drou04}. 
In addition, EXD calculations have been reported for two electrons in a double
quantum dot \cite{melni06,nie02.2}. 

It is of interest to note that the EXD approach is also used in other fields,
but under different names. In particular, the term ``shell model calculations''
is used in nuclear theory, while the term ``full configuration interaction''
is employed in quantum chemistry.  

\begin{figure}[t]
\centering\includegraphics[width=10cm]{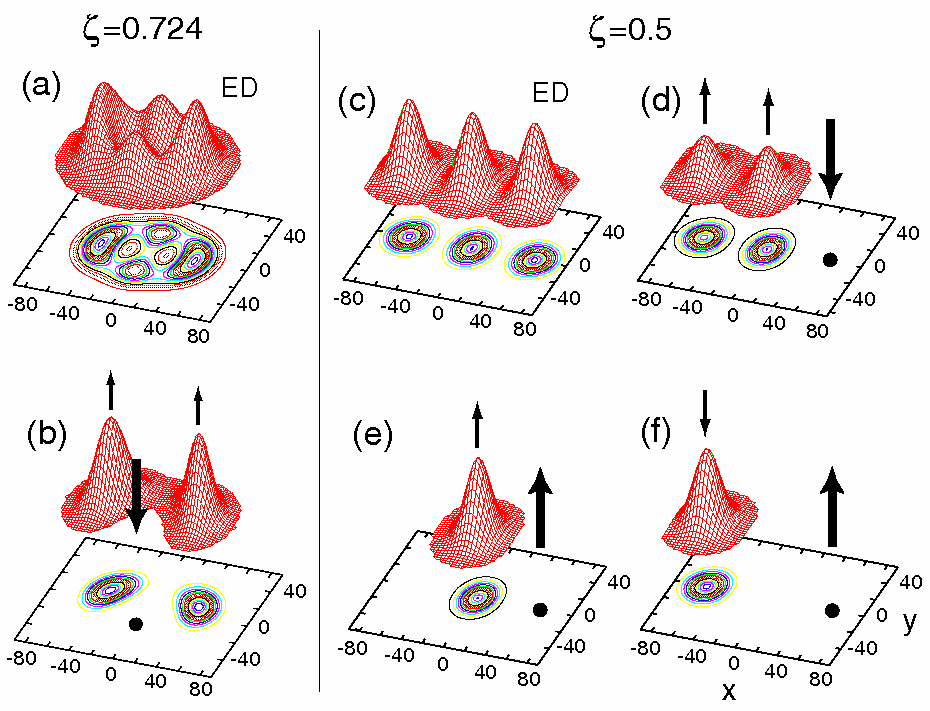}
\caption{(Color online) Exact-diagonalization electron densities (EDs) and
spin-resolved CPDs for $N = 3$ electrons in an anisotropic quantum dot
at zero magnetic field ($B=0$) and in the case of a strong Coulomb repulsion 
with a dielectric constant $\kappa=1$.
(a-b) Results for the ground state (which has total spin $s=1/2$ and spin 
projection $S_z=1/2$) when $\hbar \omega_x = 4.23$ meV and 
$\hbar \omega_y = 5.84$ meV (i.e., for an intermediate anisotropy 
$\zeta=0.724$). (c-f) Results for the first excited state (which has also 
total spin $s=1/2$ and spin projection $S_z=1/2$) when 
$\hbar \omega_x = 3.137$ meV and $\hbar \omega_y = 6.274$ meV (i.e., for a
strong anisotropy $\zeta=0.5$). The thick arrows and solid dots in the CPDs 
indicate the spin direction $\sigma_0$ and position ${\bf r}_0$ of the fixed 
electron [see \eref{sponcpd}]. The thin arrows indicate the spin direction of 
the remaining two electrons. The effective mass is $m^*=0.070m_e$ for the 
intermediate anisotropy (a-b) and $m^*=0.067m_e$ for the strong anisotropy 
(c-f). Lengths are in nanometers. The vertical axes are in arbitrary units.
}
\label{exd3e}
\end{figure}

\subsubsection{An example involving spin-resolved CPDs.}

Here we present an example of an EXD calculation exhibiting formation 
of a Wigner molecule in quantum dots. The case we chose is that of 
$N=3$ electrons under zero magnetic field in an 
anisotropic quantum dot with $\hbar \omega_x=4.23$ meV and $\hbar \omega_y=5.84$
meV (i.e., with an intermediate anisotropy $\zeta=\omega_x/\omega_y=0.724$) and
dielectric constant $\kappa=1$ (strong interelectron repulsion).
In particular, \fref{exd3e} displays results for the ground state of the
three electrons with total spin $s=1/2$ and spin projection $S_z=1/2$.

The electron density in \fref{exd3e}(a) has the shape of a diamond and suggests
formation of a Wigner molecule resonating between two isosceles triangular 
isomers (which are the mirror image of each other). The detailed interlocking
of the two triangular configurations is further revealed in the spin-resolved 
CPD that is displayed in \fref{exd3e}(b). It can be concluded that one triangle 
is formed by the points ${\bf R}_1 = (0,-20)$ nm, ${\bf R}_2 = (-43,10)$ nm, 
and ${\bf R}_3 = (43,10)$ nm, while the second one (its mirror with respect
to the $x$-axis) is formed by the points ${\bf R^\prime}_1 = (0,20)$ nm,
${\bf R^\prime}_2 = (-43,-10)$ nm, and ${\bf R^\prime}_3 = (43,-10)$ nm.

The two-triangle configuration discussed for three electrons above may be seen 
as the embryonic precursor of a quasilinear structure of two intertwined 
``zig-zag'' crystalline chains. Such double zig-zag crystaline chains may
also be related to the single zig-zag Wigner-crystal chains discussed recently 
in relation to spontaneous spin polarization in quantum wires 
\cite{klir06,piac04}.

For strong anisotropies (e.g., $\zeta \leq 1/2$), the three electrons form a 
straightforward linear Wigner molecule [see the electron density in 
\fref{exd3e}(c)], and the spin-resolved CPDs can be used
to demonstrate \cite{li07} formation of prototypical entangled states, like 
the so-called $W$ states \cite{cira00,woot00}. From the CPDs [displayed in 
\fref{exd3e}(d), \fref{exd3e}(e) and \fref{exd3e}(f)] of the first excited 
state (having $s=1/2$ and $S_z=1/2$), one can infer that its intrinsic spin 
structure is of the form $| \uparrow  \uparrow  \downarrow \;\rangle 
-| \downarrow  \uparrow  \uparrow \;\rangle$. 
The ground state (not shown) of this linear Wigner molecule has
also a total spin $s=1/2$ and spin projection $S_z=1/2$, and its intrinsic
spin structure corresponds to a form $2 | \uparrow  \downarrow  \uparrow 
\;\rangle - | \uparrow  \uparrow  \downarrow \;\rangle - | \downarrow
 \uparrow  \uparrow \;\rangle$ \cite{li07}.  

\subsection{Particle localization in Monte-Carlo approaches}

Quantum Monte-Carlo (MC) approaches \cite{qmcbook} have been successfully used 
in many areas of condensed-matter physics; they are divided in two main 
branches, path-integral MC (PIMC) and variational/diffusion MC (V/DMC). 
Unlike the exact diagonalization, quantum MC approaches cannot calculate
excited states, and they are restricted to the description of ground-state 
properties

Applied to the case of circular quantum dots at zero magnetic field, PIMC 
calculationss \cite{egger99,loz00,hart00,reus03,egger05} have been able to 
reproduce and describe electron localization with increasing $R_W$ and 
formation of Wigner molecules. In 2D quantum dots, a focus of the PIMC studies 
\cite{egger99,reus03} has been the determination of the critical value, 
$R_W^{\rm cr}$, for the Wigner parameter at which the phase transition from a 
Fermi liquid to a Wigner molecule occurs. Naturally, only an estimate of 
this critical value can be determined, since the phase transition is 
not sharp, but smooth, due to the finite size of the quantum dot. Obtaining a 
precise value of $R_W^{\rm cr}$ is also hampered by the variety of criteria 
employed by different researchers in the 
determination of this transition (e.g., height of localized density humps,  
appearance of a hump at the center of the dot, etc.). 

In the literature of PIMC studies \cite{egger99,reus03}, one finds the 
critical value $R_W^{\rm cr} \sim 4$, which is in agreement with 
exact-diagonalization studies \cite{ront06}. This is also in general agreement
with the estimate $R_W^{\rm cr} \sim 1$ based on the abrupt onset of spatial 
symmetry breaking in unrestricted Hartree-Fock \cite{yl99}. Of course
the unrestricted-Hartree-Fock estimate has to be refined through the 
subsequent step of symmetry restoration. We believe that it is most 
appropriate to consider these two estimates mentioned above as the lower and 
upper limit of a transition region. The important conclusion is that the 
transition to Wigner crystallization in quantum dots takes place for much 
higher electron densities compared to the infinite two-dimensional electron 
gas (for which a value $R_W^{\rm cr} \sim 37$ \cite{cepe89} has been 
reported.)\footnote[7]{
Often the Wigner-Seitz radius $r_s$, in units of the effective Bohr radius
$a^*_B=\hbar^2\kappa/(m^*e^2)$ of the quantum dot, is used instead of
the Wigner parameter $R_W$ (denoted some times by $\lambda$). In these units,
one has $r_s \approx R_W$.}

A disadvantage of the PIMC method is that the case of an applied magnetic
field cannot be easily incorporated in its formalism, and therefore related
studies have not been reported. Other well known difficulties are the
fermion sign problem and the nonconservation of total spin 
\cite{ront06,egger99}.

Commenting on the other main branch of quantum Monte Carlo, i.e., the
variational/diffusion MC, we wish to stress the crucial role played by the 
general form of the trial wave function used. Indeed, an early V/DMC study 
\cite{pede00} using a single configurational state 
function (i.e., a primitive combination of products of Slater determinants for
the two spin directions that is an eigenstate of the total angular momentum 
$\hat{L}$, the square of the total spin $\hat{\bf S}^2$, and the total-spin 
projection $S_z$) was unable to describe the formation of Wigner molecules in 
quatum dots at zero magnetic field. Another V/DMC study \cite{nie02.1} managed 
to demonstrate electron localization, but at the cost of using a single product
of two Slater determinants (multiplied by a Jastrow factor) which violated the 
conservation of both the total angular momentum and total spin (without the 
possibility of further corrections related to symmetry restoration). 

Most recently, more sophisticated trial wave functions involving a large 
number of configurational state functions with good total angular momentum and
total spin have been employed, which enabled eventually confirmation of 
the formation of Wigner molecules via V/DMC methods, both at
zero \cite{umri06} and high magnetic field \cite{umri05}.

There are, however, disagreements between the V/DMC results \cite{umri07} 
and those from PIMC and EXD calculations concerning the details 
of Wigner-molecule formation in circular quantum dots in the absence of
an applied magnetic field. In particular, these disagreements focus on 
the density scale for the cross-over and the strength of azimuthal and radial
electron correlations as a function of $R_W$. 

Such disagreements remain an open question for two reasons: \\
(i) The criterion of lowest energy (evoked
by the V/DMC approaches) is not sufficient to guarantee the quality of the
variational many-body wave function. A counterexample to this lowest-energy 
criterion was presented by us for the case of the Laughlin wave functions in 
Refs.\ \cite{yl02.2,yl04.2} (see also \sref{remvslaughlin}). Most recently,
this point was also illustrated within the framework of variational Monte
Carlo calculations \cite{wain07}.\\
(ii) The V/DMC studies for larger $N$ \cite{umri06,umri07} have presented only
calculations for CPDs. However, due to the presence of dummy integrations in 
\eref{cpds} (which result in an averaging over the remaining $N-2$ particles),
the ability of the CPDs to portray the intrinsic crystalline structure of the 
Wigner molecule diminishes with increasing $N$. As a result, higher-order 
correlation functions, like $N$-point correlations, may be required. The fact 
that higher-order correlation functions reflect the crystalline correlations 
more accurately than the CPDs was illustrated for the case of rotating boson 
molecules in Ref.\ \cite{baks07} (see also \sref{exdbos}). 

A detailed comparison between ground-state energies calculated with quantum
MC and exact-diagonalization methods can be found in Ref.\ \cite{ront06}.
For a comparison between variation-before-projection (see \sref{restsym}) and 
V/DMC total energies, see Ref.\ \cite{degi07}.

\section{The strongly correlated regime in two-dimensional quantum dots:
The two-electron problem and its significance}
\label{stcor2e}

In section 2 and section 3, we focused on the general principles and the
essential theoretical framework of the method of symmetry breaking and of 
subsequent symmetry restoration for finite condensed-matter systems. In 
addition, in section 4, we presented the basic elements of the 
exact-diagonalization approach. In the following four sections, we will focus 
on specific applications and predictions from these methods in the field of 
semiconductor quantum dots and of ultracold bosons in harmonic traps,
in particular regarding the emergence and properties of Wigner molecules under
various circumstances. At the same time we will continue to elaborate and 
further expand on more technical aspects of these methods. 

In this section, we start by concentrating on the description of two-electron
molecules in QDs. A discussion on the importance of the two-electron problem 
is given in \sref{histsig}.

\subsection{Two-electron elliptic dot at low magnetic fields}
\label{2eelldot}

Here, we present an exact diagonalization and an approximate
(generalized Heitler-London, GHL) microscopic treatment for two electrons in a
{\it single\/} elliptic QD specified by parameters that correspond to a 
recently fabricated experimental device \cite{elle06}.

The two-dimensional Hamiltonian for the two interacting electrons is given by 
\begin{equation}
{\cal H} = H({\bf r}_1)+H({\bf r}_2)+ \gamma e^2/(\kappa r_{12}),
\label{ham}
\end{equation}
where the last term is the Coulomb repulsion, $\kappa$ (12.5 for GaAs) is the
dielectric constant, and $r_{12} = |{\bf r}_1 - {\bf r}_2|$. The prefactor
$\gamma$ accounts for the reduction of the Coulomb strength due to the finite
thickness of the electron layer in the $z$ direction and for any additional
screening effects due to the gate electrons. $H({\bf r})$ is the
single-particle Hamiltonian for an electron in an external perpendicular
magnetic field ${\bf B}$ and an appropriate confinement potential \eref{hsp}.
For an elliptic QD, the external potential is written as
\begin{equation}
V(x,y) = \frac{1}{2} m^* (\omega^2_{x} x^2 + \omega^2_{y} y^2).
\end{equation}
Here the effective mass is taken to be $m^*=0.07m_0$. In the Hamiltonian 
(\ref{hsp}), we neglect the Zeeman contribution due to the negligible value 
($g^* \approx 0$) of the effective Land\'{e} 
factor in our sample \cite{sali01}.

\subsubsection{Generalized Heitler-London approach.}

The GHL method for solving the Hamiltoninian (\ref{ham}) consists of two steps.
In the first step, we solve selfconsistently the ensuing
unrestricted Hartree-Fock equations allowing for lifting of the
double-occupancy requirement (imposing this requirement gives the
{\it restricted\/} HF method, RHF). For the $S_z=0$ solution, this step 
produces two single-electron orbitals $u_{L,R}({\bf r})$ that are localized 
left $(L)$ and right $(R)$ of the
center of the QD [unlike the RHF method that gives a single doubly-occupied
elliptic (and symmetric about the origin) orbital].
At this step, the many-body wave function is a single Slater
determinant $\Psi_{\rm{UHF}} (1\uparrow,2\downarrow) \equiv
| u_L(1\uparrow)u_R(2\downarrow) \rangle$ made out of the two occupied UHF
spin-orbitals $u_L(1\uparrow) \equiv u_L({\bf r}_1)\alpha(1)$ and
$u_R(2\downarrow) \equiv u_R({\bf r}_2) \beta(2)$, where
$\alpha (\beta)$ denotes the up (down) [$\uparrow (\downarrow)$] spin.
This UHF determinant is an eigenfunction of the projection $S_z$ of the total
spin $\hat{S} = \hat{s}_1 + \hat{s}_2$, but not of $\hat{S}^2$ (or the parity
space-reflection operator).

In the second step, we restore the broken parity and total-spin symmetries by
applying to the UHF determinant the projection operator \eref{prjp}. For two
electrons, this operator reduces to 
${\cal P}_{\rm spin}^{s,t}=1 \mp \varpi_{12}$, where the
operator $\varpi_{12}$ interchanges the spins of the two electrons;
the upper (minus) sign corresponds to the singlet.
The final result is a generalized Heitler-London two-electron wave function
$\Psi^{s,t}_{\rm{GHL}} ({\bf r}_1, {\bf r}_2)$ for the ground-state singlet
(index $s$) and first-excited triplet (index $t$), which uses
the UHF localized orbitals,
\begin{equation}
\Psi^{s,t}_{\rm{GHL}} ({\bf r}_1, {\bf r}_2) \propto
{\bf (} u_L({\bf r}_1) u_R({\bf r}_2) \pm u_L({\bf r}_2) u_R({\bf r}_1) {\bf )}
\chi^{s,t},
\label{wfghl}
\end{equation}
where $\chi^{s,t} = (\alpha(1) \beta(2) \mp \alpha(2) \beta(1))$ is the spin
function for the 2$e$ singlet and triplet states.
The general formalism of the 2D UHF equations and of the subsequent restoration
of broken spin symmetries was presented in section 2.2.

The use of {\it optimized\/} UHF orbitals in the generalized Heitler-London
method is suitable for treating {\it single elongated\/} QDs \cite{yl02.3}, 
including the special case of elliptically deformed ones discussed in this
section. The GHL is equally applicable to double QDs with
arbitrary interdot-tunneling coupling \cite{yl02.3}. In contrast,
the Heitler-London (HL) treatment \cite{hl} (known also as the simple
Valence bond), where non-optimized ``atomic'' orbitals of two isolated QDs are
used, is appropriate only for the case of a double dot with small 
interdot-tunneling coupling \cite{burk99}.

The orbitals $u_{L,R}({\bf r})$ are expanded in a real Cartesian
harmonic-oscillator basis, i.e.,
\begin{equation}
u_{L,R}({\bf r}) = \sum_{j=1}^K C_j^{L,R} \varphi_j ({\bf r}),
\label{uexp}
\end{equation}
where the index $j \equiv (m,n)$ and $\varphi_j ({\bf r}) = X_m(x) Y_n(y)$,
with $X_m(Y_n)$ being the eigenfunctions of the one-dimensional oscillator in the
$x$($y$) direction with frequency $\omega_x$($\omega_y$). The parity operator
${\cal P}$ yields ${\cal P} X_m(x) = (-1)^m X_m(x)$, and similarly for $Y_n(y)$.
The expansion coefficients $C_j^{L,R}$ are real for $B=0$ and complex for finite
$B$. In the calculations we use $K=54$ and/or $K=79$, yielding convergent 
results.

\subsubsection{Exact diagonalization.}
\label{exdell}

In the EXD method, the many-body wave function is written as a linear
superposition over the basis of non-interacting two-electron 
determinants, i.e.,
\begin{equation}
\Psi^{s,t}_{\rm{EXD}} ({\bf r}_1, {\bf r}_2) =
\sum_{i < j}^{2K} {\cal A}_{ij}^{s,t} | \psi(1;i) \psi(2;j)\rangle,
\label{wfexd}
\end{equation}
where $\psi(1;i) = \varphi_i(1 \uparrow)$ if $1 \leq i \leq K$ and
$\psi(1;i) = \varphi_{i-K}(1 \downarrow)$ if $K+1 \leq i \leq 2K$ [and
similarly for $\psi(2,j)$].
The total energies $E^{s,t}_{\rm{EXD}}$ and the coefficients
${\cal A}_{ij}^{s,t}$ are obtained through a ``brute force'' diagonalization of
the matrix eigenvalue equation corresponding to the Hamiltonian in 
\eref{ham}. The exact-diagonalization wave function does not
immediately reveal any particular form, although, our calculations below
show that it can be approximated by a GHL wave function in the case of the
elliptic dot under consideration.

\subsubsection{Results and comparison with measurements.}

To model the experimental quantum dot device, we take, following Ref.\
\cite{elle06}, $\hbar \omega_x=4.23$ meV, $\hbar \omega_y=5.84$ meV, 
$\kappa=12.5$, and $\gamma=0.862$. The corresponding anisotropy is 
$\omega_y/\omega_x=1.38$, indicating that the quantum dot considered here is 
closer to being circular than in other experimental systems 
\cite{zumb04,kyri02}. 

\begin{figure}[tbhp]
\centering\includegraphics[width=8cm]{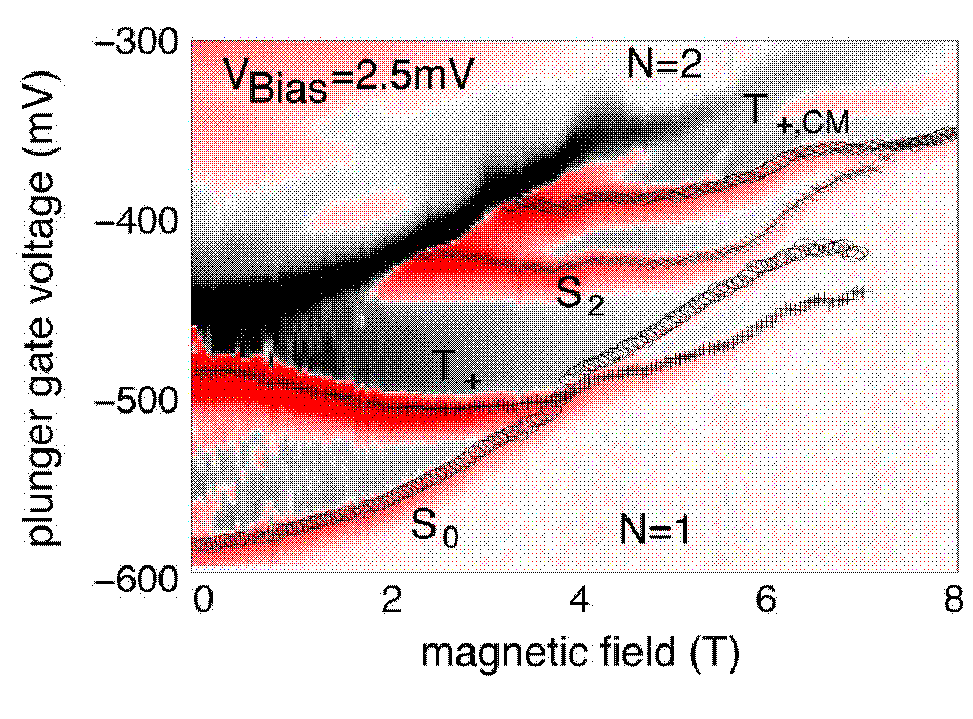}
\caption{(Color online) Differentiated current $\rmd I/\rmd V_\mathrm{pg}$ at
$V_\mathrm{bias}=2.5$~mV (the subscript ${\rm pg}$ denotes the plunger gate). 
Gray striped regions (red online) marked by symbols correspond to positive
(peaks) $dI/dV_\mathrm{pg}$.
The dark black region (also black online) corresponds to negative
$dI/dV_\mathrm{pg}$. Electron numbers $N$ are indicated. Transitions between 
the one-electron ground state and the 2$e$ spin-singlet ground (excited for
$B > 3.8$ T) state ($S_0$), spin-triplet excited (ground for $B > 3.8$ T) 
state ($T_+$), spin-singlet excited state ($S_2$), and
spin-triplet plus center-of-mass excited state ($T_{+,{\rm CM}}$) are labeled.
}
\label{ethex}
\end{figure}
\begin{figure}[tbhp]
\centering\includegraphics[width=6.5cm]{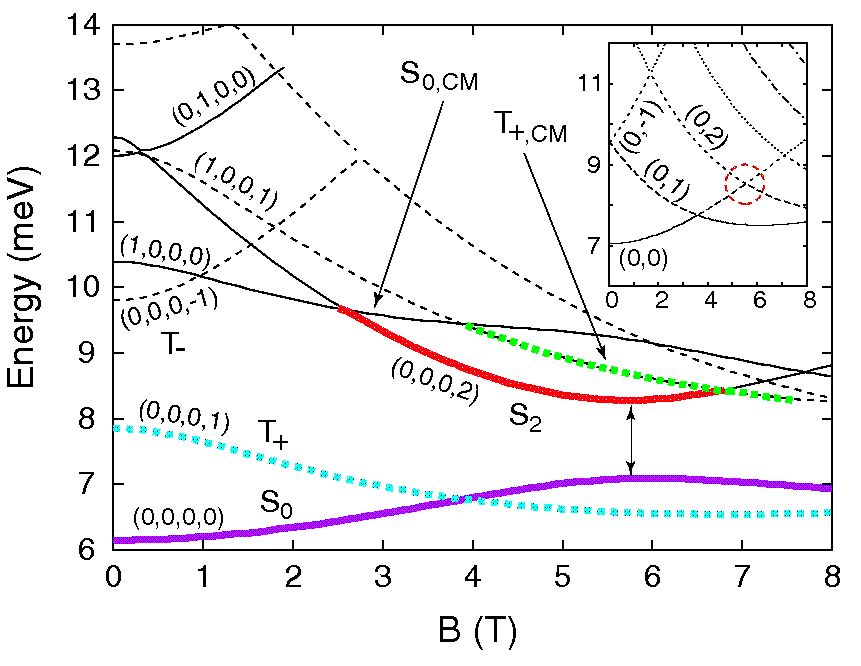}
\caption{(Color online) Calculated exact-diagonalization energy spectrum
in a magnetic field, referenced to $2 \hbar \sqrt{\omega_0^2 + \omega_c^2/4}$, 
of a 2$e$ dot with anisotropic harmonic confinement 
(for the dot parameters, see text). 
We have adopted the notation $(N_x,N_y,n,m)$, where $(N_x,N_y)$ refer to the
CM motion along the $x$ and $y$ axes and $(n,m)$ refer to the number of radial 
nodes and angular momentum of the relative motion in the corresponding 
{\it circular\/} dot.
Inset: The EXD spectrum of the corresponding circular dot. Only the $(n,m)$ 
indices are shown, since $N_x=N_y=0$ for all the plotted curves. 
Solid lines denote singlets. Dashed lines denote triplets.
}
\label{ethth}
\end{figure}

As shown in Ref.\ \cite{elle06}, the experimental findings can be 
quantitatively interpreted by comparing to the results of the EXD calculations 
for two electrons in an anisotropic harmonic confinement potential with the
parameters listed above. All the states observed in the measured spectra
(as a function of the magnetic field) can be unambiguously identified 
\cite{elle06} with calculated ground-state and excited states of the 
two-electron Hamiltonian (compare \fref{ethex} and \fref{ethth}).

\begin{figure}[t]
\centering\includegraphics[width=6.0cm]{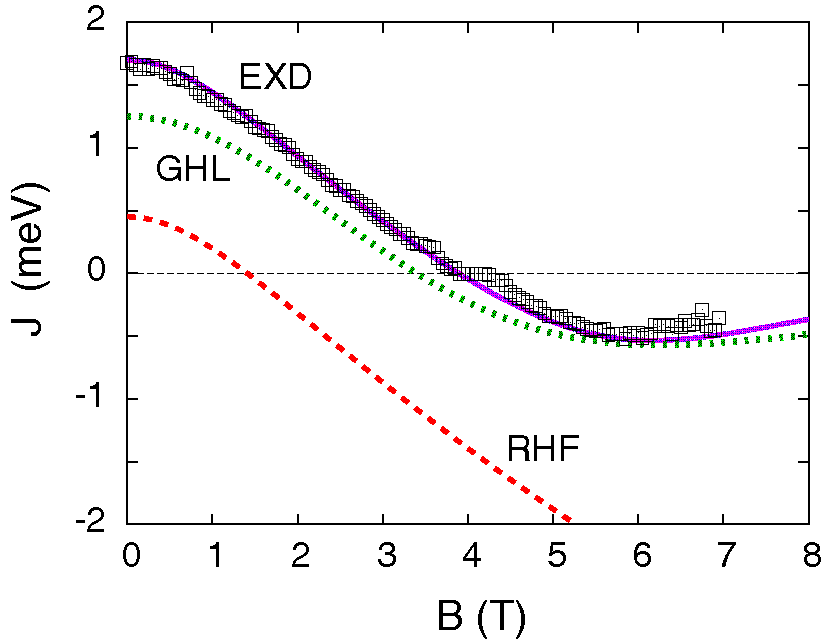}
\caption{(Color online) Comparison between the lowest-triplet/lowest-singlet 
energy splitting [$J(B)$] calculated with different
methods and the experimental results (open squares). 
Solid line (online magenta):
EXD. Dotted line (online green): GHL. Dashed line (online red): RHF.
For the parameters used in the calculation to model the anisotropic QD,
see text.}
\label{fig1_th_hmfw}
\end{figure}

Moreover, the calculated magnetic-field-dependent 
energy splitting, $J_{\rm{EXD}}(B)=E^t_{\rm{EXD}}(B)-E^s_{\rm{EXD}}(B)$,
between the two lowest singlet ($S_0$) and triplet ($T_+$) states is 
found to be in remarkable agreement with the experiment [see 
\fref{fig1_th_hmfw}]. 

\begin{figure}[t]
\centering\includegraphics[width=6.0cm]{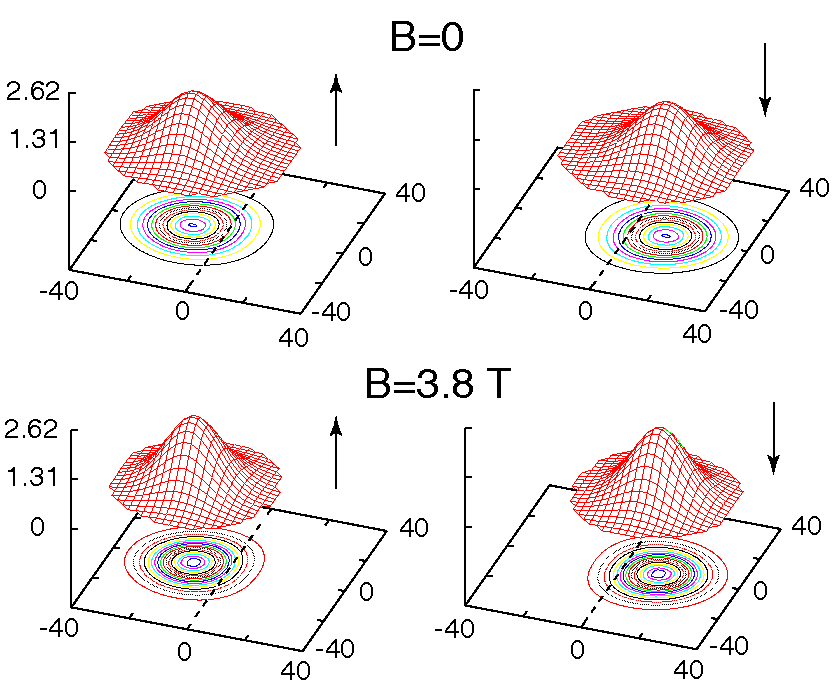}
\caption{(Color online) 
Single-particle UHF orbitals (modulus square) that are used in the 
construction of the GHL wave function in \eref{wfghl}.
Arrows indicate up and down spins.
For the parameters used in the calculation to model the anisotropic QD,
see text. Lengths in nm and orbital densities in 10$^{-3}$ nm$^{-2}$.}
\label{fig2_th_hmfw}
\end{figure}

A deeper understanding of the structure of the many-body wave function can
be acquired by comparing the measured $J(B)$ with that calculated within the 
GHL and RHF approximations. To facilitate the comparisons, the calculated 
$J_{\rm{GHL}}(B)$ and $J_{\rm{RHF}}(B)$ curves are plotted also in 
\fref{fig1_th_hmfw}, along with the exact-diagonalization 
result and the measurements. 
Both the RHF and GHL schemes are appealing intuitively, because they minimize 
the total energy using single-particle orbitals. It is evident, however, from 
\fref{fig1_th_hmfw} that the RHF method, which assumes that both electrons 
occupy a common single-particle orbital, is not able to reproduce the 
experimental findings. On the contrary, the generalized Heitler-London
approach, which allows the two
electrons to occupy two spatially separated orbitals, appears to be a good 
approximation. Plotting the two GHL orbitals [see \fref{fig2_th_hmfw}] for the
singlet state clearly demonstrates that the two electrons do not occupy the 
same orbital, but rather fill states that are significantly spatially 
separated.

The UHF orbitals from which the GHL singlet state is constructed
[see \eref{wfghl}] are displayed on \fref{fig2_th_hmfw} for 
both the $B=0$ and $B=3.8$ T cases. The spatial shrinking of these orbitals 
at the higher $B$-value illustrates the ``dissociation'' of the electron dimer 
with increasing magnetic field. The asymptotic convergence (beyond the ST point)
of the energies of the singlet and triplet states, [i.e., $J(B) \rightarrow 0$ 
as $B \rightarrow \infty$] is a reflection of the dissociation of the 2$e$ 
molecule, since the ground-state energy of two fully spatially separated 
electrons (zero overlap) does not depend on the total spin. We stress again
that the RHF, which corresponds to the more familiar physical picture of a
QD-Helium atom, fails to describe this dissociation, because $J_{\rm{RHF}}(B)$
diverges as the value of the magnetic field increases.

\begin{figure}[t]
\centering\includegraphics[width=6.0cm]{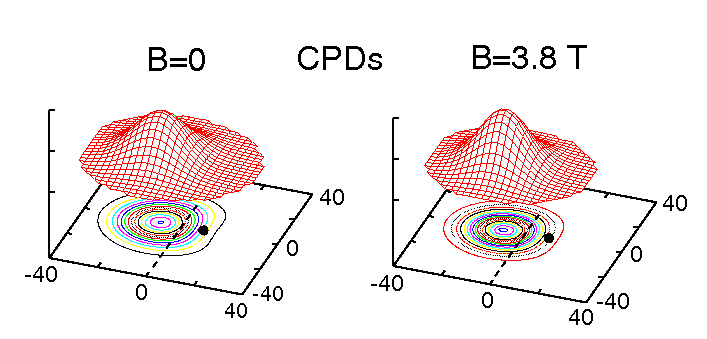}
\caption{(Color online) 
CPDs extracted from the exact-diagonalization wave 
function for the singlet state for $B=0$
and $B=3.8$ T. The CPD expresses the conditional probability for finding the
second electron at position ${\bf r}$ given that the first electron is 
located at ${\bf r}_0$ (denoted by a heavy solid dot).
For the parameters used in the calculation to model the anisotropic QD,
see text. Lengths in nm and CPDs in arbitrary units.}
\label{fig3_th_hmfw}
\end{figure}

In contrast to the RHF, the GHL wave function is able to capture the importance
of correlation effects. Further insight into the inportance of correlations
in our QD device can be gained through inspection \cite{elle06} of the 
conditional probability distributions [see \eref{cpds}] associated with
the EXD solutions; see an illustration in  \fref{fig3_th_hmfw}.
Indeed, already at zero magnetic field, the calculated CPDs 
provide further support of the physical picture of two localized electrons 
forming a state resembling an H$_2$-type \cite{yl02.1,yl02.3,elle06} Wigner 
molecule \cite{yl99,egger99}.

\subsubsection{Degree of entanglement.}
\label{secentan}

Further connections between the strong correlations found in our microscopic
treatment and the theory of quantum computing \cite{burk99} can be made
through specification of the degree of entanglement between the two
localized electrons in the molecular dimer. For two electrons, we can
quantify the degree of entanglement by calculating a well-known measure
of entanglement such as the von Neumann entropy \cite{yl06.tnt,you01} for
{\it indistinguishable\/} particles.
To this effect, one needs to bring the EXD wave function into a diagonal form 
(the socalled ``canonical form'' \cite{you01,schl01}), i.e.,
\begin{equation}
\Psi^{s,t}_{\rm{EXD}} ({\bf r}_1, {\bf r}_2) =
\sum_{k=1}^M z^{s,t}_k | \Phi(1;2k-1) \Phi(2;2k) \rangle,
\label{cano}
\end{equation}
with the $\Phi(i)$'s being appropriate spin orbitals resulting from a unitary
transformation of the basis spin orbitals $\psi(j)$'s [see \eref{wfexd}];
only terms with $z_k \neq 0$ contribute. The upper bound $M$ can be
smaller (but not larger) than $K$ (the dimension of the
single-particle basis); $M$ is referred to as the Slater rank.
One obtains the coefficients of the canonical expansion from the fact that
the $|z_k|^2$ are eigenvalues of the hermitian matrix 
${\cal A}^\dagger {\cal A}$
[${\cal A}$, see \eref{wfexd}, is antisymmetric]. The von Neumann
entropy is given by ${\cal S} = -\sum_{k=1}^M |z_k|^2 \log_2(|z_k|^2)$ with the
normalization $\sum_{k=1}^M |z_k|^2 =1$.

\begin{figure}[t]
\centering\includegraphics[width=5cm]{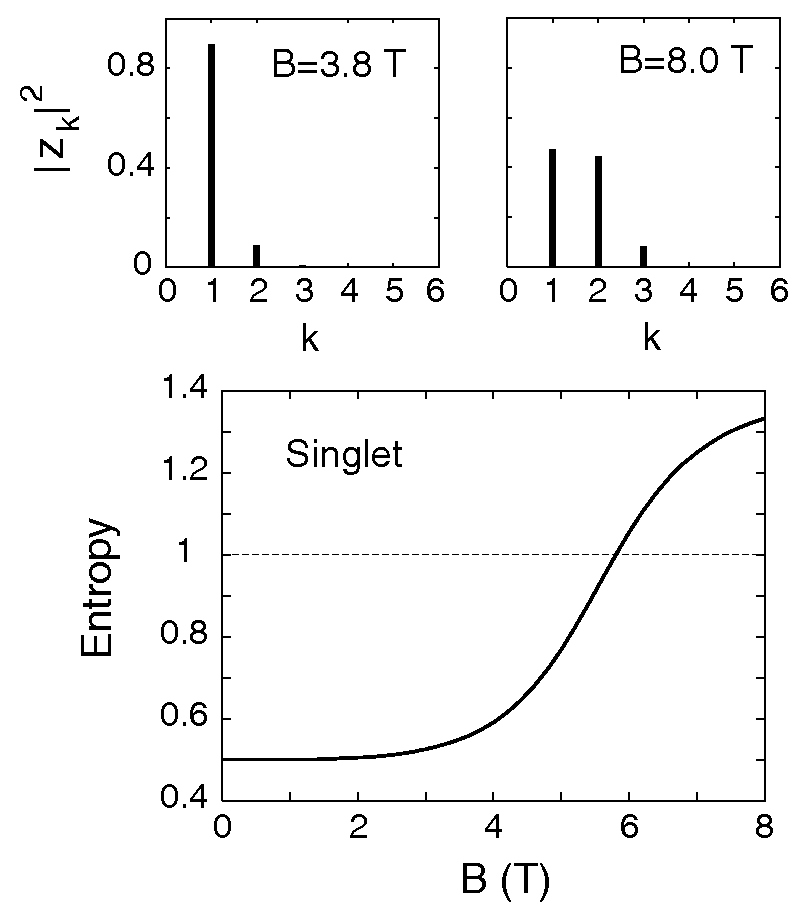}
\caption{
Von Neumann entropy for the lowest singlet EXD state of the elliptic dot as a 
function of the magnetic field $B$. 
On the top, we show histograms for the $|z_k|^2$
coefficients [see \eref{cano}] of the singlet state at $B=3.8$ T (left)
and $B=8.0$ T (right) illustrating the dominance of two determinantal 
configurations (in agreement with the generalized Heitler-London 
picture). Note the small third coefficient $|z_3|^2=0.081$ for $B=8.0$ T.
For the parameters used to model the experimental device, see text.
}
\label{fig4_th_hmfw}
\end{figure}

The EXD singlet has obviously a Slater rank $M > 2$.
The von Neumann entropy for the EXD singlet (${\cal S}^s_{\rm{EXD}}$) is
displayed in \fref{fig4_th_hmfw}. It is remarkable that  
${\cal S}^s_{\rm{EXD}}$ increases with increasing $B$, but remains close to 
unity for large $B$, although the 
maximum allowed mathematical value is $\log_2(K)$ [for example, for
$K=79$, $\log_2(79)=6.3$]. The saturation
of the entropy for large $B$ to a value close to unity reflects the dominant
(and roughly equal at large $B$) weight of two configurations in the canonical 
expansion [see \eref{cano}] of the exact-diagonalization
wave function, which are related \cite{yl06.tnt} to the two terms in the 
canonical expansion of the GHL singlet. This 
is illustrated by the histograms of the $|z_k^s|^2$ coefficients for $B=3.8$ T 
and $B= 8.0$ T in \fref{fig4_th_hmfw} (top). Notice that the ratio
$|z_2|^2/|z_1|^2$ reflects the extent of the overlap between the two GHL
orbitals \cite{yl06.tnt}, with the ratio increasing for smaller overlaps
(corresponding to a more complete dissociation of the Wigner molecule).

The above discussion illustrates that microscopic calculations that are
shown to reproduce experimental spectra \cite{elle06} can be used
to extract valuable information that allows assessment of the suitability
of a given device for quantum computations.

\subsection{Two-electron circular dots at zero magnetic field}
\label{2ecirdot}

In \sref{restsym}, we illustrated the formation of ``rotating electron 
molecules'' in the case of a two-electron {\it circular\/} QD, where one needs
to consider restoration of the rotational symmetry as well, in addition to
the restoration of the total spin. There, we focused on properties of the
ground state ($L=0$). 

\begin{figure}[t]
\centering\includegraphics[width=12cm]{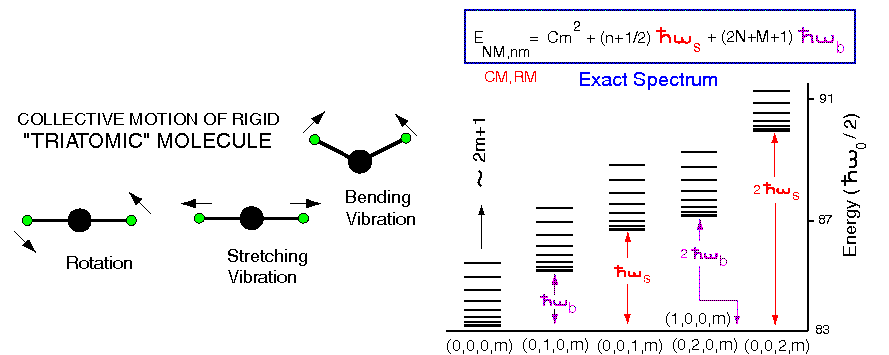}
\caption{(Color online) 
The calculated spectrum of a two-electron circular parabolic quantum dot, with 
$R_W=200$. The quantum numbers are $(N, M, n, m)$ with $N$ corresponding to 
the number of radial nodes in the center of mass (CM) wavefunction, and $M$ is
the CM azimuthal quantum number. The integers $n$ and $m$ are the 
corresponding quantum numbers for the electrons' relative motion (RM) and the 
total energy is given by $E_{NM,nm} = E^{\rm CM}_{NM} + E^{\rm RM}(n,|m|)$. 
The spectrum may be summarized by the ``spectral rule'' given in the figure, 
with an effective rigid moment of inertia $C = 0.037$ (corresponding to an 
angular momentum ${\cal L}=\hbar m$), the phonon for the stretching
vibration $\hbar \omega_s = 3.50$, and the phonon for the 
bending vibration coincides with that of the CM motion, i.e., $\hbar \omega_b 
= \hbar \omega_0 = 2$. The quantum numbers $(N_0,M_0,n_0,m)$ specifying each 
rotational band are given at the bottom, with $m=0,1,2,\ldots$ (the levels
$m=0$ and $m=1$ in each band may not be resolved on the scale of the figure). 
We note that the energy separation between levels in a given rotational 
band increases as $(2m+1)$ with increasing $m$, which is a behavior 
characteristic of a rigid rotor. 
All energies are in units of $\hbar \omega_0/2$, where $\omega_0$ is the 
parabolic confinement frequency.
}
\label{tnt_2005_fig1}
\end{figure}

In this section, we will further examine the excitation spectra of a 
two-electron QD by using the rather simple exact solution of this problem 
provided through separation of the center-of-mass and inter-electron 
relative-distance degrees of freedom \cite{yl00.2}. The spectrum obtained 
for $R_W = 200$ (\fref{tnt_2005_fig1}), exhibits features that are 
characteristic of a collective rovibrational dynamics, akin to that of a 
natural ``near-rigid'' triatomic linear molecule with an infinitely heavy 
middle particle representing the center of mass of the dot. This spectrum 
transforms to that of a ``floppy'' molecule for smaller value of $R_W$ 
(i.e., for stronger confinements characterized by a larger value 
of $\omega_0$, and/or for weaker inter-electron repulsion), ultimately 
converging to the independent-particle picture associated with the circular 
central mean-field of the QD. 

\begin{figure}[t]
\centering\includegraphics[width=6cm]{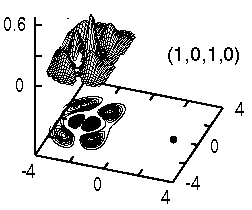}
\caption{CPD of the excited multi-vibrational state (1,0,1,0) of a $2e$
circular parabolic QD with $R_W=200$ (see text for details). 
The solid dot portrays the position
of the reference point ${\bf r}_0=(d_0,0)$, where $d_0=2.6$ is half the
interelectron distance at the ground state $(0,0,0,0)$. Distances are in
units of $l_0 \sqrt{2}$; the scale of the vertical axis is arbitrary. 
}
\label{2ecircpd}
\end{figure}

Further 
evidence for the formation of the electron molecule and the emergence of
a rovibrational spectrum was found through examination \cite{yl00.2} of the 
conditional probability distributions for various states $(N,M,n,m)$ (see
the caption of \fref{tnt_2005_fig1} for the precise meaning of these
quantum numbers labeling the spectra). As an example, we display in
\fref{2ecircpd} the CPD for the bottom state ($m=0$) of the rotational band
$(1,0,1,m)$ (not shown in \fref{tnt_2005_fig1}); it reveals that this 
state corresponds to a vibrational motion of the electron molecule both
along the interelectron axis (one excited stretching-mode phonon; see
\fref{tnt_2005_fig1}) and perpendicularly to this axis (two excited
bending-mode phonons; see \fref{tnt_2005_fig1}).

It is instructive to note here certain 
similarities between the formation of a ``two-electron molecule'' in man-made 
quantum dots, and the collective (rovibrational) features observed in the 
electronic spectrum of doubly-excited helium atoms \cite{kell80,kell97,berr89}.

\subsection{Historical significance of the two-electron problem}
\label{histsig}

In spite of being the simplest many-body system, the significance of the 
problem of two interacting electrons confined in an external potential cannot 
be overstated. Historically it played a central role in the
development of the quantum theory of matter through the failure of the
Bohr-type semiclassical models to account for the natural He atom.
Most recently it has influenced the development of several fields like 
nonlinear physics, atomic physics, semiconductror quantum dots, and 
quantum computing.

It is instructive to make here a historical detour. Indeed, the failure 
of Bohr-type semiclassical models, based on the orbiting of spatially
correlated (antipodal) electrons in conjunction with the Bohr-Sommerfeld 
quantization rule, to yield a reasonable estimate of the
ground state of the He atom signaled a looming crisis in physics in the
1920's, which Bohr himself, as well as others, had been keenly aware of, as
summarized succintly by Sommerfeld: ``All attempts made hitherto to solve
the problem of the neutral helium atom have proved to be unsuccessful''
\cite{somm23}; see also the 10th chapter entitled 
``It was the Spring of hope, it was the Winter of despair'' 
in the book by Pais \cite{pais},
the review by Van Vleck \cite{vleck} and the book by Born \cite{born}.

While, since, numerical solutions of the two-electron Schr\"{o}dinger equation
provided a quantitative resolution to the problem, the first successful 
semiclassical treatment of the three-body Coulomb system awaited till 1980
\cite{leop80,tann92}.

Furthermore, based on rather general
group-theoretical arguments arising from the observation of hierarchies with
lower symmetry in the excited spectra, and motivated by ideas originating
in nuclear-physics spectroscopy, it has been discovered in the late 1970's
and early 1980's that electron correlations in doubly excited He lead to
quantization of the spectrum much like in a linear triatomic molecule,
$e$-He$^{2+}$-$e$. This molecular picture, with near rigidity and 
separability, results in ``infinite sequences of vibrational levels, 
on each of which is built an infinite sequence of rotational levels''
\cite{kell80,ezra83,cdlin86}.

The two previous \sref{2eelldot} and \sref{2ecirdot} describing the 
formation and properties of a 2$e$ Wigner molecule in a {\it single\/} QD may 
be viewed as the culmination of this historical background.
Interestingly, as in the aforementioned semiclassical treatments, the
collinear configuration plays a special role in the molecule-like model,
serving perhaps as ``partial vindication'' of the geometry considered 
originally by Niels Bohr.

\section{Rotating electron molecules in two-dimensional quantum dots under a 
strong magnetic field: The case of the lowest Landau level 
($\omega_c/2\omega_0 \rightarrow \infty$)}
\label{rotellll}

\subsection{REM analytic trial wave functions}
\label{antrial}

In the last ten years, and in particular since 1999 (when it was explicitly
demonstrated \cite{yl99} that Wigner crystallization for small systems is 
related to symmetry breaking at the {\it unrestricted\/} Hartree-Fock 
mean-field level), the number of publications addressing the formation and 
properties of Wigner (or electron) molecules in 2D QDs and quantum dot 
molecules has grown steadily 
\cite{yl99,yl00.1,yl02.1,yl02.2,mk96,yl00.2,gra01,sba03,maks96,%
maks00,mik02.3,rua99,man00,egger99,loz00,nie02.1,%
cre99,gra00,pee01,loz01,loz02,mik02.1,mik02.2,cre00,ront02}.
A consensus has been reached that rotating electron molecules are 
formed both in zero \cite{yl00.1,yl02.1,yl02.2,yl02.3,yl00.2,gra01,mik02.3,%
man00,egger99,loz00,nie02.1,cre99,gra00,%
pee01,loz01,loz02,mik02.1,mik02.2} 
and high \cite{yl03.2,mk96,yl02.3,sba03,maks96,sek96,maks00,rua99,cre00,%
ront02} magnetic fields.

At $B=0$, in spite of considerable differences explored in this report (see
next paragraph), formation of REMs in 
quantum dots is driven by the same physical 
factors as Wigner crystallization in infinite 2D 
media, i.e., when the strength of the interelectron repulsion relative 
to the zero-point kinetic energy ($R_W$) exceeds a certain critical value, 
electrons spontaneously crystallize around sites forming geometric
molecular structures. At high magnetic fields, the formation of Wigner 
molecules may be thought of as involving a two-step crystallization process: 
(I) the localization of electrons results from the shrinkage of the orbitals
due to the increasing strength of the magnetic field; (II) then, 
even a weak interelectron Coulomb repulsion is able to arrange the localized
electrons according to geometric molecular structures (thus this process is 
independent of the value of $R_W$). It has been found 
\cite{yl02.2,yl03.2,mk96,maks96}
that the molecular structures at high $B$ coincide with the equilibrium 
configurations at $B=0$ of $N$ classical point charges 
\cite{lozo87,beda94,kong02,bol93}. 

Due to the small finite number, $N$, of electrons, however, 
there are two crucial differences between the REM and the bulk Wigner crystal.
Namely, (I) the crystalline structure is that of the equilibrium 2D 
configuration of $N$ classical point charges, and thus consists of nested 
polygonal rings, in contrast to the well known hexagonal bulk crystal;   

(II) in analogy  with the case of 3D natural molecules, the Wigner molecules 
rotate as a whole (collective rotations); they behave, however, as highly
floppy (non-rigid) rotors.

A most striking observation concerning the REMs is that their formation
and properties have been established with the help of traditional {\it ab 
initio\/} many-body methods, i.e., exact diagonalization,
\cite{yl00.2,maks96,sek96,mik02.3,rua99,cre99,ront02}  quantum Monte Carlo 
\cite{umri06,egger99,nie02.1,umri05,loz01}, and the systematic 
controlled hierarchy \cite{yl99,yl01,yl02.1,yl02.2,yl03.1,mk96,sba03,loz02} of 
approximations involving the UHF and subsequent post-Hartree-Fock
methods. This contrasts with the case of the Jastrow/Laughlin \cite{laug83.1} 
and composite-fermion \cite{jain89,jain90} wave functions, 
which were constructed through ``intuition-based guesswork.''

In spite of its appearance in the middle nineties and its firm foundation in 
many-body theory, however, the REM picture had not, until recently, 
successfully competed with the CF/JL picture; indeed many research papers
\cite{nie02.2,jain97,gr97,yang98,tau00,man01,rez02} and 
books \cite{haw98} describe the physics of quantum dots 
in high magnetic fields following exclusively
notions based on CF/JL functions, as expounded in 1983 (see Ref.\ 
\cite{laug83.1}) and developed in detail in 1995 in Ref.\ 
\cite{jain95} and Ref.\ \cite{jain96}. 
One of the main obstacles for more frequent use of the REM picture had been 
the lack of analytic correlated wave functions associated with this picture. 
This situation, however, changed with the recent explicit derivation of such 
REM wave functions \cite{yl02.2}.

The approach used in Ref.\ \cite{yl02.2} for constructing the analytic REM 
functions in high $B$ consists of two-steps: First the
breaking of the rotational symmetry at the level of the single-determinantal
unrestricted Hartree-Fock approximation yields states representing electron 
molecules. Subsequently the rotation of the electron molecule is 
described through restoration of the circular symmetry via post Hartree-Fock 
methods, and in particular Projection Techniques \cite{rs}. 
The restoration of symmetry goes beyond the single determinantal mean-field 
description and yields multi-determinantal wave functions. 

In the zero and low-field cases, the broken symmetry UHF orbitals need to be 
determined numerically, and, in addition, the restoration of the total-spin
symmetry needs to be considered for unpolarized and partially polarized cases.
The formalism and mathematical details of this procedure at $B=0$ have been 
elaborated in previous sections.

In the case of high magnetic fields, the spins of the electrons are fully 
polarized. Furthermore, one can specifically consider the limit
when the confining potential can be neglected compared to the confinement
induced by the magnetic field, so that the Hilbert space is restricted to the
lowest Landau level. Then, assuming a symmetric gauge, the UHF 
orbitals can be represented \cite{yl02.2,mz83} by displaced Gaussian analytic 
functions, centered at different positions $Z_j \equiv X_j+ \rmi Y_j$ 
according to the equilibrium configuration of $N$ classical point charges 
\cite{lozo87,beda94,kong02,bol93} arranged at the vertices of nested regular 
polygons (each Gaussian representing a localized electron). Such displaced 
Gaussians in the lowest Landau level are written as 
\begin{equation}
u(z,Z_j) = (1/\sqrt{\pi}) \exp[-|z-Z_j|^2/2] \exp[-\rmi (xY_j-yX_j)],
\label{gaus}
\end{equation}
where the phase factor is due to the gauge invariance. $z \equiv x+\rmi y$, 
and all lengths are in dimensionless units of 
${l_B}\sqrt{2}$ with the magnetic length being $l_B=\sqrt{\hbar c/eB}$. 
Note that expression \eref{gaus} is a special case of the more general
expression \eref{uhfo} for a displaced Gaussian which corresponds to
situations with smaller magnetic fields when the restriction to the 
lowest Landau level breaks down. The notation 
$z \equiv x + \rmi y$ is associated with positive angular
momenta for the single-particle states in the lowest Landau level.
Ref.\ \cite{yl02.2} used $z \equiv x - \rmi y$ and negative
single-particle angular momenta in the lowest Landau level. The final
expressions for the trial wave functions do not depend on these choices.

\begin{table}[b]
\caption{\label{remvsexd}%
Comparison of yrast-band energies obtained from REM and EXD calculations for
$N=6$ electrons in the lowest Landau level, that is in the
limit $B \rightarrow \infty$. In this limit the external confinement can
be neglected and only the interaction energy contributes to the yrast-band
energies. Energies in units of $e^2/(\kappa l_B)$. For the REM results, the
(1,5) polygonal-ring arrangement was considered. The values of the fractional
filling may be obtained for each $L$ as $\nu=N(N-1)/(2L)$.
}
\begin{indented}
\item[]\begin{tabular}{rccc||rccc}
\br
$L$  & REM & EXD  & Error (\%) & $L$  & REM & EXD  & Error (\%)  \\
\mr
 70  & 2.3019  & 2.2824  & 0.85  & 140  & 1.6059  & 1.6006  & 0.33 \\
 75  & 2.2207  & 2.2018  & 0.85  & 145  & 1.5773  & 1.5724  & 0.31 \\
 80  & 2.1455  & 2.1304  & 0.71  & 150  & 1.5502  & 1.5455  & 0.30 \\
 85  & 2.0785  & 2.0651  & 0.65  & 155  & 1.5244  & 1.5200  & 0.29 \\
 90  & 2.0174  & 2.0054  & 0.60  & 160  & 1.4999  & 1.4957  & 0.28 \\
 95  & 1.9614  & 1.9506  & 0.55  & 165  & 1.4765  & 1.4726  & 0.27 \\
100  & 1.9098  & 1.9001  & 0.51  & 170  & 1.4542  & 1.4505  & 0.26 \\
105  & 1.8622  & 1.8533  & 0.48  & 175  & 1.4329  & 1.4293  & 0.25 \\
110  & 1.8179  & 1.8098  & 0.45  & 180  & 1.4125  & 1.4091  & 0.24 \\
115  & 1.7767  & 1.7692  & 0.42  & 185  & 1.3929  & 1.3897  & 0.23 \\
120  & 1.7382  & 1.7312  & 0.40  & 190  & 1.3741  & 1.3710  & 0.23 \\
125  & 1.7020  & 1.6956  & 0.38  & 195  & 1.3561  & 1.3531  & 0.22 \\
130  & 1.6681  & 1.6621  & 0.36  & 200  & 1.3388  & 1.3359  & 0.21 \\
135  & 1.6361  & 1.6305  & 0.34  &      &         &         &      \\
\br
\end{tabular}
\end{indented}
\end{table}

Ref.\ \cite{yl02.2} used these analytic orbitals to first construct 
the broken symmetry UHF determinant, $\Psi^{\rm{UHF}}_N$, and then proceeded
to derive analytic expressions for the many-body REM wave functions by 
applying onto $\Psi^{\rm{UHF}}_N$ an appropriate projection operator 
${\cal P}_L$ (see \sref{remtheo}) that restores the circular symmetry and 
generates {\it correlated\/} wave functions with good total angular momentum 
$L$. These REM wave functions can be easily written down \cite{yl02.2} in 
second-quantized form for any classical polygonal 
ring arrangement $(n_1,n_2,\ldots,n_r)$ by following certain simple rules for 
determining the coefficients of the determinants $D(l_1,l_2,\ldots,l_N) \equiv 
{\rm{det}}[z_1^{l_1},z_2^{l_2}, \cdot \cdot \cdot, z_N^{l_N}]$,
where the $l_j$'s denote the angular momenta of the individual electrons.

The REM functions associated with the $(0,N)$ and $(1,N-1)$ ring 
arrangements, respectively [here $(0,N)$ denotes a regular polygon with $N$
vertices, such as an equilateral triangle or a regular hexagon, and $(1,N-1)$
is a regular polygon with $N-1$ vertices and one occupied site in its center],
are given by 
\begin{eqnarray}
\fl \Phi_L^{(0,N)} (z_1,z_2,\ldots,z_N) && =
 \sum^{l_1 + \cdot \cdot \cdot +l_N=L}%
_{0 \leq l_1<l_2< \cdot \cdot \cdot <l_N}
\left( \prod_{i=1}^N l_i! \right)^{-1}  
\left( \prod_{1 \leq i < j \leq N} 
\sin \left[\frac{\pi}{N}(l_i-l_j)\right] \right)  \nonumber \\
&& \times \; D(l_1,l_2,\ldots,l_N)
\exp(-\sum_{i=1}^N z_i z_i^*/2),
\label{phi1}
\end{eqnarray} 
with 
\begin{equation}
L=L_0+Nm, \;\; m=0,1,2,3,\ldots,
\label{lwm0n}
\end{equation}
and
\begin{eqnarray}
\fl \Phi_L^{(1,N-1)}(z_1,z_2,\ldots,z_N)&& = 
\sum^{l_2+ \cdot \cdot \cdot +l_N=L}%
_{1 \leq l_2 < l_3 < \cdot\cdot\cdot < l_N}
\left( \prod_{i=2}^N l_i! \right)^{-1} 
\left( \prod_{2 \leq i < j \leq N} 
\sin \left[\frac{\pi}{N-1}(l_i-l_j)\right] \right)  \nonumber \\
 \nonumber \\
&& \times \; D(0,l_2,\ldots,l_N)
\exp(-\sum_{i=1}^N z_i z_i^*/2),
\label{phi2}
\end{eqnarray}
with 
\begin{equation}
L=L_0+(N-1)m, \;\; m=0,1,2,3,\ldots,
\label{lwm1nm1}
\end{equation}
where $L_0=N(N-1)/2$ is the minimum allowed total angular momentum for
$N$ (fully spin polarized) electrons in high magnetic fields.

Notice that the REM wave functions [\Eref{phi1} and \Eref{phi2}]
vanish identically for values of the total angular momenta outside the 
specific values given by the sequences \eref{lwm0n} and \eref{lwm1nm1}, 
respectively; these sequences are termed as magic angular momentum
sequences.

\begin{figure}[t]
\centering{\includegraphics[width=12.cm]{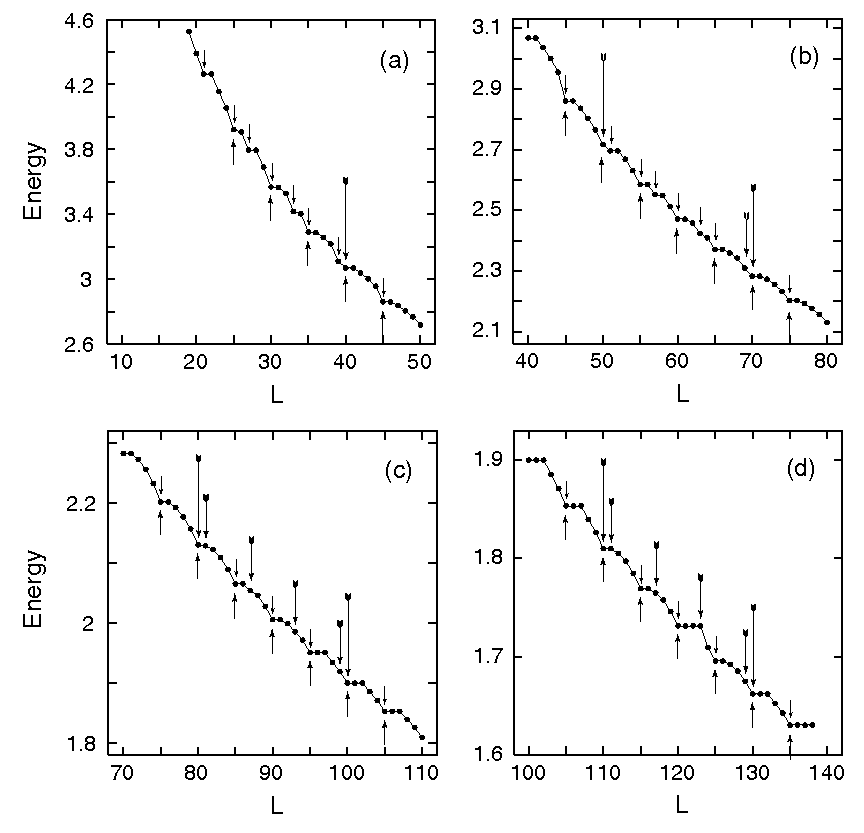}}
\caption{
Total interaction energy from exact-diagonalization calculations as a function
of the total angular momentum $(19 \leq L \leq 140)$ for $N=6$ electrons
in the lowest Landau level. The upwards pointing arrows indicate the magic
angular momenta corresponding to the classically most stable (1,5) polygonal
ring arrangement of the Wigner molecule. The short downwards pointing arrows
indicate successful predictions of the composite-fermion model. 
The medium-size downwards pointing arrows indicate predictions of the 
composite-fermion model that fail to materialize as magic angular momenta.
The long downward arrows indicate EXD magic angular momenta not predicted by 
the composite-fermion model. 
Energies in units of $e^2/\kappa l_B$, where $\kappa$ is the dielectric
constant.
}
\label{exdarrows}
\end{figure}

We remark that, while the original REM analytic wave function was derived
in the context of a high magnetic field (that is in the fractional quantum
Hall effect regime), it is valid for any circumstance where the spectrum
consists of a degenerate manifold of LLL-like states (even with no 
magnetic field present). Indeed a wave function having the form of the
REM wave function discussed by us above has been employed recently for
graphene quantum dots with a zig-zag boundary condition and in the
absence of a magnetic field \cite{guin07}.

In the remaining of this section, we continue discussing the properties of 
analytic REM wave functions associated with fully spin polarized electrons.
However, we mention here that, following the methodology of Ref.\ 
\cite{yl02.2} for fully spin polarized REMs. Dai \etal \cite{dai07} and Shi 
\etal \cite{shi07} have most recently presented analytic trial wave functions 
for rotating electron molecules with partial spin polarizations. 

\subsection{Yrast rotational band in the lowest Landau level}

As an accuracy test, we compare in \tref{remvsexd} REM and 
exact-diagonalization results for the interaction energies of the yrast band 
associated with the magic angular momenta $L_m$ [see \eref{lmeq}] of $N=6$ 
electrons in the lowest Landau level. An yrast\footnote[8]{
The word {\it yrast\/} is the superlative of the Swedish {\it yr\/},
which means dizzy \cite{bm1}. The term yrast is widely used in nuclear
spectroscopy.}
state is defined as the lowest-energy state for a given angular momentum $L$. 
As a result, the yrast band represents excitations with purely rotational 
motion; no other excitations, like center-of-mass or vibrational modes, are 
present.  

As seen from \tref{remvsexd}, the REM wave functions offer an excellent 
approximation to the EXD ones, since the relative error of the REM energies is 
smaller than 0.3\%, and it decreases steadily for larger $L$ values. Of course,
a small difference in the energies between approximate and 
exact-diagonalization results is only
one of several tests for deciding whether a given trial wave function is a 
good approximation. As will be discussed below, comparison of conditional 
probability distrubutions is an equally (if not more) important test.

\subsection{Inconsistencies of the composite-fermion view for semiconductor
quantum dots.}

Before the development of the REM approach, electrons in the 
lowest Landau level in two-dimensional quantum dots 
were thought as being well approximated by composite 
fermion trial wave functions. However, results obtained with the REM and
exact-diagonalization calculations led researchers to examine inconsistencies
and discrepancies of the CF approach in the context of quantum dots.
This section focuses on these issues.

For $N=6$, \fref{exdarrows} displays (in four frames) the total interaction 
energy from exact-diagonalization as a function of the total angular momentum 
$L$ in the range $ 19 \leq L \leq 140 $. [The total kinetic energy in the
Hamiltonian (\ref{hlll}), being a constant, can be disregarded.] One can 
immediately observe the appearance of downward cusps, implying states of 
enhanced stability, at certain magic angular momenta. 

For the CF theory, the magic angular momenta can be determined by 
\begin{equation}
L=L^*+mN(N-1)=L^*+2mL_0.
\label{mamcf}
\end{equation}
Namely, for $N=6$, if one knows the non-interacting $L^*$'s, the CF magic 
$L$'s in any filling-factor interval $1/(2m-1) \geq \nu \geq 1/(2m+1)$ 
[corresponding to the angular-momentum interval 
$15(2m-1) \leq L \leq 15(2m+1)]$, $m=1,2,3,4,\ldots$, can be found by adding 
$2mL_0=30m$ units of angular momentum to each of the $L^*$'s. 
To obtain the non-interacting $L^*$'s, one needs first to construct
\cite{yl03.2,sek96,jain95} the compact Slater determinants.
Let $N_n$ denote the number of electrons in the $n$th Landau level with 
$\sum_{n=0}^t N_n=N$; $t$ is the index of the highest occupied Landau level 
and all the lower Landau levels with $n \leq t$ are assumed to be occupied. 
The compact determinants are defined as those in 
which the $N_n$ electrons occupy 
contiguously the single-particle orbitals (of each $n$th Landau level) having
the lowest angular momenta $l=-n,-n+1,\ldots,-n+N_n-1$. The compact Slater 
determinants are usually denoted as $[N_0,N_1,\ldots,N_t]$; see Refs.\
\cite{yl03.1,jain95} for details.

The compact determinants and the corresponding non-interacting $L^*$'s 
for $n=6$ are listed in Table \ref{cfcom}. 

\begin{table}[b]
\caption{\label{cfcom} Compact Slater determinants 
and associated angular momenta $L^*$ for $N=6$ electrons according to
the CF presciption. Both $L^*=-3$ and $L^*=3$ are associated with two compact
states each, the one with lowest energy being the preferred one.}
\begin{indented}
\item[]\begin{tabular}{cr}
\br
 Compact state   &   $L^*$   \\ 
\mr
$[$1,1,1,1,1,1$]$    &   $-$15   \\
$[$2,1,1,1,1$]$      &   $-$9    \\
$[$2,2,1,1$]$        &   $-$5    \\
$[$3,1,1,1$]$        &   $-$3    \\
$[$2,2,2$]$          &   $-$3    \\
$[$3,2,1$]$          &   0       \\
$[$4,1,1$]$          &   3       \\
$[$3,3$]$            &   3       \\
$[$4,2$]$            &   5       \\
$[$5,1$]$           &   9       \\
$[$6$]$              &   15      \\
\br
\end{tabular}
\end{indented}
\end{table}

There are nine different values of $L^*$'s, and thus the CF theory for $N=6$ 
predicts that there are always nine magic numbers in any interval 
$15(2m-1) \leq L \leq 15(2m+1)$ between two consecutive angular momenta of
Jastrow/Laughlin states, $15(2m-1)$ and $15(2m+1)$, $m=1,2,3,\ldots$ (henceforth
we will denote this interval as ${\cal I}_m$). For example, using 
Table \ref{cfcom} and \eref{mamcf}, the CF magic numbers in the 
interval $15 \leq L \leq 45$ ($m=1$) are found to be the following 
nine
\begin{equation}
15,\;21,\;25,\;27,\;30,\;33,\;35,\;39,\;45.
\label{cfmam1}
\end{equation}
On the other hand, in the interval $105 \leq L \leq 135$ ($m=4$), the CF 
theory predicts the following set of nine magic numbers,
\begin{equation}
105,\;111,\;115,\;117,\;120,\;123,\;125,\;129,\;135.
\label{cfmam4}
\end{equation}

An inspection of the total-energy-vs-$L$ plots in \fref{exdarrows} reveals 
that the CF prediction misses the actual magic angular momenta specified by 
the exact-diagonalization calculations as those associated with the downward 
cusps. It is apparent that the number of downward cusps in any interval
${\cal I}_m$ is always different from 9. Indeed, there are 10 cusps in 
${\cal I}_1$ [including that at $L=15$, not shown in \fref{exdarrows}(a)], 
10 in ${\cal I}_2$ [see \fref{exdarrows}(b)], 7 in 
${\cal I}_3$ [see \fref{exdarrows}(c)], and 7 in ${\cal I}_4$ 
[see \fref{exdarrows}(d)]. In detail, the CF theory fails in the
following two aspects: (I) There are exact magic numbers that are consistently
missing from the CF prediction in every interval; with the exception of the 
lowest $L=20$, these {\it exact\/} magic numbers (marked by a long downward 
arrow in the figures) are given by $L=10(3m-1)$ and $L=10(3m+1)$, 
$m=1,2,3,4,\ldots$; 
(II) There are CF magic numbers that do not correspond to downward cusps in
the EXD calculations (marked by medium-size downward arrows in the figures). 
This happens because cusps associated with $L$'s whose difference from $L_0$ 
is divisible by 6 (but not simultaneously by 5) progressively weaken and 
completely disappear in the intervals ${\cal I}_m$ with $m \geq 3$; only cusps 
with the difference $L-L_0$ divisible by 5 survive.  On the other hand, the CF
model predicts the appearance of four magic numbers with $L-L_0$ divisible 
solely by 6 in every interval ${\cal I}_m$, at $L=30m\mp9$ and $30m\mp3$, 
$m=1,2,3,\ldots$ The overall extent of the inadequacy of the CF model can be 
appreciated better by the fact that there are six false predictions (long and 
medium-size downward arrows) in every interval ${\cal I}_m$ with $m \geq 3$, 
compared to only five correct ones [small downward arrows, see 
\fref{exdarrows}(c) and \fref{exdarrows}(d)].

In contrast to the CF model, the magic angular momenta in the REM theory
are associated with the polygonal ring configurations of $N$ classical 
point charges. This is due to the fact that the enhanced stability of
the downward cusps results from the coherent collective rotation of the
regular-polygon REM structures. Due to symmetry requirements, such collective 
rotation can take place only at magic-angular-momenta values. The in-between
angular momenta require the excitation of additional degrees of freedom (like
the center of mass and/or vibrational modes), which raises the total energy
with respect to the values associated with the magic angular momenta.

For $N=6$, the ring configuration of lowest energy is the (1,5), while there 
exists a (0,6) isomer \cite{kong02,bol93} with higher energy. As a result, 
our exact-diagonalization calculations \cite{yl03.2} (as well as earlier ones 
\cite{sek96,maks00,rua99} 
for lower angular momenta $L \leq 70$) have found that there exist two 
sequences of magic angular momenta, a {\it primary\/} one $(S_p)$ with 
$L=15+5m$ [see \Eref{lwm1nm1}], associated with the most stable (1,5) 
classical molecular configuration, and a {\it secondary\/} one $(S_s)$ with 
$L=15+6m$ [see \Eref{lwm0n}], associated  with the metastable $(0,6)$ 
ring arrangement. Furthermore, our calculations (see also Refs.\  
\cite{maks00,rua99}) show that the secondary sequence $S_s$ contributes 
only in a narrow range of the lowest angular momenta; in the region of
higher angular momenta, the primary sequence $S_p$ is the only one that
survives and the magic numbers exhibit a period of five units of angular
momentum. It is interesting to note that the initial competition between the
primary and secondary sequences, and the subsequent prevalence of the
primary one, has been seen in other sizes as well \cite{rua99} i.e., $N=5,7,8$.
Furthermore, this competition is reflected in the field-induced molecular phase
transitions associated with broken symmetry UHF solutions in a parabolic QD. 
Indeed, Ref.\ \cite{li06} demonstrated recently that, as a function of 
increasing $B$, the UHF solutions for $N=6$ first depict the transformation of
the maximum density droplet \cite{mac93} (see definition in \sref{remtheo}) 
into the (0,6) molecular configuration; then (at higher $B$) the (1,5) 
configuration replaces the (0,6) structure as the one having the lower HF 
energy.

The extensive comparisons in this section lead to the conclusion
that the composite-fermion model does not explain the systematic trends 
exhibited by the magic angular momenta in 2D quantum dots 
in high magnetic fields. 
These trends, however, were shown to be a natural consequence of the formation
of REMs and their metastable isomers. 

These results motivated a reexamination of the original composite-fermion 
approach (the mean-field CF) and led to a reassessment of the significance
of the residual interaction, neglected in the mean-field CF theory.
Initially, it has been reported that some CF functions away from
the main fractions [e.g., for $N=19$ and $L=1845$ and $N=19$ and $L=3555$] may 
reproduce the aforementioned crystalline patterns \cite{jeon04.1}.

Subsequently, Jain and coworkers have found that inclusion of the residual
interaction is absolutely necessary to account for the full range of 
inconsistencies of the mean-field CF theory \cite{jeon07}. However, this 
latter development was achieved with the trade off of abandoning the 
fundamental nature of the composite fermion as an elementary, independent and 
weakly interacting quasi-particle. Indeed, the revised \cite{jeon07} CF picture
amounts to an exact diagonalization method which uses a correlated basis
set (made out of CF wave functions).

Another attempt to update the CF theory in order to account for 
crystallization consists of combining the REM analytic wave function
$\Phi^{\rm REM}_{L}(z_1,z_2,\ldots,z_N)$ (see \sref{antrial}) with Jastrow 
prefactors \cite{jeon05}, namely one uses a variational wave function of the 
form
\begin{equation}
\Psi^{{2p,}{\rm CFC}}_L = \prod_{i < j} (z_i-z_j)^{2p} \Phi^{\rm REM}_{L^*},
\label{wfcfc}
\end{equation}
with $L=L^*+pN(N-1)$ and $p$ serves as an additional variational parameter.
Obviously, the crystalline patterns in such an approach originate from the REM
wave function and the Jastrow prefactors simply increase the variational
freedom, leading to a numerical improvement. Although this approach is 
a straightforward variational improvement of the analytic REM method
\cite{yl02.2}, it is being referred to \cite{shi07,jeon05} as a 
composite-fermion crystal (CFC). 

More direct variational improvements of the analytic REM wave functions can be
devised in the spirit of the variation-after-projection method. For example, 
one can use angular-momentum conserving variational parameters in front of
the sine coefficients in the REM expansion \cite{guin07}.

\begin{figure}[t]
\centering\includegraphics[width=6.5cm]{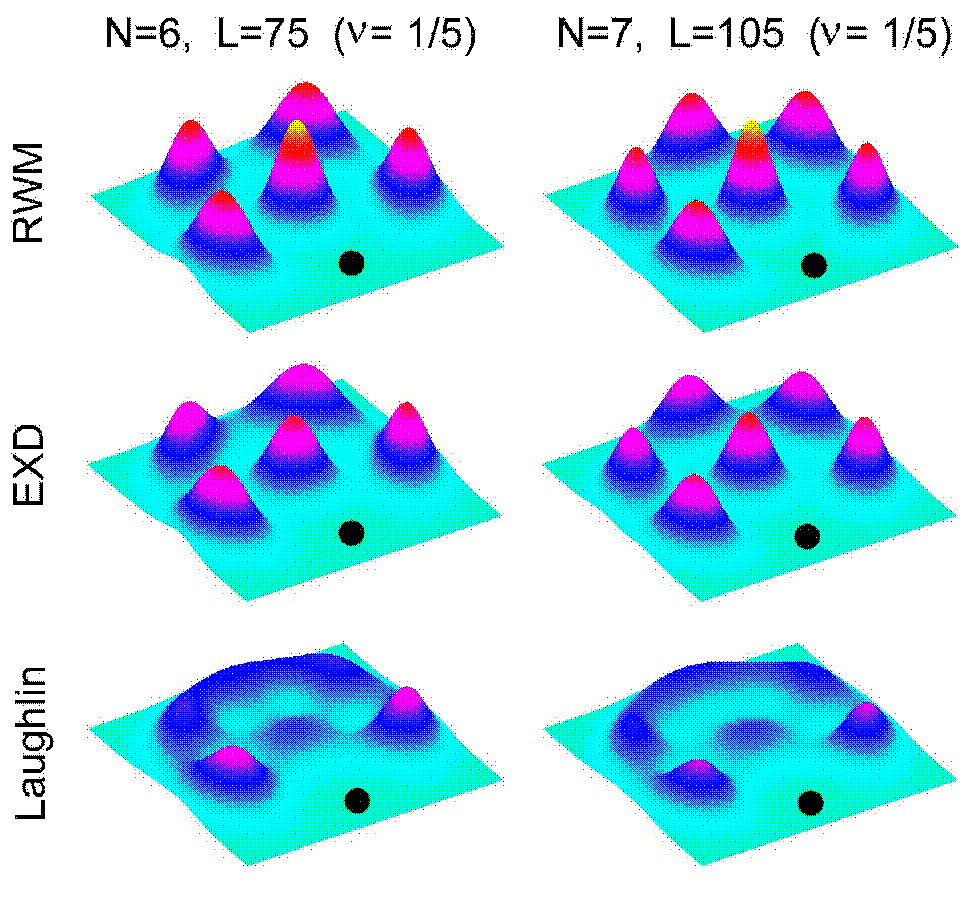}
\caption{(Color online) 
Conditional probability distributions at high $B$ for $N=6$ electrons and 
$L=75$ ($\nu=1/5$, left column) and for $N=7$ electrons and $L=105$ (again 
$\nu=1/5$, right column). Top row: REM case. Middle row: The case of exact 
diagonalization. Bottom row: The Jastrow/Laughlin case.
The exact diagonalization and REM wave functions have a pronouned crystalline
character, corresponding to the (1,5) polygonal configuration of the REM for 
$N=6$, and to the (1,6) polygonal configuration for $N=7$. In contrast, the 
Jastrow/Laughlin wave functions exhibit a characteristic liquid profile that 
depends smoothly on the number $N$ of electrons. The observation point 
(identified by a solid dot) is located at $r_0 = 5.431 l_B$ 
for $N=6$ and $L=75$ and $r_0=5.883 l_B$ for $N=7$ and $L=105$.
The EXD Coulomb interaction energies (lowest Landau level)
are 2.2018 and 2.9144 $e^2/\kappa l_B$ for $N=6,L=75$ and $N=7,L=105$, 
respectively. The errors relative to the corresponding exact-diagonalization 
energies and the overlaps of the trial functions with the EXD ones are:
(I) For $N=6,L=75$, REM: 0.85\%, 0.817; JL: 0.32\%, 0.837. 
(II) For $N=7,L=105$, REM: 0.59\%, 0.842; JL: 0.55\%, 0.754. 
}
\label{cpdsn6n7}
\end{figure}

\begin{figure}[t]
\centering\includegraphics[width=4.cm]{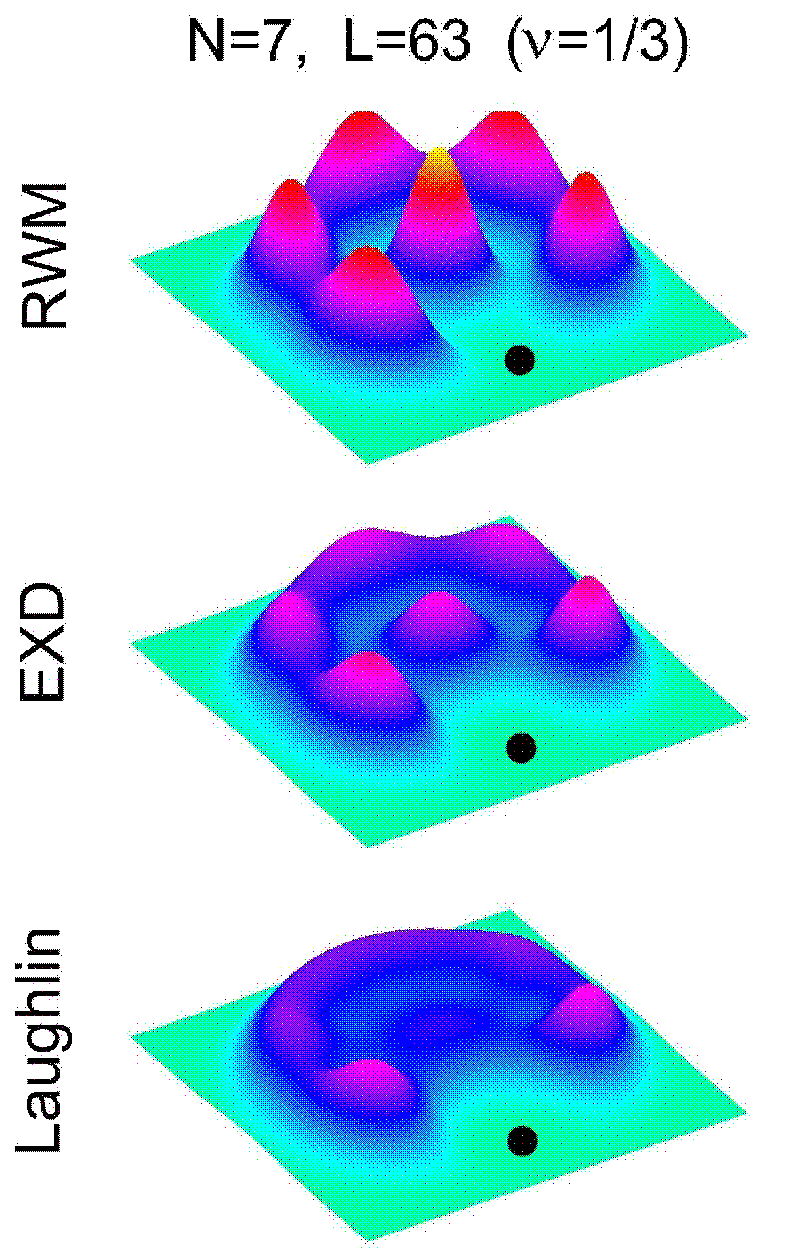}
\caption{(Color online) 
CPDs at high $B$ for $N=7$ and $L=63$ ($\nu=1/3$). 
Top: REM case; Middle: EXD case; 
Bottom: JL case. Unlike the JL CPD (which is liquid), the CPDs for
the exact-diagonalization and REM wave functions exhibit a well developed
crystalline character [corresponding to the (1,6) polygonal 
configuration of the REM for $N=7$ electrons]. The observation point 
(identified by a solid dot) is located at $r_0=4.568 l_B$.
}
\label{cpdsn7nu13}
\end{figure}

\subsection{REM versus Laughlin wave functions: Conditional probability 
distributions and multiplicity of zeroes}
\label{remvslaughlin}

Recent extensive numerical calculations \cite{yl02.2,yl04.2} have revealed 
major disagreements between the intrinsic structure of the Jastrow/Laughlin
trial wave functions \cite{laug83.1} for the main fractions $\nu=1/(2m+1)$
and that of the exact-diagonalization and REM wave functions. Indeed, it
was found that both EXD and REM wave functions exhibit crystalline
correlations, while the Jastrow/Laughlin ones are liquid-like as originally
described in Ref.\ \cite{laug83.1}. 

To illustrate the differences between the intrinsic structure of the REM and
EXD states in the lowest Landau level versus the familiar Jastrow/Laughlin 
ones, we display
in \fref{cpdsn6n7} the CPDs for cusp states corresponding to a low filling 
factor $\nu = 1/5$ and for two different sizes, i.e., for $N=6$ 
electrons ($L=75$, left column) and $N=7$ electrons ($L=105$, right column). 
In \fref{cpdsn6n7}, the top row depicts the REM case; 
the EXD case is given by the middle row, while the CF case [which reduces to 
the JL wave functions for fractions $1/(2p+1)$] are given by the bottom row. 

There are three principal conclusions that can be drawn from an inspection
of \fref{cpdsn6n7} (and the many other cases studied in Ref.\ \cite{yl03.2}).

(I) The character of the exact-diagonalization states is unmistakably 
crystalline with the EXD 
CPDs exhibiting a well developed molecular polygonal configuration [(1,5) for
$N=6$ and (1,6) for $N=7$, with one electron at the center], in agreement with 
the explicitly crystalline REM case. 

(II) For all the examined instances covering the low fractional fillings 
1/9, 1/7, and 1/5, the Jastrow/Laughlin wave functions fail to capture 
the intrinsic crystallinity of the exact-diagonalization
states. In contrast, they represent
``liquid'' states in agreement with an analysis that goes back to 
the original papers \cite{laug83.1,laug87} by Laughlin. In particular, Ref.\ 
\cite{laug87} investigated the character of the JL states through the use 
of a pair correlation function [usually denoted by $g(R)$] that determines the 
probability of finding another electron at the absolute relative distance 
$R=|{\bf r} - {\bf r}_0|$ from the observation point ${\bf r}_0$. 
Our anisotropic CPD of \Eref{cpds} is of course more general (and more 
difficult to calculate) than the $g(R)$ function of Ref.\ \cite{laug87}. 
However, both our $P({\bf r},{\bf r}_0)$ (for $N=6$ and $N=7$ electrons) and 
the $g(R)$ (for $N=1000$ electrons, and for $\nu=1/3$ and $\nu=1/5$) in Ref.\ 
\cite{laug87} reveal a similar characteristic liquid-like and
short-range-order behavior for the JL states, eloquently described in Ref.\ 
\cite{laug87} (see p. 249 and p. 251). Indeed, we remark that only the 
first-neighbor electrons on the outer rings can be distinguished as separate 
localized electrons in our CPD plots of the JL functions 
[see \fref{cpdsn6n7}]. 

(III) For a finite number of electrons, pronounced crystallinity of the 
exact-diagonalization states occurs already at the $\nu=1/5$ value 
(see \fref{cpdsn6n7}). This finding is particularly interesting in light of 
expectations \cite{jeon04.1,jeon04.2} (based on comparisons 
\cite{laug83.1,laug87,lam84} between the JL states and the 
static bulk Wigner crystal) that a liquid-to-crystal phase transition may take
place only at lower fillings with $\nu \leq 1/7$. 

Of interest also is the case of $\nu=1/3$. Indeed, for this
fractional filling, the liquid JL function is expected to provide the best
approximation, due to very high overlaps (better than 0.99) with the exact 
wave function \cite{fqhe1,jainbook,tsip01}. 
In \fref{cpdsn7nu13}, we display the CPDs
for $N=7$ and $L=63$ ($\nu=1/3$), and for the three cases of REM, EXD, and JL
wave functions. Again, even in this most favorable case, the CPD of the
JL function disagrees with the EXD one, which exhibits clearly a (1,6) 
crystalline configuration in agreement with the REM CPD. 

Similar crystalline correlations at higher fractions were also found for 
quantum dots of larger sizes, e.g., $N=8$, and $N=9$ 
electrons. As illustrative examples 
for these additional sizes [see also the EXD CPD for $N=12$ electrons in
\fref{n12cpd} below (in \sref{yrana})], we displayed in figure 5 
of Ref.\ \cite{yl03.2} the CPDs for $N=8$ and $L=91$ ($1/5 < \nu=4/13 < 1/3$)
and for $N=9$ and $L=101$ ( $1/3 < \nu=36/101 < 1$). Again, the CPDs
(both for the REM and the EXD wave functions)
exhibit a well developed crystalline character in accordance with the 
(1,7) and (2,7) polygonal configurations of the REM, appropriate for $N=8$ and
$N=9$ electrons, respectively.

Another area of disagreement between REM and Laughlin wave functions
concerns the properties of the zero points. In this respect, we recall that
the Jastrow/Laughlin trial functions for $N$ electrons have the form
\begin{equation}
\Phi^{\rm JL}(z_1,z_2, \ldots, z_N) = \left( \prod_{1 \leq i<j \leq N} 
(z_i-z_j)^{2m+1} \right) \exp(-\sum_{i=1}^N z_i z_i^*/2).
\label{wfjl}
\end{equation}

Due to the Jastrow factors $(z_i-z_j)^{2m+1}$, it is apparent that the
Laughlin expressions \eref{wfjl} (as a function of a given $z_i$) have $N-1$ 
zero points, each of order $2m+1$, which are bound to the positions of the 
remaining $N-1$ electrons. In contrast, as discussed in Ref.\ \cite{yl02.2}, 
the analytic REM wave functions do not have zeroes with order higher than 
unity. In particular, only $N-1$ of the REM zeroes are bound to the positions 
of the remaining electrons, while the rest of them are free. Recently, it has 
been shown through extensive numerical studies \cite{anis04} that the 
properties of REM zeroes are in agreement with the behavior of the zeroes in 
exact-diagonalization wave functions; this is another indication of the 
superiority of the REM picture compared to the Laughlin theory.  

Before exiting this discussion, we remark about discrepancies of the Laughlin
quasihole theory in the context of quantum dots. In particular, we recall
that the Laughlin quasihole, with $N$ additional units of angular
momentum, has been conjectured to be the first excited state. However, LLL
exact-diagonalization calculations for $N$ 
electrons in a quantum dot have revealed
that this is not the case. Instead, the first excited state corresponds to an
increment in the total angular momentum which varies as the number of
electrons localized on one of the rings of the rotating electron molecule,
usually the outermost one; see \fref{gapn7} in \sref{impther} below.

\section{Rotating electron molecules in two-dimensional quantum dots under a 
strong, but finite external  magnetic field ($\omega_c/2\omega_0 > 1$)}
\label{rotelfinb}

\begin{figure}[t]
\centering\includegraphics[width=6.3cm]{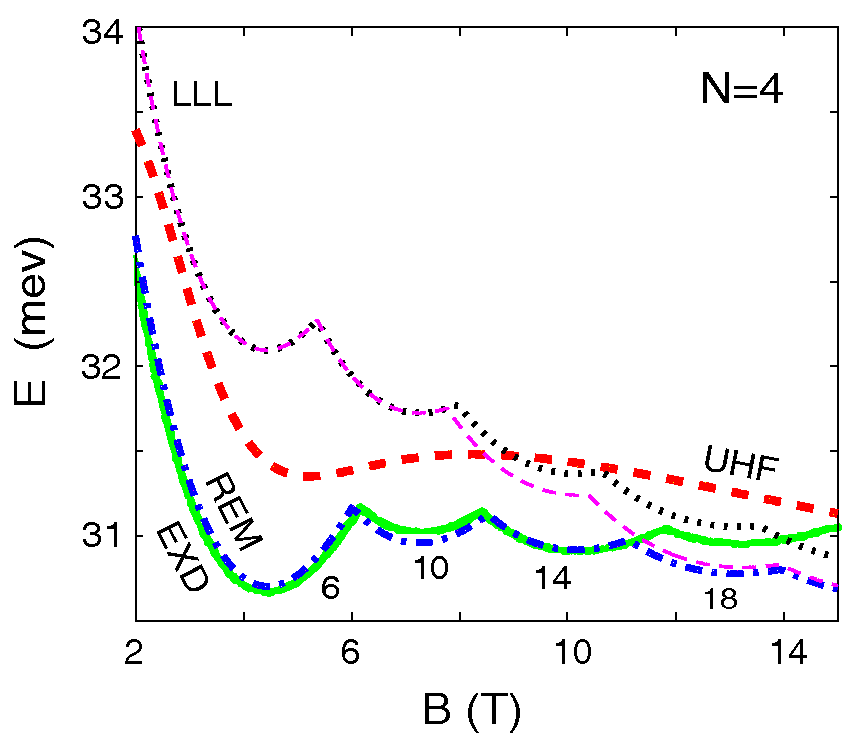}
\caption{(Color online)
Two-step-method versus exact-diagonalization calculations: 
Ground-state energies for $N=4$
electrons (referenced to $4 \hbar \tilde{\omega}$) as a function of the 
magnetic field $B$.
Thick dashed line (red): broken-symmetry UHF (static electron molecule).
Solid line (green): EXD (from Ref.\ \cite{rua99}).
Thick dashed-dotted line (blue): REM.
Thin dashed line (violet, marked LLL): the commonly used approximate energies
$\tilde{E}^{\rm{EXD}}_{\rm{LLL}}(B)$ (see text for details).
Thin dotted line (black): $\tilde{E}^{\rm{REM}}_{\rm{LLL}}(B)$ (see text).
For $ B < 8$ T, the $\tilde{E}^{\rm{EXD}}_{\rm{LLL}}(B)$ and
$\tilde{E}^{\rm{REM}}_{\rm{LLL}}(B)$ curves coincide;
we have checked that these curves approach each other also at larger
values of $B$, outside the plotted range. Numbers near the bottom curves
denote the value of magic angular momenta [$L_m$, see \eref{lmeq}]
of the ground state. Corresponding fractional filling factors are specified by
$\nu = N(N-1)/(2L_m)$.
Parameters used: confinement $\hbar \omega_0=3.60$ meV, 
dielectric constant $\kappa=13.1$, effective mass $m^*=0.067 m_e$. 
}
\label{remvsexdf}
\end{figure}

\subsection{Ground-state energies in medium and high magnetic field}

The general form \eref{uhfo} for the displaced Gaussian orbitals [in
conjunction with the projected REM wave function \eref{wfprj1}] enables us to 
calculate REM ground-state energies for moderately-high $B$, when corrections 
arising from higher Landau levels must be taken into consideration. Unlike the
lowest-Landau-level case, where the azimuthal integration can be carried out 
analytically, the energies \eref{eproj} (and corresponding CPDs) associated 
with the general REM wave function \eref{wfprj1} require {\it numerical\/}
integration over the azimuthal angles $\gamma_q$.

Before proceeding with the presentation of results for $N > 10$,
we demonstrate the accuracy of the two-step method embodied in \Eref{wfprj1}
through comparisons with existing exact-diagonalization results for smaller 
sizes. In \fref{remvsexdf}, our REM calculations for the ground-state energies
as a function of $B$ are compared to EXD calculations \cite{rua99} for $N=4$ 
electrons in an external parabolic confinement. The thick dotted line (red) 
represents the broken-symmetry UHF approximation (first step of our method), 
which naturally is a smooth curve lying above the EXD one [solid line 
(green)]. The results obtained after restoration of symmetry [dashed-dotted 
line (blue); marked as REM] agree very well with the EXD one in the whole 
range 2 T$ < B <$ 15 T. We recall here that, for the parameters of the quantum
dot, the electrons form in the intrinsic frame of reference a square about
the origin of the dot, i.e., a (0,4) configuration, with the zero indicating
that no electron is located at the center. According to \eref{lmeq},
$L_0=6$, and the magic angular momenta are given by $L_m=6+4k$, $k=0$, 
1, 2, $\ldots$ Note that the REM energies are slightly lower than the EXD 
ones in several subranges. According to the Rayleigh-Ritz variational theorem,
this indicates that the hyperspherical-harmonics calculation (equivalent to an
exact-diagonalization approach) of \cite{rua99} did not converge fully in 
these subranges.

To further evaluate the accuracy of the two-step method, we also display 
in \fref{remvsexdf} [thin dashed line (violet)] ground-state energies 
$\tilde{E}^{\rm{EXD}}_{\rm LLL}(B)$ calculated with the commonly used 
approximate LLL Hamiltonian \cite{maks96,jain95,jeon04.2,yang93}  
\begin{equation}
\tilde{\cal H}_{\rm LLL}=
N \hbar \tilde{\omega} + \hbar (\tilde{\omega} - \frac{\omega_c}{2}) L 
+ \sum_{j>i=1}^N \frac{e^2}{\kappa r_{ij}},
\label{hlll2}
\end{equation}
where $\tilde{\omega} = \sqrt{\omega_0^2 +\omega_c^2/4}$. The LLL Hamiltonian 
$\tilde{\cal H}_{\rm LLL}$ reduces to the previously introduced Hamiltonian
${\cal H}_{\rm LLL}$ [see \Eref{hlll}] in the limit $B \rightarrow \infty$.
Both Hamiltonians restrict the many-body wave functions within the lowest
Landau level, and they both accept the same set of eigenstates as solutions.
Indeed the term $\hbar (\tilde{\omega} - \omega_c/2) L$ is proportional
to the total angular momentum, and thus its presence influences only the
eigenvalues, but not the composition of the eigenstates.
$\tilde{\cal H}_{\rm LLL}$ corresponds to a situation where the external 
harmonic confinement is added to ${\cal H}_{\rm LLL}$ as a perturbation
(see section II.B in Ref.\ \cite{li06}). As a result, (i) the degeneracy of
the single-particle levels in the lowest Landau level 
is lifted and (ii) there is an
eigenstate with minimum energy (the ground state) at each value of $B$ 
(expressed through the cyclotron frequency $\omega_c$). Naturally,
the LLL levels used in the exact diagonalization of $\tilde{\cal H}_{\rm LLL}$
are given by expression \eref{splll}, but with 
$\Lambda = \tilde{l} = \sqrt{\hbar/(m^*\tilde{\omega})}$.

We find that the energies $\tilde{E}^{\rm EXD}_{\rm LLL} (B)$ 
tend to substantially overestimate the REM (and EXD) energies for lower values
of $B$ (e.g., by as much as 5.5\% at $B \sim 4$ T). On the other hand, for 
higher values of $B$ ($>$ 12 T), the energies 
$\tilde{E}^{\rm{EXD}}_{\rm{LLL}}(B)$ tend to agree rather well with 
the REM ones. We stress that the results labelled simply as EXD correspond
to exact diagonalizations without any restrictions on the Hilbert space,
i.e., the full Darwin-Fock single-particle spectrum is considered at a given
$B$.

A behavior similar to $\tilde{E}^{\rm EXD}_{\rm LLL} (B)$ is exhibited also by
the $\tilde{E}^{\rm{REM}}_{\rm{LLL}}(B)$ ground-state energies 
[which are calculated using the Hamiltonian \eref{hlll2} and the LLL analytic 
REM wave functions in \sref{antrial} with lengths in units of 
$\sqrt{\hbar/(m^*\tilde{\omega})}$ instead of $l_B \sqrt{2}$; 
dotted line (black)]. 
A similar agreement between REM and EXD results, and a similar inaccurate 
behavior of the LLL approximate Hamiltonian \eref{hlll2} was found by
us also for $N=3$ electrons in the range 2 T $< B <$ 16 T shown in figure 2 of
Ref.\ \cite{li06} (the exact-diagonalization calculation in this figure was 
taken from Ref.\ \cite{haw93}). 

In all cases, the total energy of the REM is lower than that of the UHF
Slater determinant (see, e.g., \fref{remvsexdf}). Indeed, a theorem 
discussed in section 3 of Ref. \cite{low62}, pertaining
to the energies of projected wave functions, guarantees that this lowering of
energy applies for all values of $N$ and $B$.

\subsection{The case of $N=11$ electrons.}

\begin{figure}[t]
\centering\includegraphics[width=6.5cm]{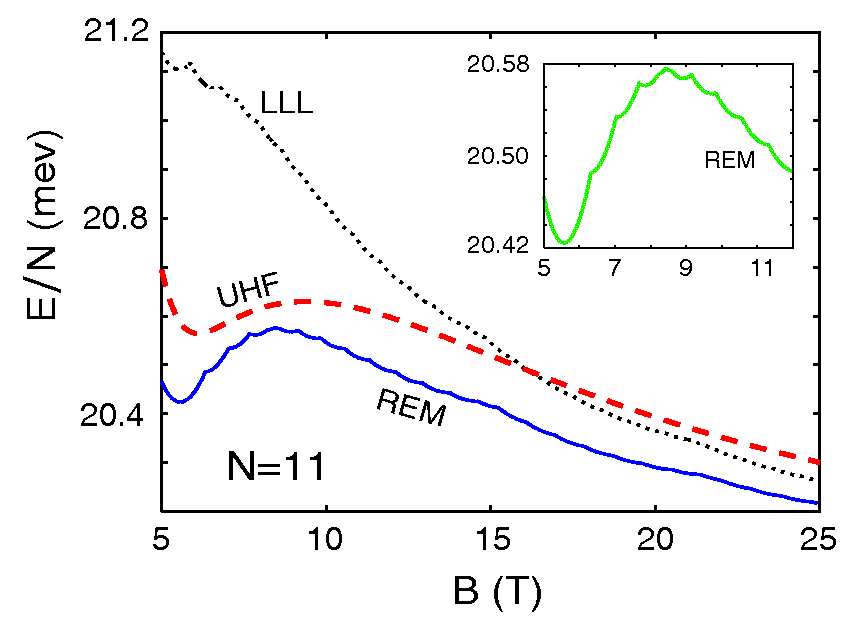}
\caption{(Color online)
Ground-state energies for $N=11$ electrons (per particle, referenced to 
$\hbar \tilde{\omega}$) as a function of the magnetic field $B$. 
Dashed line (red): UHF (static electron molecule). 
Solid line (blue): REM. Dotted line (black): 
Approximate energies 
$\tilde{E}^{\rm{REM}}_{\rm{LLL}}(B)$ (see text). 
Parameters used: confinement $\hbar \omega_0=3.60$ meV, 
dielectric constant $\kappa=13.1$, effective mass $m^*=0.067 m_e$. 
The inset shows a magnification of the REM curve in the range
5 T $<B<$ 12 T.
\label{n11eon}
}
\end{figure}

\begin{table}[b]
\caption{\label{k1k2}%
Ground-state magic angular momenta and their decomposition $\{ k_1,k_2\}$
for $N=11$ in the nagnetic-field range 5 T $ \leq B \leq $ 25 T. 
The results correspond to the REM (see lower curve in \fref{n11eon}). 
The parameters used are as in \fref{n11eon}.
}
\begin{indented}
\item[]\begin{tabular}{rrr|rrr}
\br
$L_m$  & $k_1$ & $k_2$ & $L_m$  & $k_1$ & $k_2$ \\
\mr
55  & 0 & 0 & 165 & 2  & 13  \\
63  & 0 & 1 & 173 & 2  & 14  \\ 
71  & 0 & 2 & 181 & 2  & 15  \\
79  & 0 & 3 & 189 & 2  & 16  \\
90  & 1 & 4 & 197 & 2  & 17  \\
98  & 1 & 5 & 205 & 2  & 18  \\
106  & 1 & 6 & 213 & 2  & 19  \\
114  & 1 & 7 & 224 & 3  & 20  \\
122  & 1 & 8 & 232 & 3  & 21  \\
130  & 1 & 9 & 240 & 3  & 22  \\
138  & 1 & 10 & 248 & 3 & 23   \\
146  & 1 & 11 & 256 & 3 & 24  \\
154  & 1 & 12 &     &   &     \\
\br
\end{tabular}
\end{indented}
\end{table}

\Fref{n11eon} presents the case for the ground-state energies of a 
quantum dot with $N=11$ electrons, which 
have a nontrivial double-ring configuration 
$(n_1,n_2)$. The most stable \cite{kong02} classical configuration is $(3,8)$, 
for which we have carried UHF (static electron molecule) and REM (projected) 
calculations in the magnetic field range 
5 T $< B <$ 25 T. \Fref{n11eon} also displays the LLL ground-state energies 
$\tilde{E}^{\rm{REM}}_{\rm{LLL}}(B)$ 
[dotted curve (black)], which, as in previous cases, overestimate
the ground-state energies for smaller $B$. The approximation
$\tilde{E}^{\rm{REM}}_{\rm{LLL}}(B)$, however, can be used 
to calculate ground-state energies for higher values of $B$.  
In keeping with the findings for smaller sizes \cite{yl04.1} [with $(0,N)$ 
or $(1,N-1)$ configurations], we found that both the UHF and the REM 
ground-state energies approach, as $B \rightarrow \infty$, the 
{\it classical\/} equilibrium energy of the (3,8) polygonal configuration 
[i.e., 19.94 meV; 4.865$E_0$ in the units of 
Ref.\ \cite{kong02}, $E_0 \equiv (m^* \omega_0^2 e^4/2\kappa^2)^{1/3}$]. 

In analogy with smaller sizes (see, e.g., \fref{remvsexdf} and 
Ref.\ \cite{li06}), the REM ground-state energies in \fref{n11eon}
exhibit oscillations as a function of $B$ (see in particular the inset). 
These oscillations are associated with magic 
angular momenta, specified by the number of electrons on each ring. 
For $N=11$ they are given by \eref{lmeq}, i.e., $L_m=55+3k_1+8k_2$, with 
the $k_q$'s being nonnegative integers. As was the case with $N=9$ electrons
\cite{li06}, an analysis of the actual values taken by the set of
indices $\{k_1,k_2 \}$ reveals several additional trends that further
limit the allowed values of ground-state $L_m$'s. In particular, starting
with the values $\{0,0 \}$ at $B=5$ T ($L_0=55$), the
indices $\{k_1,k_2 \}$ reach the values $\{3,24 \}$ at $B=25$ T ($L_m=256$). 
As seen from \tref{k1k2}, the outer index $k_2$ changes faster than 
the inner index $k_1$. This behavior minimizes the total kinetic energy of 
the independently rotating rings; indeed, the kinetic energy of the inner ring
(as a function of $k_1$) rises faster than that of the outer ring (as a 
function of $k_2$) due to smaller moment of inertia (smaller radius) of the 
inner ring [see \eref{eclkin}]. 

\begin{figure}[t]
\centering\includegraphics[width=7.0cm]{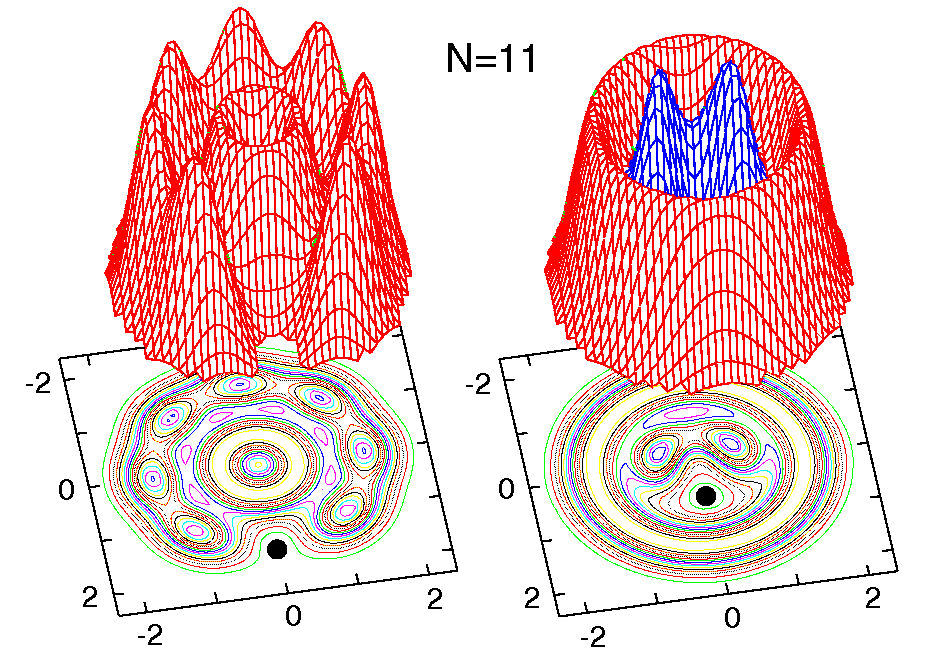}
\caption{(Color online)
Conditional probability distributions for the REM ground state of $N=11$ 
electrons at $B=10$ T ($L=106$). The electrons are arranged in a (3,8) 
structure. The observation point (solid dot) is placed on (left) the outer 
ring at $r_0=1.480 R_0$, and (right) on the inner ring at $r_0=0.557 R_0$. 
Parameters used: confinement $\hbar \omega_0=3.60$ meV, 
dielectric constant $\kappa=13.1$, effective mass $m^*=0.067 m_e$.
Lengths in units of $R_0=(2e^2/m^* \kappa \omega_0^2)^{1/3}$. 
CPDs (vertical axes) in arbitrary units.
}
\label{n11cpd}
\end{figure}

In addition to the overestimation of the ground-state energy values 
for smaller magnetic fields (see \fref{n11eon} and our discussion above), 
there are additional shortcomings of the lowest-Landau-level approximation 
pertaining to the ground-state ring configurations.
In particular, for $N=11$, we find that according to the LLL approximation the 
ground-state angular momentum immediately after 
the maximum density droplet ($L_0=55$) is $L_m=66$,
i.e., the one associated with the $(0,N)$ vortex-in-the-center configuration. 
This result, erroneously stated in Refs.\ \cite{tore04,tore06} as the ground 
state, disagrees with the correct result that includes the full effect of the
confinement and is listed in \tref{k1k2}, where the ground-state angular 
momentum immediately following the maximum density droplet is $L_m=63$. This 
angular-momentum value corresponds to the classicaly most stable (3,8) ring 
configuration, that is, a configuration with no vortex at all
(see also the case of $N=9$ electrons in Ref.\ \cite{li06}).

\Fref{n11cpd} displays the REM conditional probability distributions for the 
ground state of $N=11$ electrons at $B=10$ T ($L_m=106$). The (3,8) ring
configuration is clearly visible. We note that when the observation point is 
placed on the outer ring (left panel), the CPD
reveals the crystalline structure of this ring only; the inner ring appears to
have a uniform density. To reveal the crystalline structure of the inner ring,
the observation point must be placed on this ring; then the outer ring
appears to be uniform in density. This behavior suggests that the two rings
rotate independently of each other, a property that is explored in the next
section to derive an approximate {\it quasiclassical} expression for the yrast
rotational spectra associated with an arbitrary number of electrons.

\subsection{Approximate analytic expression for the yrast-band spectra}
\label{yrana}

\begin{figure}[b]
\centering\includegraphics[width=8.cm]{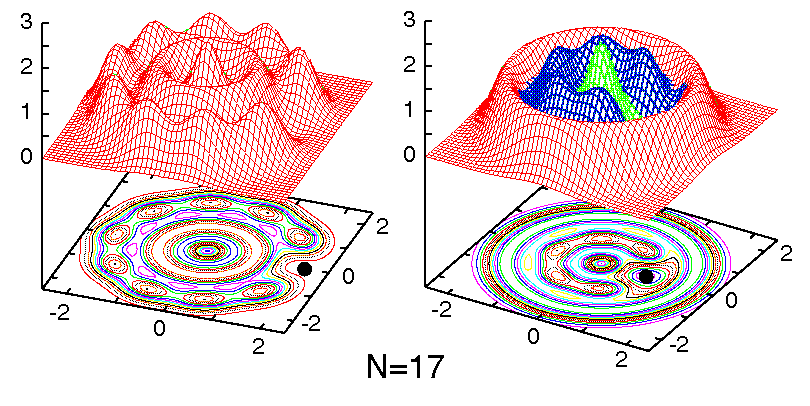}
\caption{(Color online)
Ground-state conditional probability distributions obtained from REM wave 
functions for the ground state of $N=17$ electrons at $B=10$ T ($L=228$).
The electrons are arranged in a (1,6,10) structure.
The observation point (solid dot) is placed on the outer ring at
$r_0=1.858R_0$ (left frame), and on the inner ring at $r_0=0.969R_0$
(right frame). The rest of the parameters are: confinement
$\hbar \omega_0 =3.6$ meV, dielectric constant $\kappa=13.1$,
effective mass $m^*=0.067m_e$.
Lengths in units of $R_0 =( 2e^2/(\kappa m^* \omega_0^2) )^{1/3}$.
CPDs (vertical axes) in arbitrary units.
}
\label{n17cpd}
\end{figure}

\begin{figure}[t]
\centering\includegraphics[width=8.0cm]{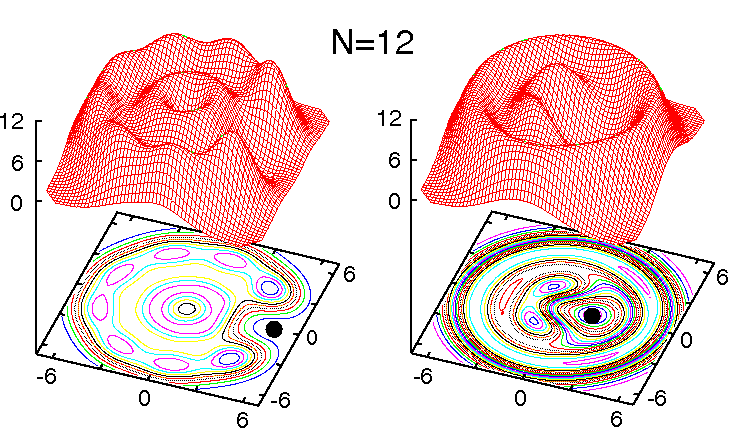}
\caption{(Color online)
CPDs for $N=12$ electrons and with angular momentum $L=132$ ($\nu=1/2$) 
calculated using exact diagonalization in the lowest Landau level.
The electrons are arranged in a (3,9) structure.
The observation point (solid dot) is placed on the outer ring at
$r_0=5.22 l_B$ (left frame), and on the inner ring at $r_0=1.87 l_B$
(right frame).  Lengths in units of $l_B$. 
CPDs (vertical axes) in arbitrary units.
}
\label{n12cpd}
\end{figure}

In \fref{n17cpd}, we display the CPD for the REM 
wave function of $N=17$ electrons.
This case has a nontrivial three-ring structure (1,6,10) \cite{kong02}
which is sufficiently complex to allow generalizations for larger numbers of
particles. The remarkable floppy character (leading to a non-classical,
non-rigid rotational inertia, see section VI of Ref.\ \cite{li06}) 
of the REM is illustrated in the CPDs of \fref{n17cpd}. Indeed, as the two CPDs
[reflecting the choice of taking the observation point [${\bf r}_0$ in
\eref{cpds}] on the outer (left frame) or the inner ring (right frame)]
reveal, the polygonal electron rings rotate {\it independently\/} of each 
other. Thus, e.g., to an observer located on the inner ring, 
the outer ring will appear
as having a uniform density, and vice versa. The wave functions obtained from
exact diagonalization exhibit also the property of independently rotating rings
[see, e.g., the $N=12$ and $L=132$ ($\nu=1/2$) case in \fref{n12cpd}], which
is a testimony to the ability of the REM wave function to capture the essential
physics of a finite number of electrons in high $B$. In particular, 
the conditional probability distribution displayed in \fref{n12cpd}
for exact-diagonalization wave functions exhibits the characteristics 
expected from the CPD evaluated using REM wave functions for the (3,9) 
configuration and with an angular-momentum decomposition into shell 
contributions [see \Eref{wfprj1} and \Eref{lmpar}]
$L_1=3+3k_1$ and $L_2=63+9k_2$ ($L_1+L_2=L_m$; for $L_m=132$ the 
angular-momentum decomposition is $L_1=6$ and $L_2=126$).


In addition to the conditional probabilities, the floppy-rotor character of the
REM is revealed in its excited rotational spectrum for a given $B$. From our
microscopic calculations based on the wave function in \eref{wfprj1}, we
have derived (see below) an approximate (denoted as ``app''), but
{\it analytic\/} and {\it parameter-free\/}, expression [see 
\eref{app} below] which reflects directly the nonrigid 
character of the REM for arbitrary size. This expression allows calculation of
the energies of REMs for arbitrary $N$, given the corresponding equilibrium
configuration of confined classical point charges.

We focus on the description of the yrast band at a given $B$.
Motivated by the aforementioned nonrigid character of the rotating electron
molecule, we consider the following kinetic-energy term corresponding to a
$(n_1,\ldots,n_q,\ldots,n_r)$ configuration (with $\sum_{q=1}^r n_q =N$):
\begin{equation}
E_{\rm{app}}^{\rm{kin}} (N)=
\sum_{q=1}^r \hbar^2 L_q^2/(2 {\cal J}_q (a_q)) - \hbar \omega_c L/2,
\label{eclkin}
\end{equation}
where $L_q$ is the partial angular momentum associated with the $q$th ring 
about the center of the dot and the total angular momentum is 
$L=\sum_{q=1}^r L_q$. ${\cal J}_q (a_q)) \equiv n_q m^* a_q^2$ is the 
rotational moment of inertia of each {\it individual\/} ring, 
i.e., the moment of inertia
of $n_q$ classical point charges on the $q$th polygonal ring of radius $a_q$.
To obtain the total energy, $E_L^{\rm{REM}}$, we include also the term
$E_{\rm{app}}^{\rm{hc}} (N) =\sum_{q=1}^r  {\cal J}_q (a_q) \tilde{\omega}^2/2$
due to the effective harmonic confinement $\tilde{\omega}$ 
(see Appendix A.1), as well as the interaction energy $E_{\rm{app}}^C$,
\begin{equation}
E_{\rm{app}}^C (N) = \sum_{q=1}^r \frac{n_q S_q}{4} \frac{e^2}{\kappa a_q} +
\sum_{q=1}^{r-1} \sum_{s > q}^r V_C(a_q,a_s).
\label{vc}
\end{equation}
The first term is the intra-ring Coulomb-repulsion energy of $n_q$ point-like 
electrons on a given ring, with a structure factor
\begin{equation}
S_q = \sum_{j=2}^{n_q}(\sin[(j-1)\pi/n_q])^{-1}.
\label{sq}
\end{equation}
The second term is the inter-ring Coulomb-repulsion energy between rings of
uniform charge distribution corresponding to the specified numbers 
of electrons on the polygonal rings. The expression fo $V_C$ is
\begin{equation}
\fl V_C(a_q,a_s)= n_q {n_s} e^2 [\kappa (a_q^2+a_s^2)^{1/2}]^{-1} \; 
{_2F_1} [3/4,1/4;1;4 a_q^2 a_s^2(a_q^2+a_s^2)^{-2}],
\label{vcc}
\end{equation}
where ${_2F_1}$ is the hypergeometric function.

For large $L$ (and/or $B$), the radii of the rings of the rotating molecule
can be found by neglecting the interaction term in the total approximate 
energy, thus minimizing only
$E_{\rm{app}}^{\rm{kin}} (N) + E_{\rm{app}}^{\rm{hc}} (N)$.
One finds 
\begin{equation}
a_q = \lambda \sqrt{L_q/n_q},
\label{rad}
\end{equation}
with $\lambda = \tilde{l} = \sqrt{\hbar /m^* \tilde{\omega}}$;
i.e., the ring radii depend on the partial angular momentum $L_q$,
reflecting the {\it lack of radial rigidity\/}. Substitution into the above
expressions for $E_{\rm{app}}^{\rm{kin}}$, $E_{\rm{app}}^{\rm{hc}}$, and
$E_{\rm{app}}^C$ yields for the total approximate energy the final
expression:
\begin{equation}
\fl E_{\rm{app},L}^{\rm{REM}}(N) =  \hbar(\tilde{\omega}-\omega_c/2) L +
\sum_{q=1}^r \frac{C_{V,q}}{L_q^{1/2}} +
\sum_{q=1}^{r-1} \sum_{s > q}^r
V_C(\lambda \sqrt{\frac{L_q}{n_q}}, \lambda \sqrt{\frac{L_s}{n_s}}),
\label{app}
\end{equation}
where the constants
\begin{equation}
C_{V,q}=0.25 n_q^{3/2} S_q e^2/(\kappa \lambda).
\label{ccvq}
\end{equation}
For simpler $(0,N)$ and $(1,N-1)$ ring configurations, \Eref{app} 
reduces to the expressions reported earlier \cite{yl04.1,maks96}.

The floppy-rotor character of the REM under strong magnetic field is reflected
in the absence in \eref{app} of a kinetic-energy term proportional to $L^2$.   
This contrasts with the rigid-rotor behavior of an electron molecule
at zero magnetic field (see \sref{2ecirdot} and Ref.\ \cite{yl04.1}).

\subsection{Possible implications for the thermodynamic limit}
\label{impther}

While our focus in this section is on the behavior of trial and exact wave 
functions in (finite) quantum dots in high magnetic fields, 
it is natural to inquire about possible implications of our findings to 
fractional-quantum-Hall-effect systems in the thermodynamic limit. 

We recall that appropriate trial wave functions for clean FQHE systems
possess a good angular momentum $L \geq L_0$, a property shared by
both the CF/JL and REM functions \cite{yl02.2,laug83.1,jain90,laug87}.
We also recall the previous finding \cite{laug83.1,laug87}
that for large fractional fillings $\nu > 1/7$, the liquid-like (and 
circularly uniform) Jastrow/Laughlin function is in the thermodynamic limit 
energetically favored compared to the broken-symmetry static Wigner crystal 
(which has no good angular momentum); for $\nu < 1/7$, the static Wigner 
crystal becomes lower in energy. This finding was enabled by the simple form 
of the JL functions, which facilitated computations of total energies as a 
function of size for sufficiently large $N$ (e.g., $N=1000$).

A main finding of the recent literature on quantum dots is that the 
{\it exact-numerical-diagonalization 
wave functions of small systems ($N \leq 12$) are crystalline in character 
for both low and high fractional fillings.\/} This finding contradicts 
earlier suggestions \cite{laug83.1,jain95,laug87} that,
for high $\nu$'s, small systems are accurately described by the liquid-like
JL wave functions and their descendants, e.g., the composite-fermion ones. 
Of course, for the same high $\nu$'s,
our small-size results cannot exclude the possibility that the CPDs of the 
exact solution may exhibit with increasing $N$ a transition from crystalline 
to liquid character, in agreement with the JL function. 
However, as of now the existence of such a transition remains an open
theoretical subject.

For the {\it low fractions\/}, the rotating-electron-molecule 
theory raises still another line of inquiry.
Due to the specific form of the REM wave functions, computational limitations
(in the so-called disk geometry that is natural to quantum dots) 
prevent us at present from making extrapolations of total energies
at a given $\nu$ as $N \rightarrow \infty$. Nevertheless, from the
general theory of projection operators, one can conclude that the REM
energies exhibit a different trend compared to the JL ones, whose energies
were found \cite{laug83.1,laug87} to be higher  
than the static Wigner crystal. Indeed the 
rotating-electron-molecule wave functions remain lower in energy than the 
corresponding {\it static\/} crystalline state for {\it all values\/} of $N$ 
and $\nu$, even in the thermodynamic limit. This is due to an ``energy gain''
theorem (see Section 3 in Ref.\ \cite{low62}) 
stating that at least one of the projected states (i.e., 
the ground state) has an energy lower than that of the original 
broken-symmetry trial function (e.g., the UHF determinant), and this
theorem applies for any number of electrons $N$ and for all values of the 
magnetic field $B$.
Naturally, the REM wave functions will be physically relevant compared to 
those of the broken-symmetry crystal at the thermodynamic limit if 
the energy gain does not vanish when $N \rightarrow \infty$; otherwise, one 
needs to consider the posssibility that the static crystal is the relevant 
physical picture.

\begin{figure}[t]
\centering\includegraphics[width=7.0 cm]{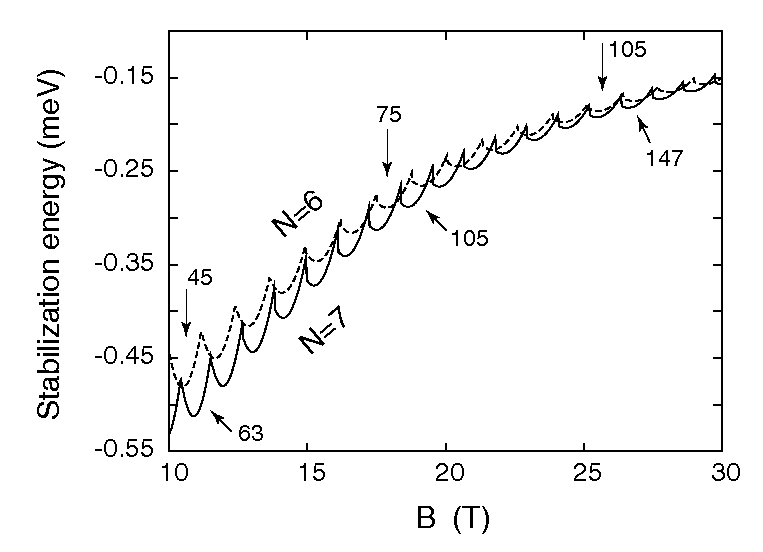}
\caption{Stabilization energies $\Delta E^{\rm{gain}}_{\rm{gs}}=
E^{\rm{gs}}_{\rm{REM}} - E_{\rm{UHF}}$
for $N=6$ (dashed curve) and 
$N=7$ (solid curve) fully polarized electrons in a parabolic QD as a function 
of $B$. The troughs associated with the major fractional fillings (1/3, 1/5, 
and 1/7) and the corresponding ground-state angular momenta 
[$L=N(N-1)/(2\nu)$] are indicated with
arrows. We have extended the calculations up to $B=120$ T (not shown), and 
verified that $\Delta E^{\rm{gain}}_{\rm{gs}}$ remains negative while its 
absolute value vanishes as $B \rightarrow \infty$.
The choice of parameters is: $\hbar \omega_0=3$ meV (parabolic confinment), 
$m^*=0.067 m_e$ (electron effective mass), and $\kappa =12.9$ (dielectric
constant).
}
\label{staben}
\end{figure}

The discussion in the above paragraph may be recapitulated by the following 
question: which state is the relevant one in the thermodynamic limit
$(N \rightarrow \infty)$ -- the broken-symmetry one (i.e., the static 
crystal) or the symmetry restored (i.e., rotating crystal) state? This
question, in the context of bulk broken-symmetry systems, has been addressed
in the early work of Anderson \cite{pwa} who concluded that 
the broken-symmetry state (here the UHF static crystalline solution)
can be safely taken as the effective ground state. In arriving at this
conclusion Anderson invoked the concept of (generalized) rigidity.
As a concrete example, one would expect a crystal to behave like a 
{\it macroscopic\/} body, whose Hamiltonian is that of a {\it heavy rigid 
rotor\/} with a low-energy excitation spectrum $L^2/2{\cal J}$, the moment of
inertia ${\cal J}$ being of order $N$ (macroscopically large when 
$N \rightarrow \infty$). The low-energy excitation spectrum of this heavy 
rigid rotor above the ground-state ($L=0$) is essentially gapless (i.e., 
continuous). Thus although the formal ground state posseses continuous
rotational symmetry (i.e., $L=0$), ``there is a manifold of other states, 
degenerate in the $N \rightarrow \infty$ limit, which can be recombined to 
give a very stable wave packet with essentially the nature'' \cite{pwa}
of the broken-symmetry state (i.e., the static Wigner crystal in our case).
As a consequence of the ``macroscopic heaviness'' as $N \rightarrow \infty$, 
one has: (I) The energy gain due to symmetry restoration (i.e., the 
stabilization energy $\Delta E^{\rm{gain}}_{\rm{gs}}=
E^{\rm{gs}}_{\rm{REM}} - E_{\rm{UHF}}$, see \fref{staben}) vanishes as 
$N \rightarrow \infty$, and (II) The relaxation of the system from the wave 
packet state (i.e., the static Wigner crystal) to the symmetrized one (i.e., 
the rotating crystal) becomes exceedingly long. This picture underlies 
Anderson's aforementioned conclusion that in the thermodynamic limit the 
broken-symmetry state may be used as the effective ground state.

Consequently, in the rest of this section we will focus on issues 
pertaining to the ``rigidity'' of the rotating electron molecule
in high magnetic fields. In particular, using our projection method and exact 
diagonalization, we have demonstrated explicitly 
\cite{yl00.2,yl04.1} that the rigid-rotor picture applies to an $N$-electron 
QD only when $B=0$. In contrast, in the presence of a high magnetic field, we 
found \cite{yl04.1,yl04.2,li06} that the electrons in the quantum dot do not 
exhibit global rigidity and therefore cannot be modeled as a macroscopic
rotating crystal. Instead, a more appropriate model is that
of a {\it highly non-rigid\/} rotor whose moment of inertia
depends strongly on the value of the angular momentum $L$. This behavior
originates from the dominance of the magnetic field over the Coulomb
repulsion.

\begin{figure}[t]
\centering\includegraphics[width=6.5 cm]{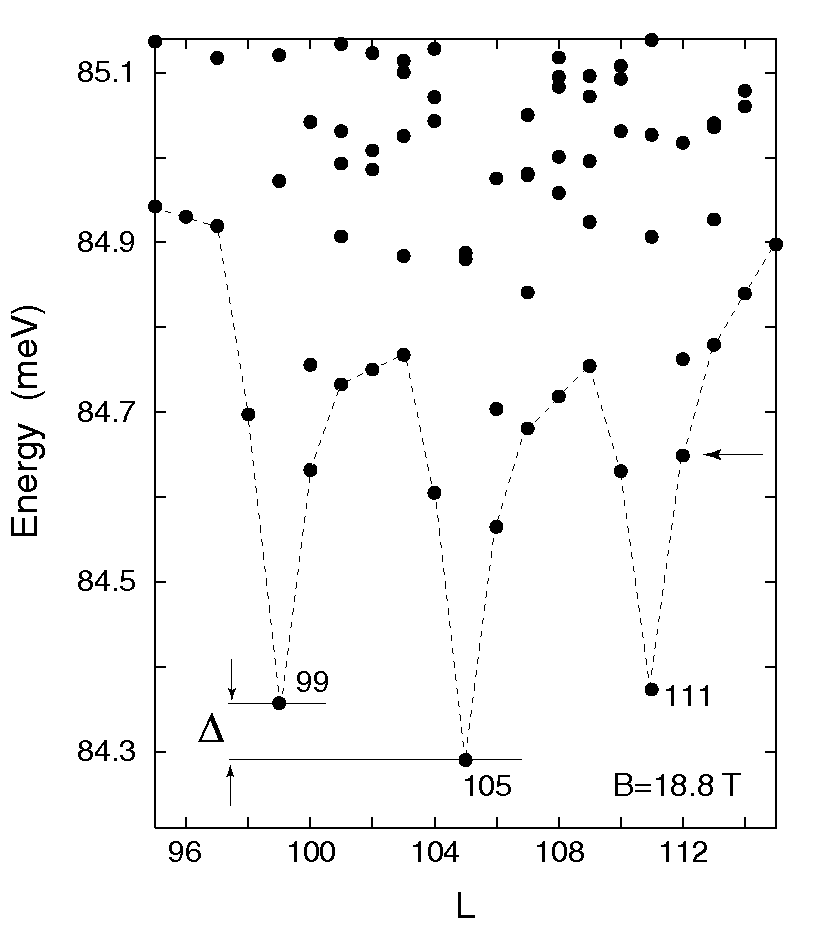}
\caption{
Low-energy part of the spectrum of the parabolic QD whose parameters
are the same as those in \fref{staben}, calculated as a function of the
angular momentum $L$ through exact diagonalization for $N=7$ electrons 
at a magnetic field $B=18.8$ T. We show here the spectrum in the interval
$95 \leq L \leq 115$ (in the neighborhood of $\nu=1/5$). 
The magic angular momentum values corresponding to cusp states are marked
(99, 105, and 111), and they are seen to be separated from the rest of the 
spectrum. For the given value of $B$, the global energy minimum (ground
state) occurs for $L_{\rm{gs}}=105$, and the gap $\Delta$ to the
first excited state ($L=99$) is indicated. The lowest energies for the
different $L$'s (the yrast band) in the plotted 
range are connected by a dashed line, as a 
guide to the eye. The zero of energy corresponds to $7 \hbar \tilde{\omega}$, 
where $\tilde{\omega}=(\omega_0^2 + \omega_c^2/4)^{1/2}$ and 
$\omega_c =eB/(m^*c)$.
The horizontal arrow denotes the energy of the Laughlin quasihole at L=112. 
It is seen that the Laughlin quasihole is not the lowest excited state,
as presumed in Ref.\ \cite{laug83.1}.
}
\label{gapn7}
\end{figure}

The non-rigid rotor at high $B$ has several unique properties: 
(I) The ground state has angular 
momentum $L_{\rm{gs}} > 0$; (II) While the rotating electron molecule 
does not exhibit {\it global\/} rigidity, it possesses {\it azimuthal\/} 
rigidity (i.e., all electrons on a given ring rotate coherently), with the 
rings, however, rotating independently of each other.
Furthermore, the radii of the rings vary for different values of $L$,
unlike the case of a rigid rotor; (III) The excitation spectra do not vary as 
$L^2$; instead they consist of terms that vary as 
$aL+\sum_{q=1}^r b_q/\sqrt{L_q}$ (with $\sum_{q=1}^r L_q=L$; 
for the precise values of the constants 
$a$ and $b$ see \sref{yrana} and Refs.\ \cite{yl04.1,li06}); 
(IV) The angular momentum values are given by the
magic values [see \sref{antrial}] $L=L_0+\sum_{q=1}^r k_q n_q$, where 
$(n_1,n_2,\ldots,n_r)$ is the polygonal ring arrangement of the static Wigner 
molecule (with $n_q$ the number of electrons on the $q$th
ring) and $k_1 < k_2 <\ldots< k_q$ are nonnegative integers. 
These magic $L$'s are 
associated with the cusp states which exhibit a relative energy gain with
respect to neighboring excitations. Thus the low-energy excitation spectrum
of the non-rigid rotor is not dense and exhibits gaps due to the occurrence of
the magic (cusp) states (see \fref{gapn7}). Furthermore, these gaps are 
reflected in the oscillatory behavior of $\Delta E^{\rm{gain}}_{\rm{gs}}$ 
(see, e.g., \fref{staben}) as a function of $B$ (or $\nu$).

As $N$ increases, more polygonal rings are successively added, and since 
the polygonal rings rotate independently of each other
(see, e.g., the case of $N=12$ in \fref{n12cpd}), we expect that the
non-rigid-rotor picture remains valid even as $N \rightarrow \infty$. As a 
result, it is plausible to conjecture the following properties at high $B$ in
the thermodynamic  limit: 
(I) the oscillatory character of $\Delta E^{\rm{gain}}_{\rm{gs}}$ will 
maintain, yielding nonvanishing  stabilization energies at the fractional
fillings $\nu$, and (II) the low-energy excitation 
spectra of the system will still exhibit gaps in the neighborhood of the magic
angular momenta (see \fref{gapn7}). 
Of course, these conjectures need to be further supported 
through numerical calculations for large $N$. Nevertheless, 
the above discussion indicates that the question of which state is physically 
relevant for low fractions in the thermodynamic limit at high $B$ -- i.e., the 
broken-symmetry static crystal or the symmetrized rotating crystal -- 
remains open, and cannot be answered solely following the path
of Anderson as described in Ref.\ \cite{pwa}.

The rotating Wigner crystal has properties characteristic of FQHE
states, i.e., it is incompressible (connected to the presence of an 
excitation gap) and carries a current 
(while the broken-symmetry static crystal is insulating). 
Thus, we may conjecture that a transition at lower fractional fillings from a 
conducting state with good circular symmetry to an insulating Wigner crystal 
cannot occur {\it spontaneously\/} for clean systems. Therefore, it should be 
possible to observe FQHE-type behavior at low fractional fillings in a clean 
system -- a prediction that could explain the observations of Ref.\ 
\cite{pan02}, where FQHE behavior has been observed for low 
fractional fillings typically associated with the formation of a
static Wigner crystal. In practice, however, impurities 
and defects may influence the properties of the rotating crystal (and its 
excitations), depending on the magnitude of the excitation gap (see, e.g., 
\fref{gapn7}). Thus one of the main challenges for observation of the
fractional quantum Hall effect at such low fillings relates to fabrication of 
high mobility (nearly impurity-free) samples \cite{wes03}.
We remark, however, that the stabilization energy and the gap $\Delta$ 
(see, e.g., \fref{gapn7}) diminish as the magnetic field increases, and as a 
result the impurities become more efficient in influencing the rotating Wigner
crystal for the lower fractional fillings (i.e., higher angular momenta).

\section{Bosonic molecules in rotating traps: Original results and 
applications}
\label{bosmol}

\subsection{Variational description of rotating boson molecules}
\label{varrbm}

Recent experimental advances in the field of trapped ultracold neutral
bosonic gases have enabled control of the strength of interatomic interactions
over wide ranges \cite{wei,par,grei,cor}, from the very weak to the
very strong. This control is essential for experimental
realizations of novel states of matter beyond the well known Bose-Einstein
condensate \cite{wei,par,grei}. In this context, the {\it linear\/} 1D 
Tonks-Girardeau regime of impenetrable trapped bosons has generated intensive 
theoretical activity \cite{gir2,dunj} and several experimental realizations of 
it have been reported most recently \cite{wei,par}.

In this section, we address the properties of strongly-repelling impenetrable 
bosons in {\it rotating\/} ring-shaped or 2D harmonic traps. It has been
found that impenetrable bosons are ``localized'' relative to each 
other \cite{roma04,roma06,wei} and exhibit nontrivial intrinsic crystalline 
correlations \cite{roma04,roma06}. 
For a small number of bosons, $N$, these crystalline
arrangements are reminiscent of the structures exibited by the well-studied
rotating electron molecules in quantum dots under high magnetic fields 
\cite{yl03.2,yl04.2,li06}. Consequently, we use in the following the term 
rotating boson molecules. A central result of our study is that the point-group
symmetries of the intrinsic crystalline structures give rise to characteristic 
regular patterns (see below) in the ground-state spectra and associated angular
momenta of the RBMs as a function of the rotational frequency for neutral
bosons (or the magnetic field for charged bosons).

An unexpected result of our studies is that the rotation of repelling bosons 
(even those interacting weakly) does not necessarily lead to formation of 
vortices, as is familiar from the case of rotating Bose-Einstein condenstates.
In particular, for small $N$, we 
will show that the Gross-Pitaevskii energies (including those corresponding to
formation of vortices) remain always higher compared to the 
ground-state energies of the RBMs. Of course, we expect that the rotating BEC 
will become the preferred ground state for sufficiently large $N$
in the case of weakly repelling neutral bosons. We anticipate, however, that
it will be feasible to test our unexpected results for small $N$ by using
rotating optical lattices, where it is established that a small finite
number of atoms can be trapped per given site \cite{grei}. 

In a non-rotating trap, it is natural to describe a localized boson (at a 
position ${\bf R}_j$) by a simple displaced Gaussian \cite{roma04}. When the 
rotation of the trap is considered, the Gaussian needs to be modified by a
phase factor, determined through the analogy between the one-boson Hamiltonian 
in the rotating frame of reference and the planar motion of a charged particle 
under the influence of a perpendicular magnetic field $B$ (described in the 
symmetric gauge). That is, the single-particle wave function of a localized 
boson is
\begin{equation} 
\varphi_j({\bf r})\equiv \varphi({\bf r},{\bf R}_j)=
\frac{1}{ \sqrt{\pi} \lambda } 
\exp \left[ 
\frac{ ({\bf r}-{\bf R}_j)^2 } {2\lambda^2} 
- \rmi {\bf r} {\bf \cdot} ({\bf Q} \times {\bf R}_j) \right],
\label{disg}
\end{equation}
with  ${\bf Q} \equiv  {\bf \hat{z}} /\Lambda^2$ 
and the width of the Gaussian $\lambda$ is a variational parameter;
$\Lambda \equiv l_B \sqrt{2} = \sqrt{2\hbar c/(eB)} = 
\sqrt{2\hbar/(m\omega_c)}$ 
for the case of a perpendicular magnetic field ${\bf B}$, and 
$\Lambda \equiv l_\Omega \sqrt{2} = \sqrt{\hbar /(m\Omega)}$
in the case of a rotating trap with rotational frequency ${\bf \Omega}$
(we recall that $\omega_c \rightarrow 2 \Omega$, see the Appendix).
Note that we consider a 2D trap, so that ${\bf r} \equiv (x,y)$ and
${\bf R} \equiv (X,Y)$.

\begin{figure}[t]
\centering{\includegraphics[width=5.50cm]{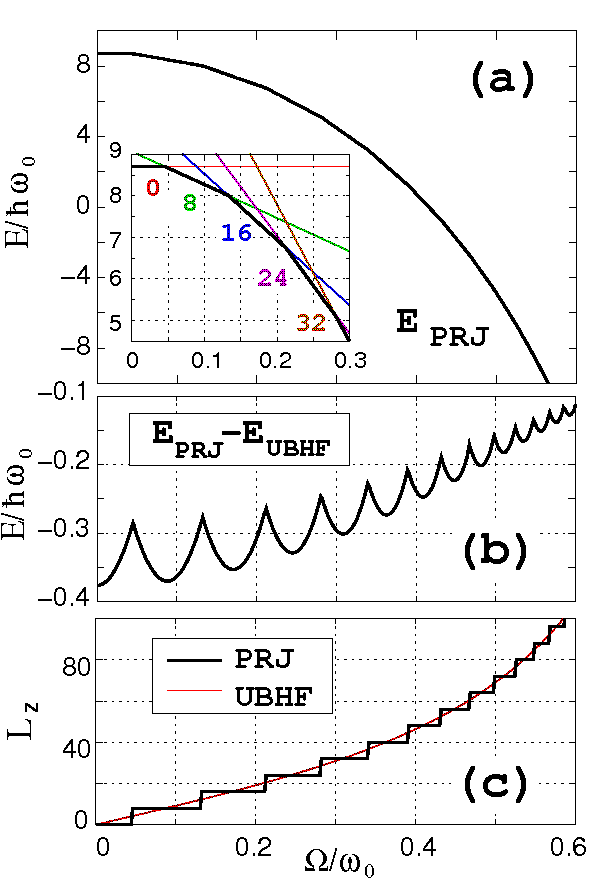}}
\caption{(Color online) 
Properties of $N=8$ neutral repelling bosons in a rotating toroidal 
trap as a function of the reduced rotational frequency $\Omega/\omega_0$.  
The confining potential is given by \eref{pot} with $n=2$ and radius 
$r_0=3 l_0$, and the interaction-strength parameter was chosen as
$R_\delta=50$. 
(a) RBM ground-state energies, $E^{\rm{PRJ}}$.
The inset shows the range $0 \leq \Omega/\omega_0 \leq 0.3$. The numbers denote
ground-state magic angular momenta.
(b) Energy difference $E^{\rm{PRJ}}-E^{\rm{UBHF}}$.
(c) Total angular momenta associated  with (i) the RBM ground states [thick
solid line (showing steps and marked as PRJ); online black] and (ii) 
the UBHF solutions (thin solid line; online red). In the figures, we may use
the symbol $L_z$, instead of simply $L$, to denote the 2D total angular
momentum.
}
\label{n8bostor}
\end{figure}
The Hamiltonian corresponding to the single-particle 
kinetic energy is given by 
\begin{equation}
H_K({\bf r})= ( {\bf p} - \hbar {\bf Q} \times {\bf r})^2/(2m),
\label{hkbosmag}
\end{equation}
for the case of a magnetic field, and by 
\begin{equation}
H_K({\bf r})= ( {\bf p} - \hbar {\bf Q} \times {\bf r})^2/(2m) 
- m \Omega^2 {\bf r}^2/2, 
\label{hkbosrot}
\end{equation}
for the case of a rotating frame of reference.\footnote[9]{
The single-particle wave function in \eref{disg} and 
the many-body projected wave function in \eref{wfprj} contain 
contributions from higher Landau levels. These wave functions belong
exclusively to the lowest Landau level only in the
limit when $\lambda=\sqrt{2} l_B$ in the case of a magnetic field,
or $\lambda=\sqrt{2} l_\Omega$ and $\Omega/\omega_0=1$ in the
case of a rotating trap.}

A toroidal trap with radius $r_0$ can be specified by the confining 
potential
\begin{equation}
V({\bf r})= \frac{\hbar \omega_0}{2} (r-r_0)^n/l_0^n,
\label{pot}
\end{equation}
with $l_0=\sqrt{\hbar/(m\omega_0)}$ being the characteristic
length of the 2D trap. For $n \gg 2$ and 
$l_0/r_0 \to 0$ this potential approaches the limit of a toroidal 
trap with zero width, which has been considered often in previous theoretical 
studies (see, e.g., Ref.\ \cite{kart}). 
In the following, we consider the case with $n=2$, which is more realistic 
from the experimental point of view. In this case, in the  limit 
$r_0=0$, one recovers a harmonic trapping potential.

To construct an RBM variational many-body wave function describing $N$
impenetrable bosons in the toroidal trap, we use $N$ displaced orbitals 
$\varphi({\bf r}, {\bf R}_i)$, $i=1,2,\ldots,N$ [see \eref{disg}] 
centered at the vertices of a regular polygon. Then, we first construct an 
{\it unrestricted\/} Bose Hartree-Fock permanent \cite{roma04,roma06}
$|\Phi^{\rm{UBHF}}_N \rangle 
\propto \sum_{P(i_m)} \varphi_1({\bf r}_{i_1}) \varphi_2({\bf r}_{i_2}) \ldots
\varphi_N({\bf r}_{i_N})$. The UBHF permanent breaks the circular symmetry of 
the many-body Hamiltonian. As was discussed in \sref{repbos},
the ``symmetry dilemma'' is resolved through a 
subsequent ``symmetry-restoration'' step accomplished via projection 
techniques \cite{yl02.1,yl02.2,py,low55,yl04.2,li06},
i.e., we construct a many-body wave function with good total angular
momentum by applying the projection operator   
$\hat{\cal P}_L = (1/2\pi) \int_0^{2\pi}\rmd \theta 
\exp[\rmi \theta (L-\hat{L})]$,
so that the final RBM wave function is given by
\begin{equation}
|\Psi^{\rm{PRJ}}_{N,L} \rangle =
\frac{1}{2 \pi} \int_0^{2\pi} 
\rmd \theta |\Phi^{\rm{UBHF}}_N (\theta) \rangle \rme^{\rmi \theta L}.
\label{wfprj}
\end{equation}
$|\Phi^{\rm{UBHF}}_N (\theta) \rangle$ is the original UBHF permanent rotated 
by an azimuthal angle $\theta$. We note that, in addition to having
good angular momenta, the projected wave function 
$|\Psi^{\rm{PRJ}}_{N,L} \rangle$ has also a {\it lower\/} energy than that 
of $|\Phi^{\rm{UBHF}}_N \rangle$
[see, e.g. $E_L^{\rm{PRJ}} - E^{\rm{UBHF}}$ in \fref{n8bostor}(b).
The projected ground-state energy is given by
\begin{equation}
E_L^{\rm{PRJ}} = 
\int_0^{2\pi} h(\theta) \rme^{\rmi \theta L} \rmd \theta \left / 
\int_0^{2\pi} n(\theta) \rme^{\rmi \theta L} \right . \rmd \theta,
\label{elprj}
\end{equation} 
where $h(\theta)= \langle \Phi_N^{\rm{UBHF}}(\theta=0)  | {\cal H} |
\Phi_N^{\rm{UBHF}}(\theta)\rangle$ and
$n(\theta)= \langle \Phi_N^{\rm{UBHF}}(\theta=0)  |
\Phi_N^{\rm{UBHF}}(\theta)\rangle$; the latter term ensures proper
normalization. 

%
\begin{figure}[t]
\centering{\includegraphics[width=5.5cm]{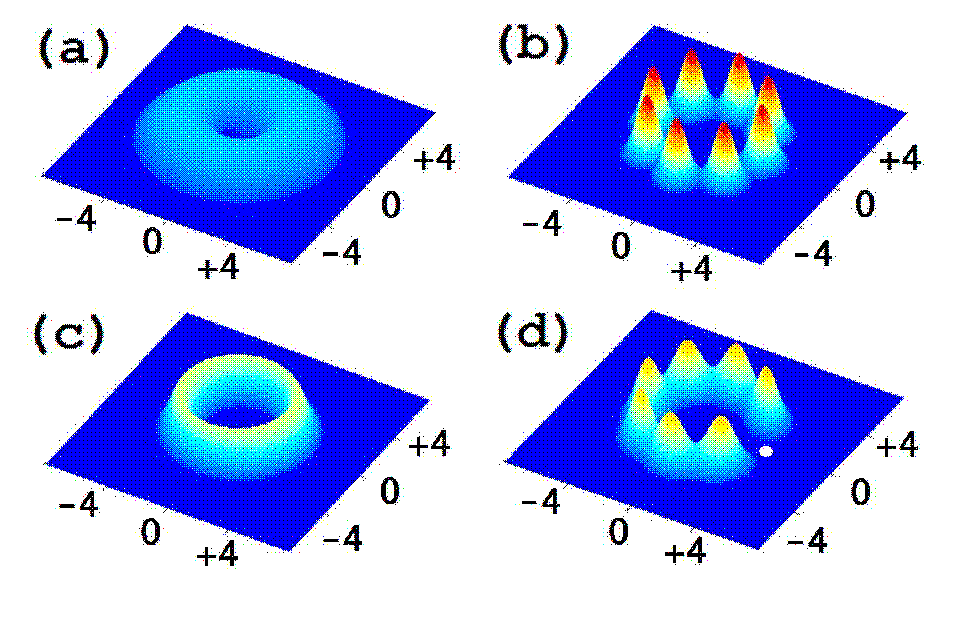}}
\caption{(Color online) 
Single-particle densities and CPDs for $N=8$ bosons in a rotating toroidal trap
with $\Omega/\omega_0=0.2$ and $R_\delta=50$. The remaining trap parameters
are as in \fref{n8bostor}. (a) Gross-Pitaevskii single-particle density. 
(b) UBHF single-particle density exhibiting breaking of the circular symmetry.
(c) RBM single-particle density exhibiting circular
symmetry. (d) CPD for the RBM wave function [PRJ wave function, see 
\Eref{wfprj}] revealing the hidden point-group symmetry in the
intrinsic frame of reference. The observation point is denoted by a white dot.
The RBM ground-state angular momentum is $L_z=16$. Lengths in units of
$l_0$. The vertical scale is the same for (b), (c), and (d), but
different for (a).
}
\label{n8bostorcpd}
\end{figure}
%

The many-body Hamiltonian in the rotating trap is given by 
\begin{equation}
{\cal H}=\sum_{i=1}^N [H_K({\bf r}_i)+V({\bf r}_i)] + \sum_{i<j}^N 
v({\bf r}_i,{\bf r}_j),
\label{hmbrot} 
\end{equation}
with the interparticle interaction being given by a contact potential 
$v_\delta({\bf r}_{i},{\bf r}_{j})=g \delta ({\bf r}_i - {\bf r}_j)$
for neutral bosons and a Coulomb potential 
$v_C({\bf r}_{i},{\bf r}_{j})=Z^2e^2/|{\bf r}_i - {\bf r}_j|$
for charged bosons. The parameter that controls the strength of the
interparticle repulsion relative to the zero-point kinetic energy is given
by $R_\delta=gm/(2\pi\hbar^2)$ \cite{roma04,roma06} for a contact potential 
and $R_W=Z^2e^2/(\hbar \omega_0 l_0)$ \cite{yl99,roma04} for a Coulomb
repulsion.

For a given value of the dimensionless rotational frequency,
$\Omega/\omega_0$, the projection yields wave functions and energies
for a whole rotational band comprising many angular momenta. In the following, 
we focus on the ground-state wave function (and corresponding angular momentum 
and energy) associated with the lowest energy in the band.  

%
\begin{figure}[t]
\centering{\includegraphics[width=5.50cm]{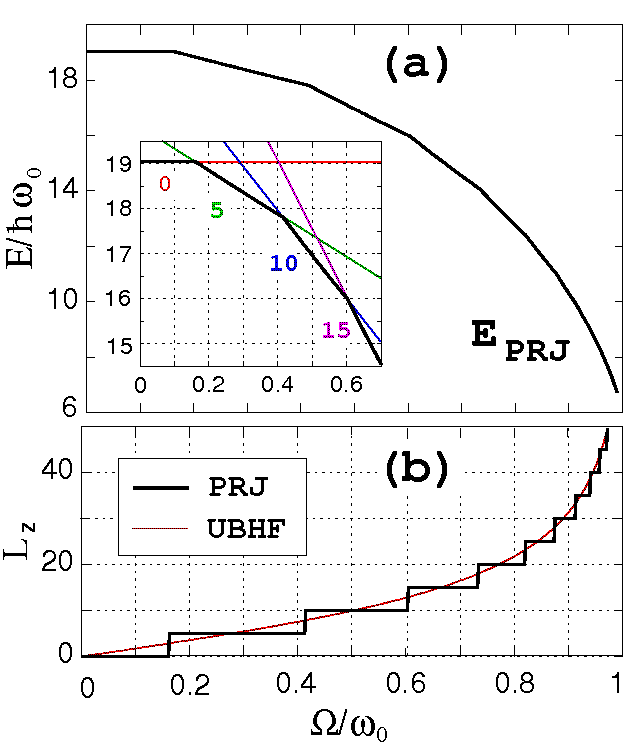}}
\caption{(Color online) 
Properties of $N=6$ neutral bosons in a rotating harmonic 
trap as a function of the reduced rotational frequency $\Omega/\omega_0$.
The confining potential is given by \eref{pot} with $n=2$ and 
$r_0=0$, and the interaction-strength parameter was chosen as $R_\delta=50$.
The intrinsic molecular structure is $(1,5)$.
(a) RBM ground-state energies, $E^{\rm{PRJ}}$. The inset shows a smaller
range. The numbers denote ground-state angular momenta.
(b) Total angular momenta associated  with (i) the RBM ground states (thick
solid line showing steps; online black) and (ii) the UBHF solutions 
(thin solid line; online red).
}
\label{n6boshar}
\end{figure}

\Fref{n8bostor}(a) displays the ground-state energy 
$E_{\rm{PRJ}}$ of $N=8$ bosons
in a toroidal trap as a function of
the dimensionless rotational frequency $\Omega/\omega_0$, with $\omega_0$
being the trap frequency. The prominent features in \fref{n8bostor}(a) 
are: (i) the 
energy diminishes as $\Omega/\omega_0$ increases; this is an effect of the
centrifugal force, and (ii) the $E_{\rm{PRJ}}$ curve consists of linear 
segments, each one associated with a given angular momentum $L$. 
Most remarkable is the regular variation of the values of $L$ with a constant 
step of $N$ units (here $N=8$) [see inset in \fref{n8bostor}(a) 
and \fref{n8bostor}(c)]. These preferred
angular momenta $L=kN$ with integer $k$, are reminiscent of the 
so called ``magic angular momenta'' familiar from studies of electrons under
high-magnetic fields in 2D semiconductor quantum dots 
\cite{yl03.2,yl04.2,li06}.

%
\begin{figure}[t]
\centering{\includegraphics[width=7.50cm]{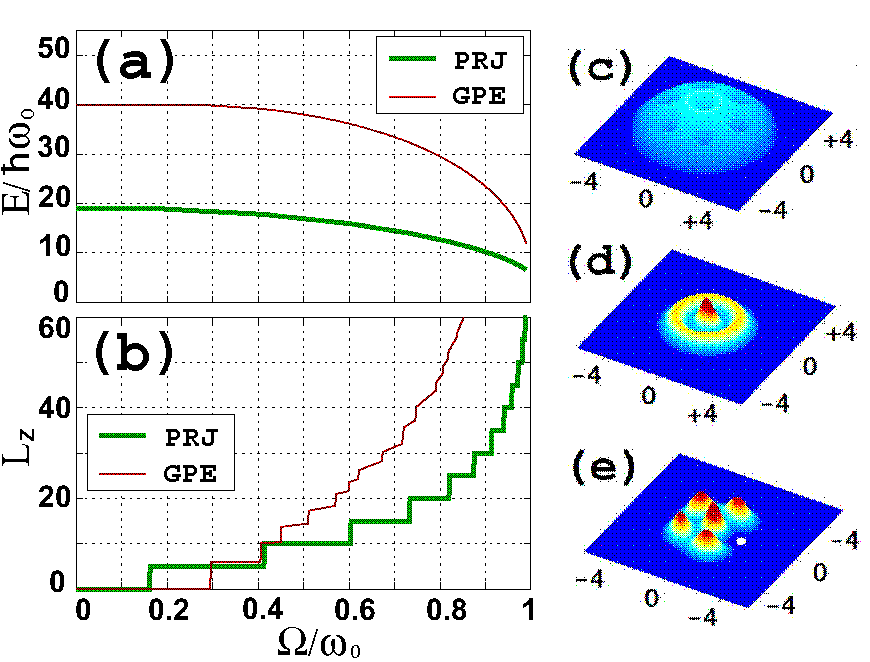}}
\caption{(Color online) 
Properties of GP solutions (thin solid line; online red) versus those of RBM 
wave functions (thick solid line; online green) for $N=6$ neutral bosons 
as a function of the reduced rotational frequency $\Omega/\omega_0$. 
A harmonic 
trap is considered, and the interaction strength equals $R_\delta=50$.
(a) Ground-state energies. (b) Associated ground-state angular momenta. (c) 
GP (BEC) single-particle density at $\Omega/\omega_0=0.65$ having 7 vortices
with a 6-fold symmetry (thus exhibiting breaking of the circular symmetry). 
(d) RBM single-particle density at $\Omega/\omega_0=0.65$ which does not break
the circular symmetry. (e) CPD of the RBM at $\Omega/\omega_0=0.65$ revealing 
the intrinsic (1,5) crystalline pattern. The white dot denotes the observation
point ${\bf r}_0$. Note the dramatic difference in spatial extent between
the GP and RBM wave functions [compare (c) with (d) and (e). Lengths in units 
of $l_0$. The vertical scale is the same for (d) and (e), 
but different for (c).
}
\label{n6bosharcpd}
\end{figure}
The preferred angular momenta reflect the intrinsic molecular structure
of the localized impenetrable bosons. We note, that the (0,8) polygonal-ring 
arrangement is obvious in the single-particle density associated with
the UBHF permanent [see \fref{n8bostorcpd}(b)]; 
(0,8) denotes no particles in the inner
ring and 8 particles in the outer one. After restoration of symmetry, however,
the single-particle density is circularly symmetric [see the PRJ 
single-particle density in \fref{n8bostorcpd}(c)] and the intrinsic 
crystallinity becomes ``hidden''; it can, however, be revealed via the 
conditional probability distribution \cite{yl99,yl04.2,li06,roma04} 
[CPD, see \fref{n8bostorcpd}(d)]. We note the Gross-Pitaevskii single-particle
density in \fref{n8bostorcpd}(a), which is clearly different from the PRJ
density in \fref{n8bostorcpd}(c). 

The internal structure for {\it charged\/} bosons in a toroidal trap (not
shown) is similar to that of neutral bosons (\fref{n8bostorcpd}), 
i.e., a (0,8) ring 
arrangement, portrayed also in the stepwise variation (in steps of 8 units)
of the total angular momenta. The internal structure is also
reflected in the variation of the ground-state total energy as a function of
the magnetic field. In contrast to the case of neutral bosons,
however, the ground-state energy curve for charged bosons is not composed of 
linear segments, but of intersecting inverted-parabola-type pieces; this is due 
to the positive contribution of the Lorentz force compared to the negative 
contribution of the centrifugal force in a rotating trap.

For RBMs in rotating {\it harmonic\/} traps,
the polygonal-ring pattern of localized bosons becomes more
complex than the simple $(0,N)$ arrangement that appears naturally in
a toroidal trap. Indeed, in harmonic traps, one anticipates the emergence of
concentric ring structures. For $N=6$ neutral bosons in a harmonic trap,
we observe that, as in the case of a toroidal trap, the ground-state energy as
a function of the reduced rotational frequency, $\Omega/\omega_0$, 
[\fref{n6boshar}(a)] is composed of linear segments, but now the corresponding
magic angular momenta [\fref{n6boshar}(b)] vary in steps of $N-1=5$ units. 
This indicates a rotating boson molecule 
consisting of {\it two\/} polygonal rings; denoted as a $(1,5)$ structure, 
with the inner ring having a single boson and the outer ring five.

%
\begin{figure}[t]
\centering{\includegraphics[width=5.50cm]{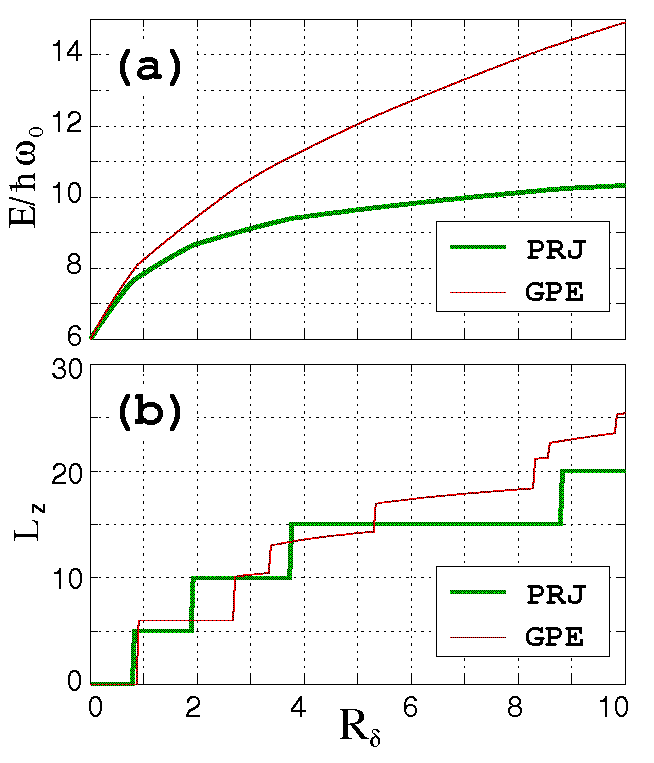}}
\caption{(Color online) 
Properties of GP solutions (thin solid line; online red) versus those of RBM
wave functions (thick solid line; online green) for $N=6$ bosons as a function
of the interaction strength $R_\delta$. A harmonic trap is considered, and 
the reduced rotational frequency equals $\Omega/\omega_0=0.85$.
(a) Ground-state energies (b) Associated ground-state angular momenta.
}
\label{n6bosrd}
\end{figure}

In \fref{n6bosharcpd}(a), we display the rotating-boson-molecule
and mean-field Gross-Pitaevskii ground-state
energies of $N=6$ strongly repelling (i.e., $R_\delta=50$) neutral bosons in a 
harmonic trap as a function of the reduced angular frequency of the trap.
The GP curve (thin solid line; online red) remains well above the RBM curve 
(thick solid line; online green) in the whole range 
$ 0 \leq \Omega/\omega_0 \leq 1$. The RBM ground-state angular momenta exhibit 
again the periodicity in steps of five units [\fref{n6bosharcpd}(b)].
As expected, the GP total angular momenta
are quantized [$L_z=0$ (no-vortex) or $L_z=6$ (one central vortex)] only for
an initial range $0 \leq \Omega/\omega_0 \leq 0.42$. For 
$\Omega/\omega_0 \geq 0.42$, the GP total angular momentum takes non-integer
values and ceases to be a good quantum number, reflecting the broken-symmetry
character of the associated mean field, with each kink signaling the
appearance of a different vortex pattern of $p$-fold symmetry ($p=1,2,3,4,\ldots$)
\cite{butt}; see an example in \fref{n6bosharcpd}(c).
The energetic superiority of the RBM wave function over the GP solution
demonstrated in \fref{n6bosharcpd}(a) was to be expected, 
since we considered the case of strongly repelling bosons. Unexpectedly, 
however, for a small number of neutral bosons the energetic advantage of the 
rotating boson molecule persists even for weakly 
repelling bosons, as illustrated in \fref{n6bosrd}(a). 
Indeed, \fref{n6bosrd}(a) displays 
the RBM (thick solid line; online green) and  GP (thin solid line; online red) 
ground-state energies for $N=6$ neutral bosons in a trap rotating with 
$\Omega/\omega_0=0.85$ as a function of the interaction parameter $R_\delta$. 
The surprising result in \fref{n6bosrd}(a) is that the GP energy remains 
above the RBM curve even for $R_\delta \rightarrow 0$. Of course the RBM wave 
function is very close to that of a BEC without vortices when 
$R_\delta \rightarrow 0$ 
(BECs {\it without\/}  vortices are approximately feasible for small $N$). 
However, for small $N$, our results show that BECs {\it with vortices\/} 
(i.e., for $L_z \geq N$) are not the preferred many-body ground states; instead, 
formation of RBMs is favored.  
Note that the energy difference $E^{\rm{GP}}-E^{\rm{PRJ}}$ increases rapidly
with increasing $R_\delta$, reflecting the fact that the RBM energies saturate 
(as is to be expected from general arguments), while the GP energies (even with
vortices fully accounted for) exhibit an unphysical divergence as 
$R_\delta \rightarrow \infty$ [\fref{n6bosrd}(a)]; 
we have checked this trend up to values of $R_\delta=100$ (not shown). 
Of interest again is the different behavior of the RBM and GP ground state
angular momenta [\fref{n6bosrd}(b)] (see also discussion 
of \fref{n6bosharcpd}(b)).

To summarize this section: We have studied the ground-state properties of 
a variational many-body wave function for repelling bosons in rotating traps 
that incorporates correlations beyond the Gross-Pitaevskii mean-field 
approximations. This variational wave function describes rotating boson 
molecules, i.e., localized bosons arranged in polygonal-ring-type patterns in 
their intrinsic frame of reference. 
For small numbers of neutral bosons, and in 
particular in the case of GP vortex formation, 
the RBM ground-state energies are
lower than those associated with the corresponding Gross-Pitaevskii BEC
solutions. Given the large differences between the properties of the RBM and
BEC wave functions (which become more pronounced for larger interaction 
parameter $R_\delta$), and the recently demonstrated ability to experimentally
control $R_\delta$ \cite{wei,par,grei,cor}, we anticipate that our results 
could be tested in experiments involving rotating optical lattices. Detection 
of rotating boson molecules could be based on a variety of approaches, 
such as the measurement of the spatial extent [contrast the RBM and BEC 
spatial extents in \fref{n6bosharcpd}(c) - \fref{n6bosharcpd}(e)], 
or the use of Hanbury Brown-Twiss-type 
experiments \cite{brtw} to directly detect the intrinsic crystalline structure 
of the RBM.

\subsection{Exact diagonalization for bosons in the lowest Landau level}
\label{exdbos}

Rotating ultracold trapped Bose condensed systems are most commonly discussed
in the context of formation of vortex lattices, which are solutions to the
Gross-Pitaevskii mean-field equation 
\cite{dalf99,legg01,ho01,fett01,peth02,pita03,peth04,baym05}. 
Such vortex lattices have indeed been found experimentally for systems 
containing a large number of bosons \cite{dali00,kett01,corn02}.
Nevertheless, several theoretical investigations 
\cite{wilk00,vief00,wilk01,regn03,regn05} of {\it rapidly\/} rotating trapped 
bosonic systems suggested formation of strongly
correlated exotic states which differ drastically from the aforementioned 
vortex-lattice states. While experimental realizations of such strongly 
correlated states have not been reported yet, there is already a significant 
effort associated with two-dimenional exact-diagonalization studies 
of a small number of particles ($N$) in the lowest Landau level; the LLL 
restriction corresponds to the regime of rapid rotation, where the rotational
frequency of the trap $\Omega$ equals the frequency of the 
confining potential. The large majority \cite{wilk00,wilk01,regn03,regn05} 
of such exact-diagonalization studies have 
attempted to establish a close connection between 
rapidly rotating bosonic gases and the physics of electrons under 
fractional-quantum-Hall-effect conditions employing the bosonic version of 
``quantum-liquid'' analytic wave functions, such as the Laughlin wave 
functions, composite-fermion, Moore-Read, and Pfaffian functions.

As described in \sref{rotellll}, the ``quantum-liquid'' picture for a small 
number of trapped electrons in the FQHE regime has been challenged in a series
of extensive studies \cite{yl02.2,yl03.2,yl06.tnt,yl04.1,yl04.2,li06} 
of electrons in 2D quantum dots under high magnetic fields. 
Such studies (both exact-diagonalization and variational) revealed 
that, at least for finite systems, the underlying physical picture governing
the behavior of strongly-correlated electrons is not that of a ``quantum 
liquid.'' Instead, the appropriate description is in terms of a ``quantum 
crystal,'' with the localized electrons arranged in polygonal concentric rings
\cite{yl02.2,yl03.2,yl04.1,yl04.2,li06,rua95,maks96,sek96}. 
These ``crystalline'' states lack \cite{yl04.2,li06} the familiar rigidity of
a classical extended crystal, and are better described
\cite{yl02.2,yl03.2,yl06.tnt,yl04.1,yl04.2,li06} as rotating electron 
(or Wigner) molecules.   

Motivated by the discovery in the case of electrons of REMs at high $B$ (and 
from the fact that Wigner molecules form also at zero magnetic field
\cite{yl99,yl03.1,elle06,yl00.2,mik02.3,egger99})  
some theoretical studies have most recently shown that analogous 
molecular patterns of localized bosons do form in the case of a small number 
of particles inside a static or rotating harmonic trap
\cite{yl06.nac,roma04,roma06,barb06,mann06}. 
In analogy with the electron case, the bosonic 
molecular structures can be referred to \cite{roma06} as {\it rotating boson 
molecules\/}; a description of RBMs via a variational wave function
built from symmetry-breaking displaced Gaussian orbitals with 
subsequent restoration of the rotational symmetry was presented in Refs.\ 
\cite{yl06.nac,roma04,roma06} and reviewed in \sref{varrbm}.

In a recent paper, Baksmaty \etal \cite{baks07} used exact diagonalization in 
the lowest Landau level to investigate the 
formation and properties of RBMs focusing on a larger number of particles than
previously studied, in particular for sizes where multiple-ring formation 
can be expected based on our knowledge of the case of 2D electrons in high $B$.
A finite number of particles ($N \le 11$) at {\it both\/} low 
($ \nu < 1/2$) and high ($ \nu \geq 1/2$) filling fractions
$\nu \equiv N(N-1)/2L$ (where $L \equiv {\cal L}/\hbar$ is the 
quantum number associated with the total angular momentum ${\cal L}$) 
was studied and both the cases of a long-range (Coulomb) and a short-range 
($\delta$-function) repulsive interaction were investigated.
In this section, we report some main results from Ref.\ \cite{baks07}.

As in the case of electrons in 2D quantum dots, we
probe the crystalline nature of the bosonic ground states by calculating the
full anisotropic two-point correlation function $P({\bf r},{\bf r}_{0})$ 
[see \Eref{cpds}] associated with the exact wavefunction 
$\Psi({\bf r}_1,{\bf r}_2,\ldots,{\bf r}_N)$.
The quantity $P({\bf r},{\bf r}_{0})$ is proportional to the probability of 
finding a boson at ${\bf r}$ given that there is another boson at the 
observation point ${\bf r}_{0}$, and it is often referred to as the 
conditional probability distribution (\sref{synopsis}).
A main finding of our studies is that consideration solely of the CPDs is not
sufficient for the boson case at high fractional fillings $\nu \geq 1/2$;
in this case, one needs to calculate even higher-order correlation functions,
e.g., the full $N$-point correlation function defined as the modulus square of
the full many-body EXD wave function, i.e., 
\begin{equation}
P({\bf r};{\bf r}_1,{\bf r}_2,\ldots,{\bf r}_{N-1})=
| \Psi({\bf r};{\bf r}_1,{\bf r}_2,\ldots,{\bf r}_{N-1})|^2,
\label{npoint_def}
\end{equation}
where one fixes the positions of $N-1$ particles and inquiries about the 
(conditional) probability of finding the $N$th particle at any position 
${\bf r}$.

The investigations in this section are also motivated by recent experimental 
developments, e.g., the realization of trapped ultracold gas assemblies 
featuring bosons interacting via a long-range dipole-dipole interaction 
\cite{pfau05,lewe02}. We expect the results presented in this section to 
be directly relevant to systems with a two-body repulsion intermediate between
the Coulomb and the delta potentials.
Additionally, we note the appearance of promising experimental techniques for 
measuring higher-order correlations in ultra-cold gases employing an atomic 
Hanbury Brown-Twiss scheme \cite{brtw} or shot-noise 
interferometry \cite{altm04,bloc05}.
Experimental realization of few-boson rotating systems can be anticipated in
the near future as a result of increasing sophistication of experiments 
involving periodic optical lattices co-rotating with 
the gas, which are capable of holding a few atoms in each site. 
A natural first step in the study of such systems is the analysis of the 
physical properties of a few particles confined in a rotating trap with open 
boundary conditions (i.e., conservation of the total angular momentum $L$).

\begin{figure}[t]
\centering{\includegraphics[width=6.8cm]{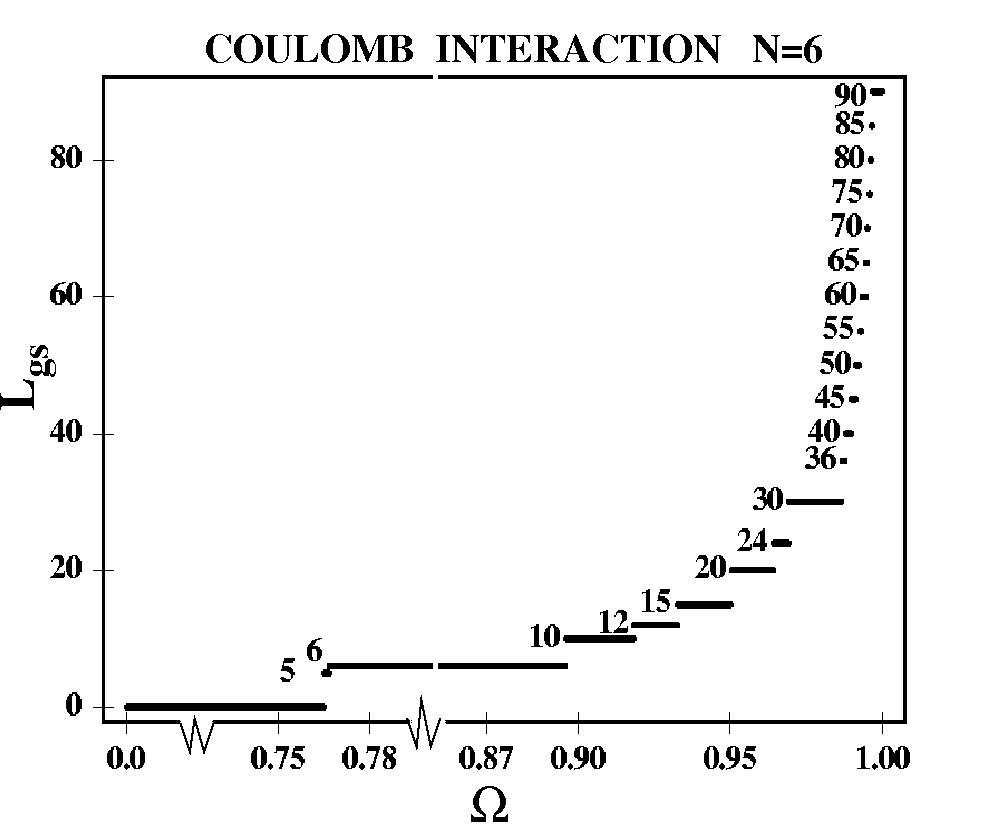}
\includegraphics[width=6.0cm]{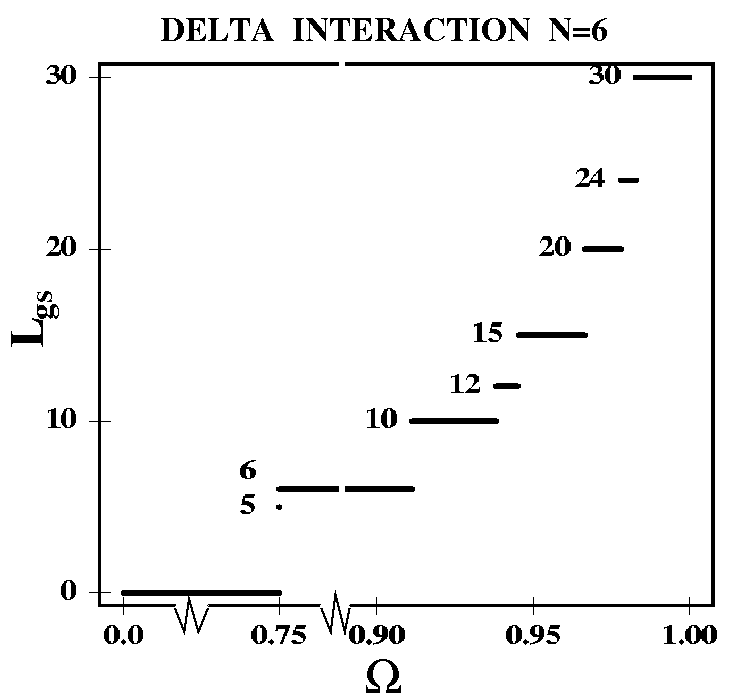}}
\caption{Ground-state angular momenta, $L_{\rm{gs}}$, for $N=6$ bosons in a 
rapidly rotating trap [described by the LLL Hamiltonian in  
\eref{H_lll}], as a function of the rotational frequency $\Omega$ expressed 
in units of $\omega_0$. The bosons interact via a Coulombic repulsion 
(left) and a delta repulsion (right), and the many-body Hilbert space is 
restricted to the lowest Landau level. 
The angular momentum associated with the first bosonic 
Laughlin state occurs at $L=30$, i.e., at $N(N-1)$. The value of 
$c=0.2 \hbar \omega_0 \Lambda$ for the Coulomb case (left) and 
the value of $g=2\pi \hbar \omega_0 \Lambda^2/N$ for the case of
a delta repulsion (right); the many-body wave functions do not depend on these 
choices.  In the delta-interaction case, the values of the angular momenta 
terminate with the value $L=30$ (the Laughlin value) at 
$\Omega/\omega_0=1$. In contrast, in the Coulomb-interaction case (left), 
the values of the ground-state angular momenta do not
terminate, but diverge as $\Omega/\omega_0 \rightarrow 1$. Note the
stepwise variation of the values of the ground-state angular momenta in both
cases, indicating the presence of an {\it intrinsic\/} point-group symmetry
associated with the (0,6) and (1,5) polygonal-ring structure of a rotating
Boson molecule.}
\label{stepbos}
\end{figure}

The main results of Ref.\ \cite{baks07} can be summarized as follows: Similar 
to the well-established (see \sref{rotellll} and \sref{rotelfinb}) emergence
of rotating electron molecules in quantum dots, rotating boson molecules 
form in rotating harmonic traps as well. The RBMs are also organized in 
concentric polygonal rings that rotate independently of each other, and the 
polygonal rings correspond to classical equilibrium configurations and/or 
their low-energy isomers. Furthermore, the degree of crystallinity increases 
gradually with larger angular momenta $L$'s (smaller filling fractions 
$\nu$'s), as was the trend \cite{yl03.2,yl04.2,li06} for the REMs and as was 
observed also for $\nu < 1/2$ in another study \cite{mann06} for rotating 
bosons in the lowest Landau level with smaller $N$ 
and single-ring structures. We finally note
that the crystalline character of the RBMs appears to depend only weakly on 
the range of the repelling interaction, for both the low (see also Ref.\ 
\cite{mann06}) and high (unlike Ref.\ \cite{barb06}) fractional fillings.

In studies of 2D quantum dots, CPDs were used some time ago in Refs.\ 
\cite{yl00.2,maks96,sek96}. For probing the intrinsic molecular structure 
in the case of ultracold bosons in 2D traps, however, they were introduced only
recently by Romanovsky \etal \cite{roma04}. The importance of using CPDs as a 
probe can hardly be underestimated. Indeed, while exact-diagonalization
calculations for {\it bosons\/} in the lowest Landau level have been reported 
earlier \cite{wilk00,vief00,wilk01,regn03,regn05}, 
the analysis in these studies did 
not include calculations of the CPDs, and consequently formation of 
rotating boson molecules was not recognized.

\begin{figure*}
\centering\includegraphics*[width=15cm]{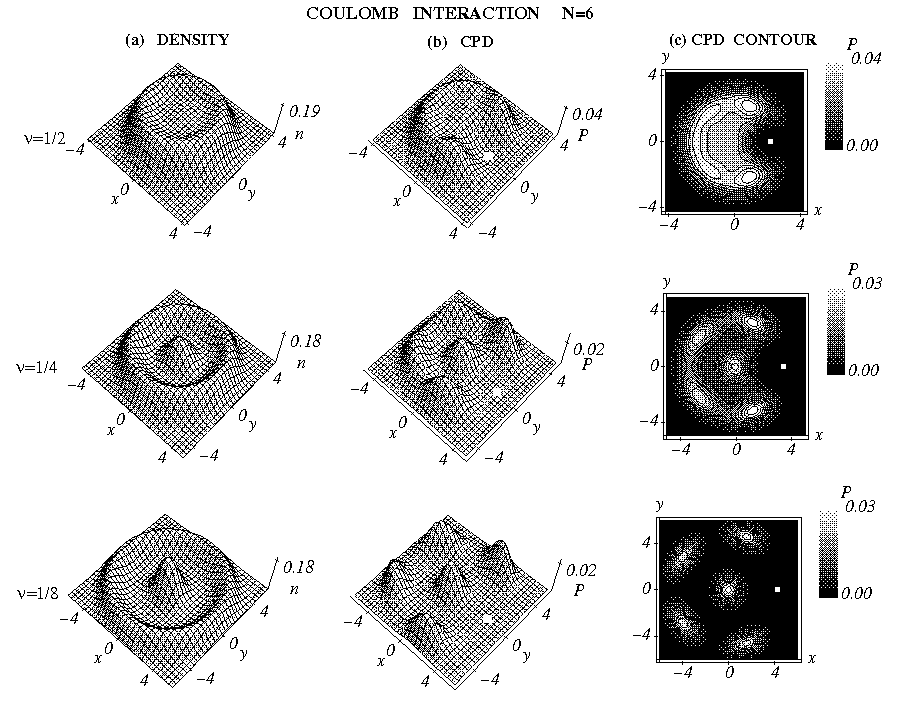}
\caption{(a) Single-particle densities [$n({\bf r})$; left column], 
(b) CPDs $\left[ P({\bf r},{\bf r}_{0}) \right]$ in 3D plots (middle column),
and (c) CPDs in contour plots (right column), portraying the strengthening 
of the crystalline RBM 
structure for $N=6$ bosons interacting via a repulsive Coulomb interaction as 
the filling fraction $\nu$ is reduced. The white dots in the CPD 
plots indicate the reference point 
${\bf r}_0$. We note in particular the gradual enhancement of the peak at the 
center of the plots, and the growth of the radius of the outer ring; the latter
reflects the nonrigid-rotor nature of the RBMs (in analogy with the findings
of Ref.\ \cite{yl04.2} regarding the properties of rotating electron
molecules). The cases of $\nu=1/4$ and $\nu=1/8$ exhibit a clear $(1,5)$
crystalline arrangement, while the case of $\nu=1/2$ (first Laughlin state) is 
intermediate between a $(1,5)$ and a $(0,6)$ pattern (see text for details).
Lengths in units of $\Lambda$. The vertical scales are in arbitrary 
units, which however do not change for the panels within the same 
column (a), (b), or (c).}
\label{6_coulomb_crystallization}
\end{figure*}

\subsubsection{The case of $N=6$ bosons in the lowest Landau level.}

As a specific example of the points discussed above in \sref{exdbos}, 
we present here results for $N=6$ bosons in the lowest Landau level.
For additional cases (e.g., $N=9$ and $N=11$), see Ref.\ \cite{baks07}. 

In analogy with the magnetic-field Hamiltonian of \Eref{hlll2}, the many-body
Hamiltonian for $N$ bosons in a rotating trap is reduced in the lowest Landau
level to the expression
\begin{equation}
\tilde{\cal H}^\Omega_{\rm{LLL}}=
N \hbar \omega_0 + \hbar(\omega_0-\Omega) L
+\sum_{i < j}^{N} v({\bf r}_{i},{\bf r}_{j}),
\label{H_lll}
\end{equation}
where $\omega_0$ specifies the 2D harmonic trap and $\Omega$ denotes the 
rotational frequency. The interparticle interaction is given
by a contact potential $v_\delta({\bf r}_{i},{\bf r}_{j})=
g \delta ({\bf r}_i - {\bf r}_j)$ for neutral bosons and a Coulomb potential 
$v_C({\bf r}_{i},{\bf r}_{j})=c/|{\bf r}_i - {\bf r}_j|$
for charged bosons.

Since $\tilde{\cal H}^\Omega_{\rm{LLL}}$ is rotationally invariant, i.e., 
$[\tilde{\cal H}^\Omega_{\rm{LLL}},L]=0$, its eigenstates $\Psi_L$ must also 
be eigenstates of the total angular momentum with eigenvalue 
$\hbar L$. For a given rotational frequency $\Omega$, the eigenstate with 
lowest energy is the ground state; we denote the corresponding angular momentum
as $L_{\rm{gs}}$.
 
We proceed to describe the EXD results for $N=6$ particles interacting via a
Coulomb repulsion by referring to \fref{stepbos}, where we plot 
$L_{\rm{gs}}$ against the angular frequency $\Omega$ of the rotating trap.
A main result from all our calculations is that 
$L_{\rm{gs}}$ increases in characteristic (larger than unity) steps that take 
only a few integer values, i.e., for $N=6$ the variations of $L_{\rm{gs}}$ are
in steps of 5 or 6. In keeping with previous work on electrons 
\cite{yl02.2,yl03.2,yl06.tnt,yl04.1,yl04.2,li06} at high $B$, and very 
recently on bosons in rotating traps 
\cite{yl06.nac,roma04,roma06,barb06,mann06}, 
we explain these {\it magic-angular-momenta\/}  patterns (i.e., for $N=6$, 
$L_{\rm{gs}} = L_0+5k$ or $L_{\rm{gs}} = L_0+6k$, with $L_0=0$) as 
manifestation of an {\it intrinsic\/} point-group symmetry associated with the
many-body wave function. This point-group symmetry emerges from the formation 
of RBMs, i.e., from the localization of the bosons at the vertices of 
concentric regular polygonal rings; it dictates that the angular momentum of a
purely rotational state can only take values $L_{\rm{gs}}=L_0+\sum_i k_i n_i$, 
where $n_i$ is the number of localized particles on the $i$th polygonal ring. 
[We remind the reader that for spin-polarized electrons in the 
lowest Landau level, the 
corresponding value is $L_0=N(N-1)/2$.] Thus for $N=6$ bosons, the series 
$L_{\rm{gs}}=5k$ is associated with an $(1,5)$ polygonal ring structure, while
the series $L_{\rm{gs}}=6k$ relates to an $(0,6)$ arrangement of particles. 
It is interesting to note that in classical calculations \cite{kong02}
for $N=6$ particles in a harmonic 2D trap, the $(1,5)$ arrangement is found to
be the global energy minimum, while the $(0,6)$ structure is the lowest
metastable isomer. This fact is apparently reflected in the smaller
weight of the $L_{\rm{gs}}=6k$ series compared to the $L_{\rm{gs}}=5k$
series, and the gradual disappearance of the former with increasing $L$.   

Magic values dominate also the ground state angular momenta of neutral bosons 
(delta repulsion) in rotating traps, as shown for $N=6$ bosons in the
right panel of \fref{stepbos}. Although the corresponding $\Omega$-ranges 
along the horizontal axis may be different compared to the Coulomb case, 
the appearance of only the two series $5k$ and $6k$ is remarkable --- pointing 
to the formation of RBMs with similar $(1,5)$ and $(0,6)$ structures in the 
case of a delta interaction as well (see also Refs.\ \cite{barb06,mann06}). 
An important difference, however, is that for the delta interaction both 
series end at $\Omega/\omega_0=1$ with the value $L=N(N-1)=30$ (for $N=6$ the 
bosonic Laughlin value at $\nu=1/2$), while for the Coulomb interaction this 
$L$ value is reached for $\Omega/\omega_0<1$ --- allowing for an infinite set 
of magic angular momenta [larger than $N(N-1)$] to develop as $\Omega/\omega_0 
\rightarrow 1$.  

Beyond the analysis of the ground-state spectra as a function of $\Omega$,
the intrinsic crystalline point-group structure can be revealed by an
inspection of the CPDs [and to a much lesser extent by an inspection of
single-particle densities]. 
Because the EXD many-body wave function is an eigenstate of the
total angular momentum, the single-particle densities are circularly symmetric
and can only reveal the presence of concentric rings through oscillations
in the radial direction. The localization of bosons within the same ring
can only be revealed via the azimuthal variations of the anisotropic CPD 
[\Eref{cpds}]. One of our findings is that for a given $N$ the crystalline 
features in the CPDs develop slowly as $L$ increases (or $\nu$ decreases). 

For $\nu < 1/2$, we find that the crystalline features are well developed for
all sizes studied by us. In \fref{6_coulomb_crystallization},
we present some concrete examples of CPDs from exact-diagonalization 
calculations associated with the ground-states of 
$N=6$ bosons in a rotating trap interacting via a repulsive Coulomb potential. 
In particular, we present the CPDs for
$L_{\rm{gs}}= 30$ (bosonic Laughlin for $\nu=1/2$), 60, and 120; these angular
momenta are associated with ground states at specific $\Omega$-ranges 
[see \fref{stepbos}]. All three of these angular momenta are divisible by 
both 5 and 6. However, only the $L_{\rm{gs}}=30$ CPD 
(\fref{6_coulomb_crystallization} top row) has a structure that is
intermediate between the $(1,5)$ and the $(0,6)$ polygonal-ring arrangements.
The two other CPDs, associated with the higher $L_{\rm{gs}}=60$ and
$L_{\rm{gs}}=120$ exhibit clearly only the $(1,5)$ structure, illustrating our
statement above that the quantum-mechanical CPDs conform to the structure of
the most stable arrangement [i.e., the $(1,5)$ for $N=6$] of classical 
point-like charges as the fractional filling decreases. 

\begin{figure*}[t]
\centering\includegraphics*[width=14.0cm]{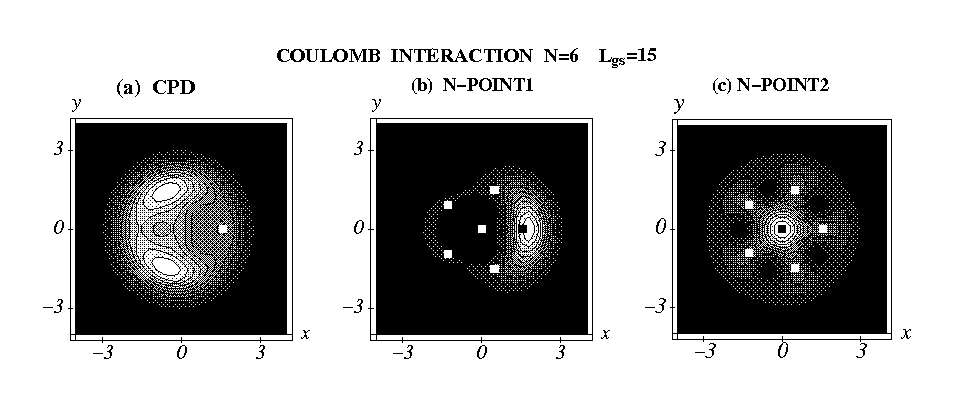}
\caption{Contour plots of the CPD (a) and $N$-point correlation function (b) 
and (c) for $N=6$ bosons with $L_{\rm{gs}}=15$ interacting via a Coulomb 
repulsion. The white squares indicate the positions of the fixed particles. 
The black square in (b) and (c) indicates the position of the 6th particle 
according to the classical $(1,5)$ molecular configuration. Note the different
arrangements of the five fixed particles, i.e., (b) one fixed
particle at the center and (c) no fixed particle at the center. Note also that 
the CPD in (a) fails to reveal the $(1,5)$ pattern, which, however, is clearly 
seen in the $N$-point correlation functions in both (b) and (c).
Lengths in units of $\Lambda$. The vertical scales are arbitrary, but
the same in (b) and (c).}
\label{6_15_npoint_clmb}
\end{figure*}

\begin{figure*}[t]
\centering\includegraphics*[width=14.0cm]{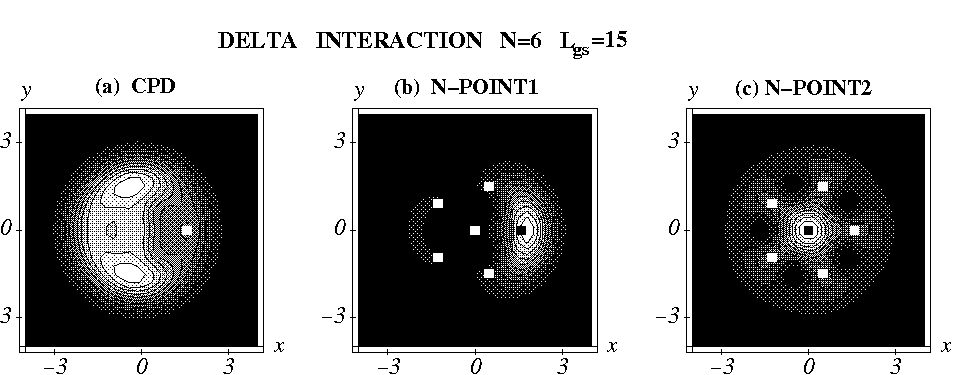}
\caption{Contour plots of the CPD (a) and $N$-point correlation function (b) 
and (c) for $N=6$ bosons with $L_{\rm{gs}}=15$ interacting via a 
$\delta$-repulsion. The white squares indicate the positions of the fixed 
particles. The black square in (b) and (c) indicates the position of the 6th 
particle according to the classical $(1,5)$ molecular configuration. Note the 
different arrangements of the five fixed particles, i.e., (b) one fixed
particle at the center and (c) no fixed particle at the center. Note also that 
the CPD in (a) fails to reveal the $(1,5)$ pattern, which, however, is clearly 
seen in the $N$-point correlation functions in both (b) and (c).
Lengths in units of $\Lambda$. The vertical scales are arbitrary, but
the same in (b) and (c).}
\label{6_15_npoint_delta}
\end{figure*}

However, for $\nu > 1/2$, the azimuthal variations may
not be visible in the CPDs, in spite of the characteristic step-like 
ground-state spectra [see \fref{stepbos} for $N=6$ bosons]. This paradox is 
resolved when one considers higher-order correlations, and in particular 
$N$-point correlations [see \Eref{npoint_def}]. 
In \fref{6_15_npoint_clmb} and  
\fref{6_15_npoint_delta}, we plot the $N$-point correlation functions for
$N=6$ bosons and $L_{\rm{gs}}=15$ for both the Coulomb interaction and 
$\delta$-repulsion, respectively, and we compare them against the 
corresponding CPDs. The value
of 15 is divisible by 5, and one expects this state to be associated with
a $(1,5)$ molecular configuration. It is apparent that the CPDs
fail to portray such fivefold azimuthal pattern. The $(1,5)$ pattern, however,
is clear in the $N$-point correlations (middle and right panels). One has two 
choices for choosing the positions of the first five particles (white dots), 
i.e., one choice places one white dot at the center and the other choice 
places all five white dots on the vertices of a regular pentagon. For both 
choices, as shown by the contour lines in the figures, the position of maximum
probability for the sixth boson coincides with the point that completes the 
$(1,5)$ configuration [see the black dots in the middle and right panels].

Note that the differences in the CPDs and $N$-point 
correlation functions between the Coulomb and the $\delta$-repulsion are
rather minimal.

\section{Summary}

This report reviewed the physics of strong correlations in two-dimensional
small finite-size condensed-matter systems, such as electrons in quantum
dots and repelling bosons in harmonic traps. It was shown that strong
correlations in such systems relate to the appearance of symmetry breaking at 
the mean-field level of description. Particular attention was given to 
the similarities of symmetry breaking in these systems despite the 
different interparticle interactions (Coulombic repulsion in quantum dots 
versus a contact potential for neutral bososns in harmonic traps).

The universal aspects of symmetry breaking in small systems (including
nuclei and molecules in quantum chemistry) have been
exploited to develop a two-step method of symmetry breaking at the 
unrestricted Hartree-Fock level and subsequent symmetry restoration via
post Hartree-Fock projection techniques. In conjunction with 
exact-diagonalization calculations, the two-step method was used to describe
a vast range of strongly-correlated phenomena associated with particle 
localization and formation of crystalline (molecular) structures of electrons 
in quantum dots and bosons in harmonic traps. Due to their finite size, these
crystalline structures are different from the familiar rigid crystals of 
extended systems; they rather resemble and exhibit similarities with 
the natural 3D molecules (e.g., ro-vibrational spectra).

It was shown that strongly-correlated phenomena emerging from symmetry
breaking include:

(I) Chemical bonding, dissociation, and entanglement in quantum dot molecules
and in electron molecular dimers formed within a single anisotropic quantum 
dot, with potential technological applications to solid-state quantum-computing
devices.

(II) Electron crystallization, with localization on the vertices of concentric
polygonal rings, and formation of rotating electron molecules in 
circular quantum dots. At zero magnetic field, the 
REMs can approach the limit of
a rigid rotor; at high magnetic field, the REMs  are highly floppy,
with the rings rotating independently of each other.

(III) In the lowest Landau level, the rotating electron molecules are described
by parameter-free analytic many-body wave functions, which are an alternative 
to the composite-fermion and Jastrow/Laughlin approaches, 
offering a new point of view of the fractional quantum Hall regime 
in quantum dots (with possible implications for the thermodynamic limit). 

(IV) Crystalline phases of strongly repelling bosons. In the case of rotating 
traps and in analogy with the REMs, such repelling bosons form rotating boson 
molecules, which are energetically favored compared to the Gross-Pitaevkii 
solutions, in particular in the regime of vortex formation.

\ack

This work is supported by the US D.O.E. (Grant No. FG05-86ER45234).
Calculations were peformed at the Georgia Institute of Technology Center
for Computational Materials Science.

\appendix
\section{}
\setcounter{section}{1}

In this Appendix, we briefly review the single-particle wave functions
and associated energy spectra of a two-dimensional circular harmonic oscillator
under the influence of a perpendicular magnetic field $B$ (relevant to
the case of quantum dots) or under rotation with angular frequency $\Omega$
(relevant to the case of trapped atomic gases in rotating harmonic traps).
These single-particle wave functions and associated spectra are known as the
Darwin-Fock states and energy levels, after the names of the authors of
two original papers \cite{darw,fock} on this subject.

\begin{figure*}[t]
\centering\includegraphics*[width=9.0cm]{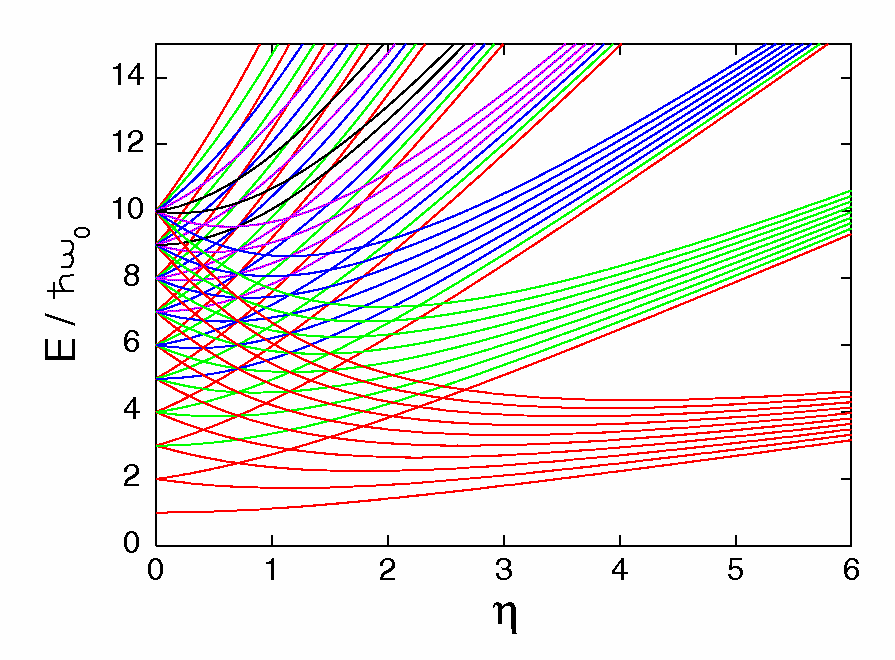}
\caption{(Color online) 
The Darwin-Fock single-particle energy levels of a 2D harmonic 
oscillator under the effect of a perpendicular magnetic field $B$
as a function of $\eta=\omega_c/\omega_0$, where $\omega_c$ is the cyclotron
frequency and $\omega_0$ is the frequency specifying the 2D harmonic
confinement. A specific color (online) indicates orbitals with the same 
number of radial nodes.
}
\label{dfmag}
\end{figure*}

\subsection{~~~~~~~~~~~~~~~Two-dimensional isotropic oscillator in a 
perpendicular magnetic field}
\label{appa1}

In this case, the Hamiltonian (for an electron of mass $m^*$) is given by:
\begin{equation}
H = \frac{1}{2m^*} ({\bf p} - \frac{e}{c} {\bf A})^2 + 
\frac{1}{2}m^*\omega_0^2 {\bf r}^2,
\label{hb}
\end{equation}
where ${\bf r}=(x,y)$ and $\omega_0$ is the frequency of the oscillator.
In the symmetric gauge, the vectror potential is given by 
${\bf A}=({\bf B} \times {\bf r})/2$, and the Hamiltonian \eref{hb} can be
rewritten in the form
\begin{equation}
H=\frac{{\bf p}^2}{2m^*} - \frac{1}{2} \omega_c \hat{l}
+ \frac{1}{2}m^*\tilde{\omega}^2 {\bf r}^2,
\label{hb2}
\end{equation}
where $\hat{l}=-\rmi \hbar (x \partial/\partial y - y \partial/\partial x)$
is the angular momentum operator of the electron (in the $z$ direction),
$\omega_c = eB/(m^*c)$ is the cyclotron frequency, and $\tilde{\omega} = 
\sqrt{\omega_0^2 + \omega_c^2 /4}$ is the effective-confinement frequency. 

The eigenfunctions of the Hamiltonian \eref{hb2} have the same functional
form as those of a 2D harmonic oscillator at B=0, but with an effective
frequency $\tilde{\omega}$, i.e., in polar coordinates
\begin{equation}
\phi_{n,l}(\rho,\theta)={\cal N}_{n,l} \rho^{|l|} \rme^{-\rho^2/2} 
\rme^{\rmi l \theta} L_n^{|l|} (\rho^2),
\label{hbwf}
\end{equation}
with $\rho = r/\tilde{l}$; the characteristic length $\tilde{l}$ is
given by $\tilde{l}=\sqrt{ \hbar/(m^* \tilde{\omega}) }$. In \eref{hbwf},
the index $n$ denotes the number of nodes in the radial direction, and
$l$ (without any subscript or tilde) denotes the angular-momentum quantum 
numbers; the $L_n^{|l|}$'s are associated Laguerre polynomials.

The single-particle energy spectrum corresponding to the Hamiltonian 
\eref{hb2} is plotted in \fref{dfmag}; the associated eigenenergies are
given by
\begin{equation}
\frac{E_{n,l}}{\hbar \omega_0} = (2n+|l|+1) \sqrt{1+\frac{\eta^2}{4}}
-\frac{l}{2}\eta,
\label{dfeb}
\end{equation}
with $\eta=\omega_c/\omega_0$. 

In the limit of $\omega_c/(2\omega_0) \rightarrow \infty$, one can neglect 
the external confinement, and the energy spectrum in \Eref{dfeb} reduces to
that of the celebrated Landau levels, i.e.,
\begin{equation}
E_{\cal M} = \hbar \omega_c ({\cal M} + \frac{1}{2}),
\label{ebll}
\end{equation}
where ${\cal M}=n+(|l|-l)/2$ is the Landau-level index.

We remark that the Landau levels are infinitely degenerate. The lowest
Landau level ${\cal M}=0$ contains all nodeless levels ($n=0$) with arbitrary 
positive angular momentum $l \geq 0$.

\begin{figure*}[t]
\centering\includegraphics*[width=9.0cm]{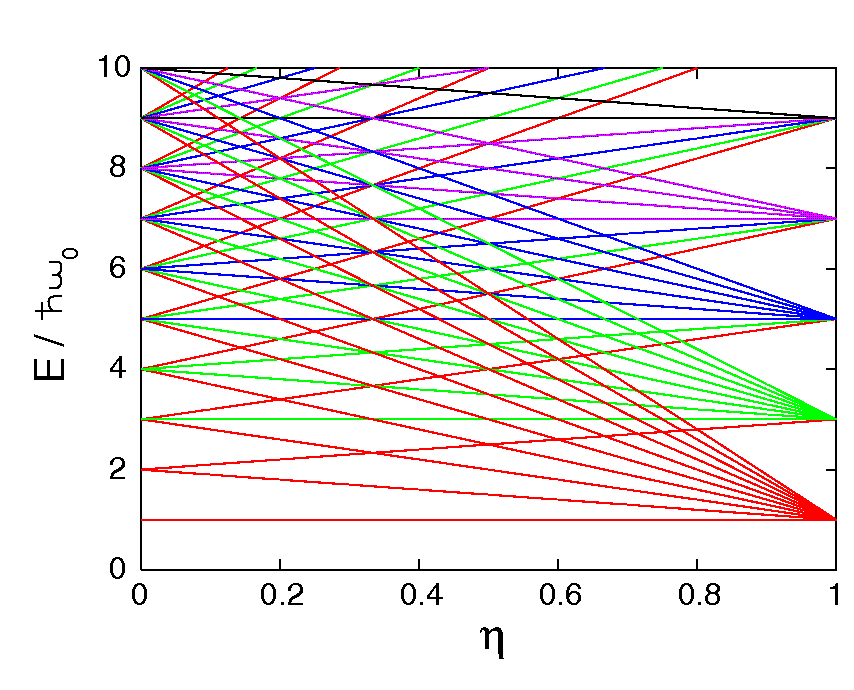}
\caption{(Color online) 
The Darwin-Fock single-particle energy levels of a 2D harmonic 
oscillator rotating with angular frequency $\Omega$
as a function of $\eta=\Omega/\omega_0$, where $\omega_0$ 
is the frequency specifying the 2D harmonic confinement.
A specific color (online) indicates orbitals with the same 
number of radial nodes.
}
\label{dfrot}
\end{figure*}

\subsection{~~~~~~~~~~~~~~Two-dimensional rotating harmonic oscillator}

In the case of a rotating isotropic oscillator, instead of the expression
\eref{hb2}, one has the following single-particle Hamiltonian:
\begin{equation}
H=\frac{{\bf p}^2}{2m} - \Omega \hat{l}
+ \frac{1}{2}m \omega_0^2 {\bf r}^2,
\label{hom2}
\end{equation}
where the mass of the particle (e.g., a bosonic or fermionic atom) is
denoted by $m$; $\Omega$ denotes the rotational frequency. 

From a comparison of the second terms in \eref{hb2} and \eref{hom2}, one 
derives the correspondence $\Omega \rightarrow \omega_c/2$. 

We note that, unlike the application of a perpendicular magnetic field, the 
rotation does not generate an effective confinement different from the 
original external one [compare the third terms between \eref{hb2} and 
\eref{hom2}]. As a result, the eigenfunctions of the Hamiltonian \eref{hom2} 
are given by the expressions \eref{hbwf}, but with $\rho=r/l_0$ where 
the characteristic length $l_0=\sqrt{\hbar/(m \omega_0)}$.

The single-particle energy spectrum corresponding to the Hamiltonian 
\eref{hom2} is plotted in \fref{dfrot} and the associated eigenenergies are
given by
\begin{equation}
\frac{E_{n,l}}{\hbar \omega_0} = (2n+|l|+1) -l \eta,
\label{dfeom}
\end{equation}
with $\eta=\Omega/\omega_0$. 

For  $\Omega/\omega_0 = 1$, the energy spectrum in \Eref{dfeom} reduces to
that of the corresponding Landau levels, i.e.,
\begin{equation}
E_{\cal M} = 2 \hbar \omega_0 ({\cal M} + \frac{1}{2}),
\label{eomll}
\end{equation}
where ${\cal M}=n+(|l|-l)/2$ is the Landau-level index.

As was the case with the perpendicular magnetic field, the Landau levels are 
infinitely degenerate, and the lowest Landau level ${\cal M}=0$ contains all 
nodeless levels ($n=0$) with arbitrary positive angular momentum $l \geq 0$.
However, unlike the magnetic-field case where $\hbar \omega_c$ depends on
$B$, the energy gap between the Landau levels in the case of rotation is
independent of $\Omega$ and equals $2\hbar\omega_0$.\\
~~~~~~~\\
~~~~~~~\\
~~~~~~~\\
~~~~~~~\\

\end{document}